\documentclass[twocolumn]{svjour2}          

\usepackage{graphicx}
\usepackage{hyperref}
\usepackage{amsmath}

\usepackage{amssymb}
\usepackage{float}
\usepackage[authoryear]{natbib}

\begin{document}

\title{Secular resonance sweeping and orbital excitation in decaying disks}


\author{Athanasia Toliou         \and
        Kleomenis Tsiganis        \and
        Georgios Tsirvoulis 
}


\institute{A. Toliou \at
              Aristotle University of Thessaloniki \\
              \email{athtolio@physics.auth.gr}           
           \and
           K. Tsiganis \at
              Aristotle University of Thessaloniki  \\
              \email{tsiganis@auth.gr}    
            \and
           G. Tsirvoulis \at
              University of Belgrade  \\
              \email{gtsirvoulis@aob.rs}       
}

\date{December 2019}

\maketitle

\begin{abstract}
 We revisit the problem of secular resonance sweeping during the dissipation of a protoplanetary disk and its possible role in exciting the orbits of primordial asteroids, in the light of recent models of solar system evolution. We develop an integrator that incorporates the gravitational effect of a uniformly (or not) depleting, axisymmetric disk with arbitrary surface density profile; its performance is verified by analytical calculations. The secular response of fictitious asteroids, under perturbations from Jupiter, Saturn and a decaying disk, is thoroughly studied. Note that the existence of a symmetry plane induced by the disk, lifts the inherent degeneracy of the two-planet system, such that the ``$s_5$'' nodal frequency can also play a major role. We examine different resonant configurations for the planets (2:1, 3:2, 5:3), disk models and depletion scenarios. For every case we compute the corresponding time paths of secular resonances, which show when and where resonance crossing occurs. The excitation of asteroids, particularly in inclination, is studied in the various models and compared to analytical estimates. We find that inclination excitation in excess of $\sim 10^{\circ}$ is possible in the asteroid belt, but the nearly uniform spread of $\Delta i\sim 20^{\circ}$ observed calls for the combined action of secular resonance sweeping with other mechanisms (e.g. scattering by Mars-sized embryos) that would be operating during terrestrial planet formation. Our results are also applicable to extrasolar planetary systems.

\keywords{secular resonance \and protoplanetary disk \and asteroid belt }
\end{abstract}

\section{Introduction}\label{sec:intro}
The sweeping of secular resonances, as a mechanism for raising the eccentricities and inclinations of small bodies in the solar system, was originally suggested by \citet{1980Icar...41...76H} and \citet{1981Icar...47..234W} who referred to it as ``scanning secular resonance theory''. The basic idea behind that theory is that, as the protoplanetary gas disk dissipates away, its gravitational influence on planets and minor bodies decreases, as does its contribution to the values of the secular precession frequencies. If, at some point in time, $t_0$, one of the fundamental frequencies of the planetray system (say, $f_i$) is equal to the precession frequency of an asteroid $f_i=f(a_0)$ for some value $a_0$ of semi-major axis, then we have a (linear) secular resonance at $a_0$. As time goes by, both the value of $f_i$ and the function $f(a)$ change, so that, for $t=t_0+\delta t$, the equality will hold for a different value of $a'=a_0+\delta a$. For a steady change of $a'(t)$, the resonance effectively ``scans'' the inner solar system. \\

This idea was later employed by \citet{1991CeMDA..52...57L}, who studied the effect of secular resonances analytically and \citet{1997Icar..129..134L}, who performed numerical simulations of test particles inside the main asteroid belt. Their results showed efficient excitation of the eccentricities of solar system asteroids but not of their inclinations. \citet{2000AJ....119.1480N} integrated the orbits of fictitious asteroids, adding also the effect of gas drag and considered three types of depletion of the protoplanetary disk; (a) uniform mass loss, exponential in time, (b) from ``inside-out'', in which the protoplanetary disk begins to dissipate closer to the Sun and progressively loses mass at larger and larger heliocentric distances, and (c) a gap opening at Jupiter's location and spreading both inwards and outwards, with a timescale longer than $3 \cdot 10^5$~yr. The surface density profile of the gas disk was adopted from the \textit{Minimum Mass Solar Nebula} (MMSN) model \citep{1977MNRAS.180...57W,1981PThPS..70...35H}. They found that, for models (b) and (c), asteroid inclinations can indeed get excited by resonance sweeping, while eccentricity excitation happens in all depletion models.  \citet{2001EP&S...53.1085N,2002Icar..159..322N} performed new simulations in which the disk was placed on the invariant plane of the planets, instead of the 'ecliptic' as 
before, and the inclination excitation of asteroids was reduced. In all these studies, the planets were considered to follow their current orbits.\\

Following the development of the Nice model \citep{2005Natur.435..459T,2005Natur.435..462M,2005Natur.435..466G}, it became widely accepted that the current orbits of the giant planets are not primordial, but rather the outcome of a dynamical instability. \citet{2007Icar..191..434O} tried to reproduce the current orbital distribution of asteroids in the main belt, starting from an initially flat disk of particles and assuming Jupiter and Saturn to follow circular and nearly co-planar orbits just beyond their mutual 2:1 mean motion resonance (MMR); they also adopted the MMSN model, which, however, is likely incompatible with the Nice model \citep{2009ApJ...698..606C}. Their results showed that only the ``inside-out'' clearing of the gas disk  would probably work, but only if we assume very long depletion timescales, which is not supported by recent astrophysical models (e.g.\ \cite{2014prpl.conf..475A}). Subsequent versions of the Nice model \citep{2007AJ....134.1790M,  2011AJ....142..152L} assumed the pre-instability configuration of the planets to be {\it multi-resonant}, in agreement with hydrodynamical simulations of the preceeding phase of gas-driven migration. \\


\citet{2011Natur.475..206W} introduced the so-called ``Grand Tack'' model, in which Jupiter and Saturn migrate extensively in their nascent gaseous disk, effectively sweeping through the main belt region, causing depletion and excitation of primordial asteroids. \citet{2016ApJ...833...40I} presented an alternative to the Grand Tack, in which Jupiter and Saturn migrate only as much as necessary to be captured on  mildly chaotic, resonant orbits. The chaotic evolution of Jupiter and Saturn results in stochastic jumps of their precession rates and makes secular resonances ``jump'' erratically through the main belt. As the authors note, this elegant mechanism is not always effective, as such a planetary state does not seem to be a generic outcome of capture simulations. \citet{2018ApJ...864...50D} studied the possible excitation of the asteroids during the very last violent episode of solar system formation: the giant planet instability phase of the Nice model. They found that, if Jupiter gets excited enough during the instability phase, the erratic variations of its forced eccentricity and inclination vectors can sufficiently excite the eccentricities and inclinations of the asteroids. This mechanism requires a rather substantial change of Jupiter's inclination to work, which, not surprisingly, happens only in a small subset of simulations; this is, unfortunately, typical in all such studies, in which one tries to simultaneously match several solar system constraints. Note that the mass of the belt was also found to decrease only by a factor of $\sim 10$, which led the authors to suggest that either the belt contained very little mass to begin with, or that another mass-depletion mechanism had acted prior to the giant planets' instability.\\

In light of these new models of solar system evolution, as well as of more refined astrophysical models of disk evolution, it seems that there is a need to revisit the secular resonance sweeping mechanism, assuming more realistic planetary configurations and disk depletion models, before we can rule on the most reasonable chain of events that likely sculpted our solar system. Moreover, the sweeping mechanism is likely relevant for extrasolar systems, especially those containing two resonant giant planets. The present paper constitutes the first part of our study, namely the understanding of how secular resonance sweeping can proceed in different disk models and for various planetary configurations; we restrict ourselves here by assuming two-planet configurations (Jupiter and Saturn). We develop a code that calculates the gravitational acceleration by a decaying, axisymmetric, 3-D disk of arbitrary surface density profile and accordingly modify the SyMBA (Symplectic Massive Body Algorithm) integrator\citep{1998AJ....116.2067D}. In  \autoref{sec:sec_freq} we test our code against analytical computations (linearized theory), using an analytically tractable galactic potential-density pair. In \autoref{sec:maps_n_paths1} we use our computations to construct maps of secular resonances. These maps show the timing and location of resonances occurrences, for every planetary system studied. In sections \ref{sec:maps_n_paths2} and \ref{sec:photoevaporation} we perform numerical simulations of test particles, under the gravitational influence of a depleting (uniformly or not) protoplanetary gas disk, computing their resulting excitation. Finally, in \autoref{sec:concl} we discuss the implications of our results for the formation of the solar system and other planetary systems.\\

\section{Secular motion in a massive disk}\label{sec:sec_freq}

In this section we will first review the basics of linear secular theory for a particle moving inside a massive, axisymmetric disk, the mass of which may decrease with time, under the additional perturbations from two (giant) planets. Subsequently, we discuss the development of our numerical code and compare results on the secular motion of test particles, obtained by said code and by the linear theory. 

\subsection{Linear secular theory}\label{subsec:linear}

We consider a particle that orbits the Sun, while also experiencing the gravitational potential of an axisymmetric, massive disk. The total potential is
\begin{equation}\label{eq:veff}
V_{\text{eff}}=V_{\odot}+V_d
\end{equation}
and the linearized motion around a circular orbit on the symmetry plane (i.e.\ the invariant plane of the system) is a linear superposition of two harmonic oscillations, with frequencies 

\begin{equation}\label{eq:kappasqr}
\kappa^2=\frac{\partial^2 V_{\text{eff}}}{\partial r^2}\bigg|_{r_0,0}  ~~~~ , ~~~~
v^2=\frac{\partial^2 V_{\text{eff}}}{\partial z^2}\bigg|_{r_0,0},
\end{equation}

\noindent 
where $(r,z)$ denote heliocentric cylindrical coordinates. The {\it epicyclic} and {\it vertical} frequencies, together with the mean motion, $n$, define the precession frequencies through
\begin{equation}\label{eq:varpi}
\dot{\varpi}=n-\kappa ~~~~ , ~~~~ \dot{\Omega}=n-v.
\end{equation}

In linear secular theory for two planets and a test particle on an interior orbit, the Laplace-Langrange solution yields the precession frequencies of the longitude of perihelion $\varpi$ and of the longitude of the ascending node $\Omega$ through

\begin{equation}\label{eq:secular_gs}
\begin{split}
g_f (a) &=n\frac{1}{4}\sum_{j=1}^{2}\frac{M_j}{M_{\odot}}\left(\frac{a}{a_j}\right)^2b^{(1)}_{3/2}(a_j)\\
s_f (a) &=-n\frac{1}{4}\sum_{j=1}^{2}\frac{M_j}{M_{\odot}}\left(\frac{a}{a_j}\right)^2b^{(1)}_{3/2}(a_j),
\end{split}
\end{equation} 
where $M_j$ is the mass of planet $j$, $a_j$ its semi-major axis, $b^{(1)}_{3/2}(a_j)$ is a Laplace coefficient and $M_{\odot}$ the mass of the Sun. Summing up all contributions, in the linear approximation, the precession frequencies for a particle are given by 

\begin{equation}\label{eq:varpi2}
\dot{\varpi}\,= g\, = \, g_f + n-\kappa ~~~~ , ~~~~ \dot{\Omega} \, = s \, = \,s_f + n-v.
\end{equation}

Following standard notation, the planetary system's fundamental frequencies are denoted by $g_{5,6}$ and $s_{5,6}$, respectively; we remind the reader that $s_5$ vanishes in the two-planets problem (no disk), due to the conservation of angular momentum. In linear theory, $g_f (a_*) \approx g_{5,6}$ or $s_f (a_*) \approx s_{5,6}$ implies an infinite amplitude of the corresponding forced term at $a=a_*$, which signifies the presence of secular resonance. \\

Note that the contribution of the disk can be also computed through \autoref{eq:secular_gs}, if we assume the disk to be composed of $j$ adjacent, circular rings of mass $M_j$ and extend the sum appropriately. If the gas disk is assumed to dissipate uniformly, such that each ring loses mass at the same rate, then the disk's contribution decreases also uniformly and the functions $g(a,t)$ and $s(a,t)$ converge to those given by the Laplace-Lagrange solution of the two-planets system, $g_f (a)$ and $s_f (a)$. In such a setting, the locations of secular resonances change with time.\\

Our goal is to build a numerical code that incorporates the gravitational effects of a massive disk, with arbitrary surface density profile and depletion type, on planets and particles. At the same time, we aim to probe the extent of applicability of linear secular theory, which would allow us to easily demonstrate how secular resonances sweep through the `asteroid belt' in different solar system models. To that end, we will compare the values of the precession frequencies of asteroids, as given by our numerical code, to the analytical estimates presented above. For this comparison we will use the well-known Miyamoto-Nagai (hereafter MN) density-potential pair \citep{1975PASJ...27..533M}, which is analytically tractable. The potential reads

\begin{equation}\label{eq:MNpot}
V_{MN}\left(r,z\right)=\frac{GM}{\sqrt{r^2+\left(\alpha+\sqrt{z^2+b^2}\right)^2}},
\end{equation}

\noindent
where $\alpha$ and $b$ are parameters that define the shape of the axisymmetric mass distribution mass. Changing their values, we can go from Plummer's \citep{1911MNRAS..71..460P} spherically symmetric potential ($\alpha=0$) to Kuzmin's \citep{1957PTarO..33...75K} razor thin disk ($b=0$). The mass parameter $M$ is chosen such that it agrees with MMSN estimates. Integrating the MMSN profile $\Sigma_{\text{MMSN}}=1700 \frac{\text{g}}{\text{cm}^2} \left(\frac{r}{1AU}\right)^{-1.5}$ up to $r=50$~AU, we derive our nominal mass parameter $M=\mathcal{M} \approx 0.017\ M_{\odot}$. 

\subsection{Numerical representation of a disk}\label{subsec:numerical}

In order to account for the gravitational effects of the disk, we developed a subroutine that adds the disk-induced accelerations in SyMBA, respecting its symplectic nature. Two options are available: (a) use of analytical expressions for the derivatives of a known disk potential, and (b) computation of disk-induced accelerations on a grid from its density profile and use of linear/splines interpolation. In the case of non-uniformly, time-varying density profiles, interpolation in time is also performed, using a set of pre-computed `snapshots' of the disk's density profile.

\begin{figure*}
 \includegraphics[width=0.45\textwidth]{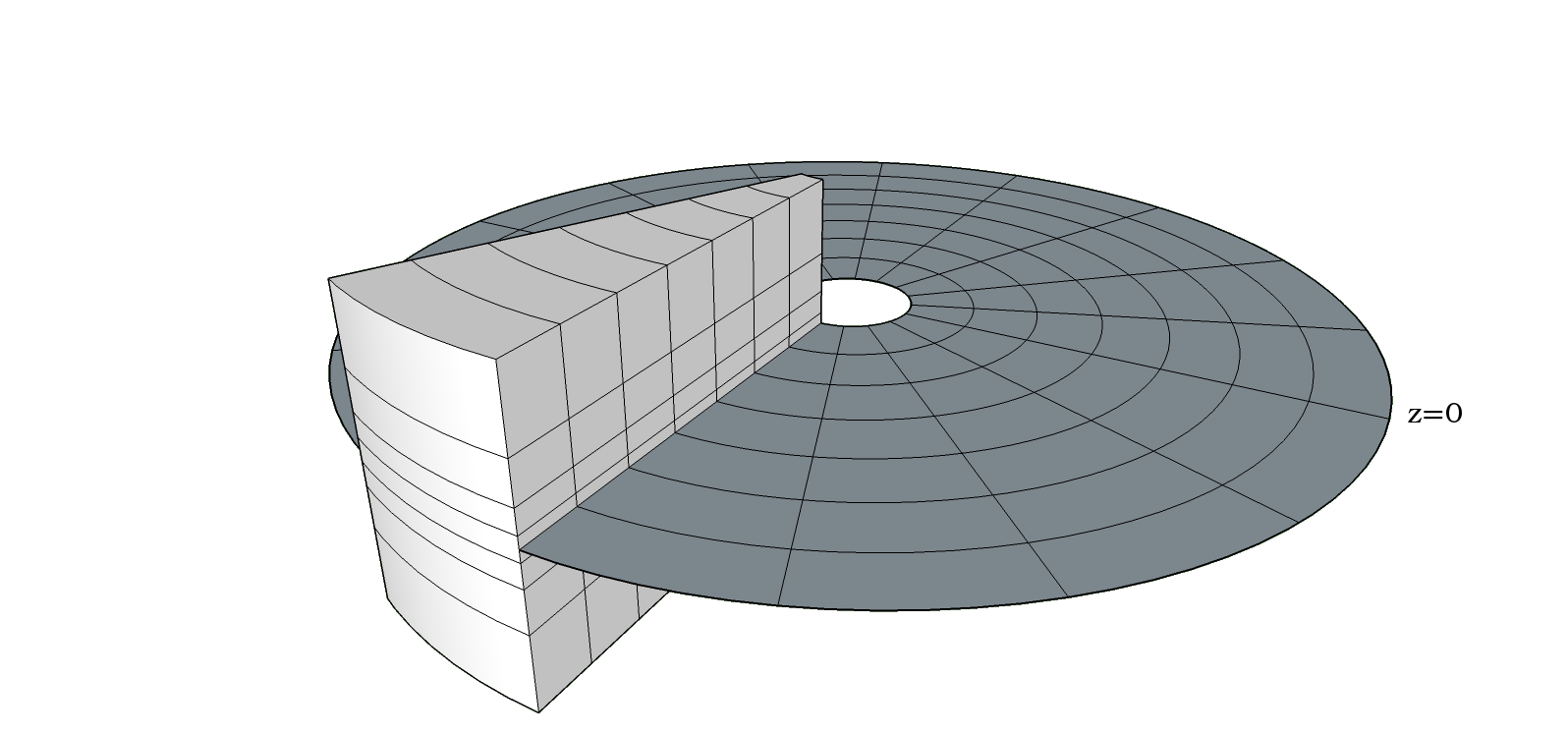}
 \includegraphics[width=0.45\textwidth]{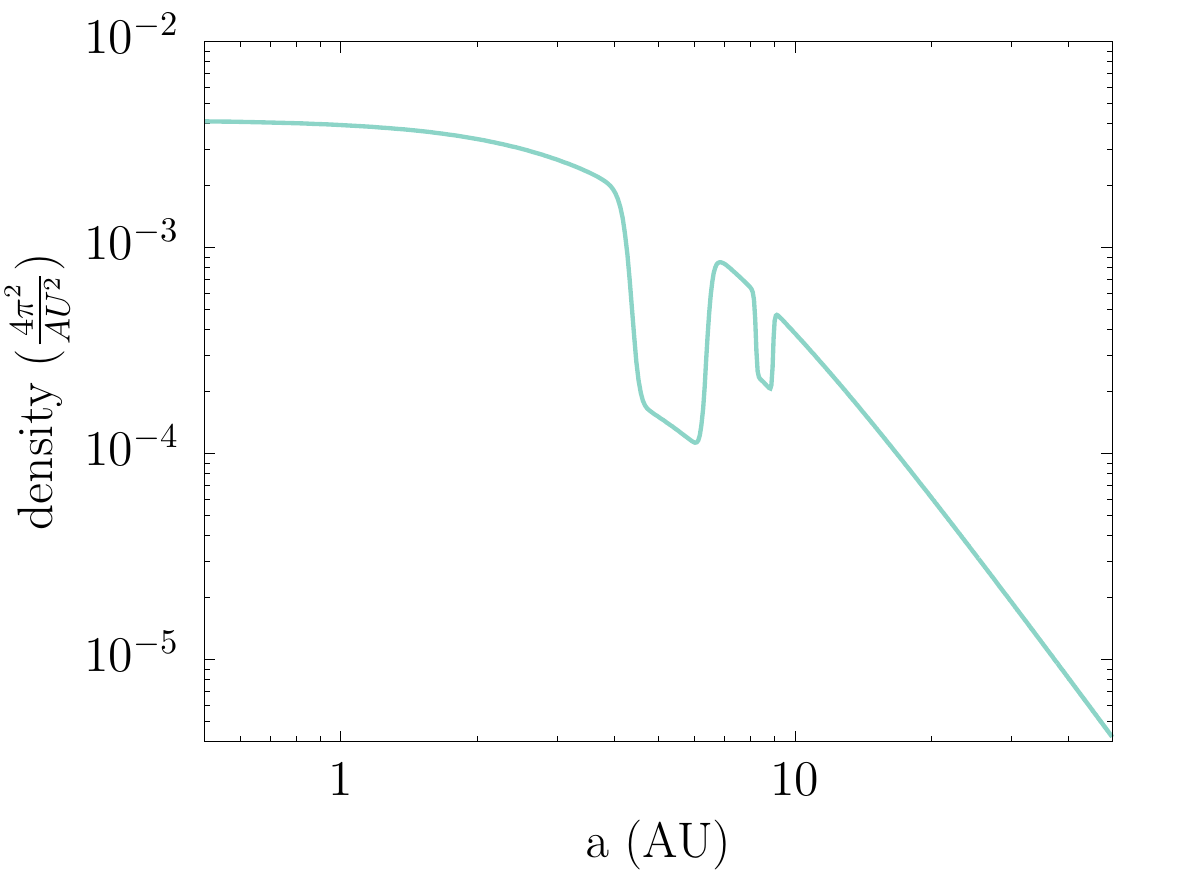}
 \caption{(left) Representation of the disk's 3-D computational grid. In this example, the cells close to $z=0$ plane have smaller width, in order to increase resolution at small $i$'s. (right) Density profile of a MN disk, multiplied by the generalized bell-shaped membership function to account for gaps at the assumed locations of Jupiter and Saturn.}
 \label{fig:grid3d}
\end{figure*}

\noindent
For (b), we define a three-dimensional grid in cylindrical coordinates $(r_i,z_i)$ that refer to the center of each cell, $i$ (see \autoref{fig:grid3d}). We choose the inner and outer boundaries of the disk in both directions, and set the accelerations outside the grid to be equal to those at the boundaries. In fact, we adjust the values at the edges of the grid, by multiplying the density with a transition function (\citet{1999MNRAS.304..793C}), for smoothness. In our computations we usually set $r_{\text{in}}=0.05$~AU and $r_{\text{out}}=50$~AU. The resolution is determined by the number of rings ($n_r$ and $n_z$), used to represent the disk. The choice of $r_{\text{out}}$ is considered realistic for a disk that photo-evaporates by the radiation of neighboring stars, in the birth cluster of our Sun \citep{2010ARA&A..48...47A}. We set  $z_{\text{max}}=\pm 0.1 r_{\text{out}}$. In the azimuthial direction, an angular width of $\Delta \phi =1^\circ$ is usually considered. \\

Protoplanetary disks are typically assumed to be `thin', i.e.\ having a dimensionless scale-height $H$ of a few per cent, and in hydrostatic equilibrium in the $z-$direction. For a given surface density profile, $\Sigma (r)$, this implies a Gaussian profile for the spatial density 
\begin{equation}
\rho=\rho_0 (r)\, \exp (-\frac{z^2}{2h^2}),
\end{equation}
where $\rho_0 (r)$ is the density at $z=0$ and $h=Hr$ is the local vertical dimension of the disk. Knowing $\rho (r,z)$ we can compute the mass contained in each cell and sum up the individual contributions from all cells, to compute the acceleration in a given cell. The singularity related to a cell's own gravity can be removed with the use of elliptic integrals, as described in \citet{2012CeMDA.114..365H} and \citet{2014CeMDA.118..299H}. Note that the interaction between massive planets and the gas disk leads to transport of angular momentum from the inner to the outer part of the disk \citep{1979MNRAS.186..799L,1980ApJ...241..425G}, which tends to open a gap at the position of the planet. We model this by multiplying the chosen surface density profiles with the generalized bell-shaped membership function 

\begin{equation}\label{eq:bell}
f(x,a,b,c)=\frac{1}{1+\left|\frac{x-c}{a}\right|^{2b}},
\end{equation}

\noindent
modified appropriately in each case, to match the position, width and depth of the gap around each planet (see \autoref{fig:grid3d}).\\

\begin{figure*}
\centering
 \includegraphics[width=0.4\textwidth]{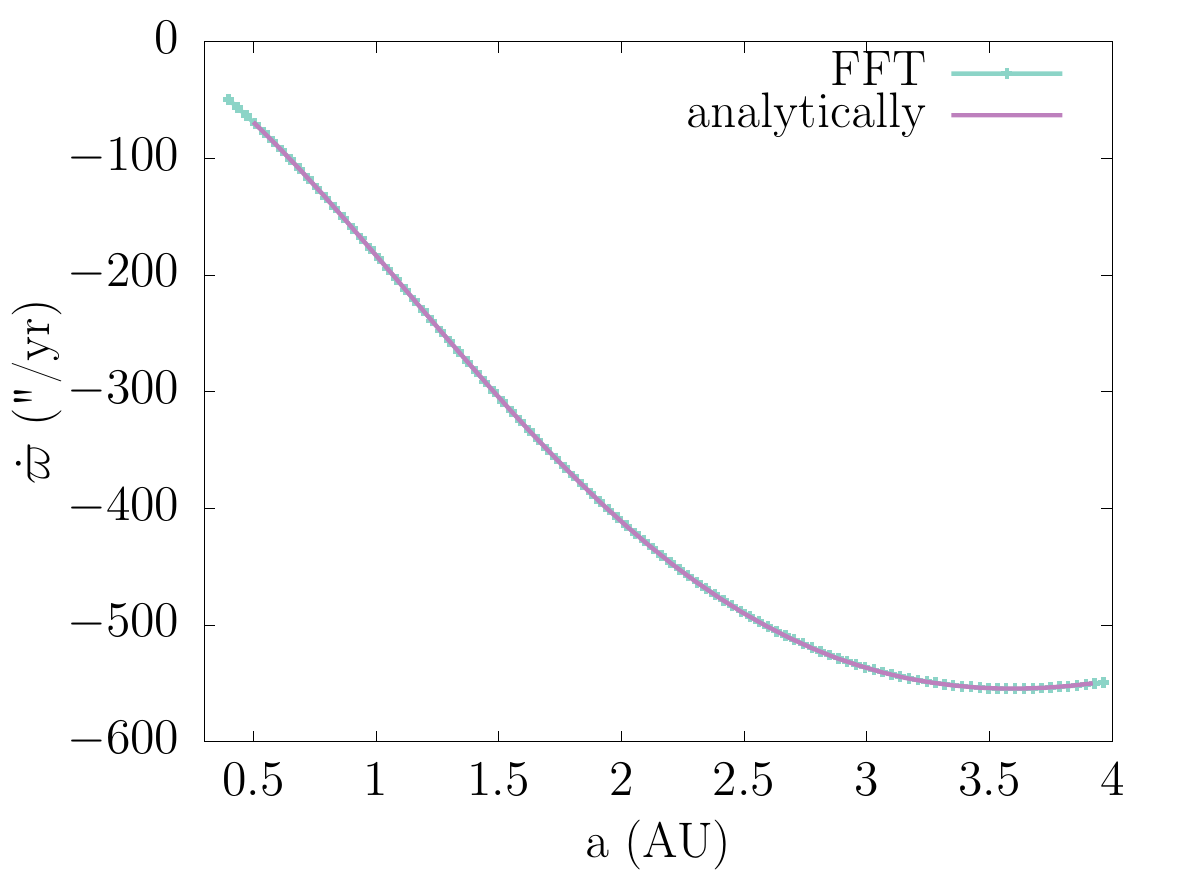}\includegraphics[width=0.4\textwidth]{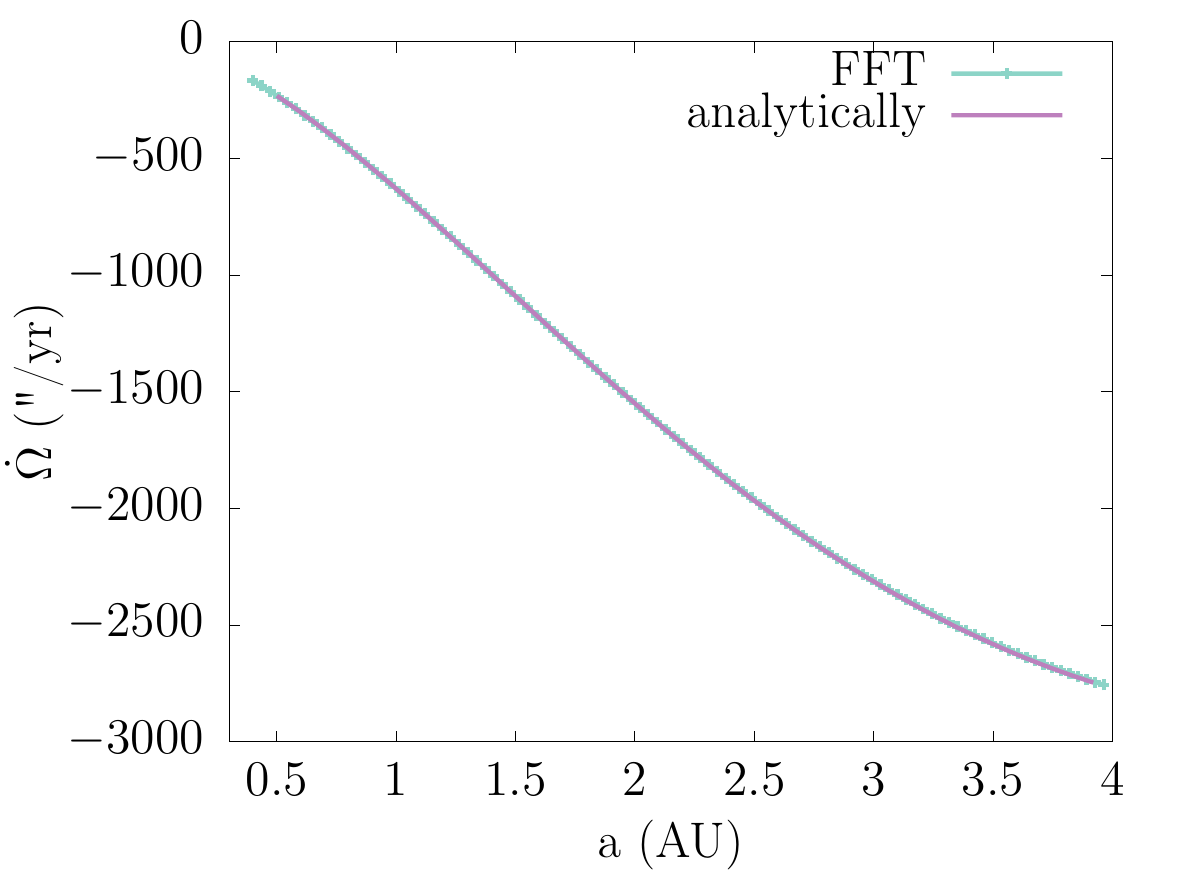}
  \includegraphics[width=0.4\textwidth]{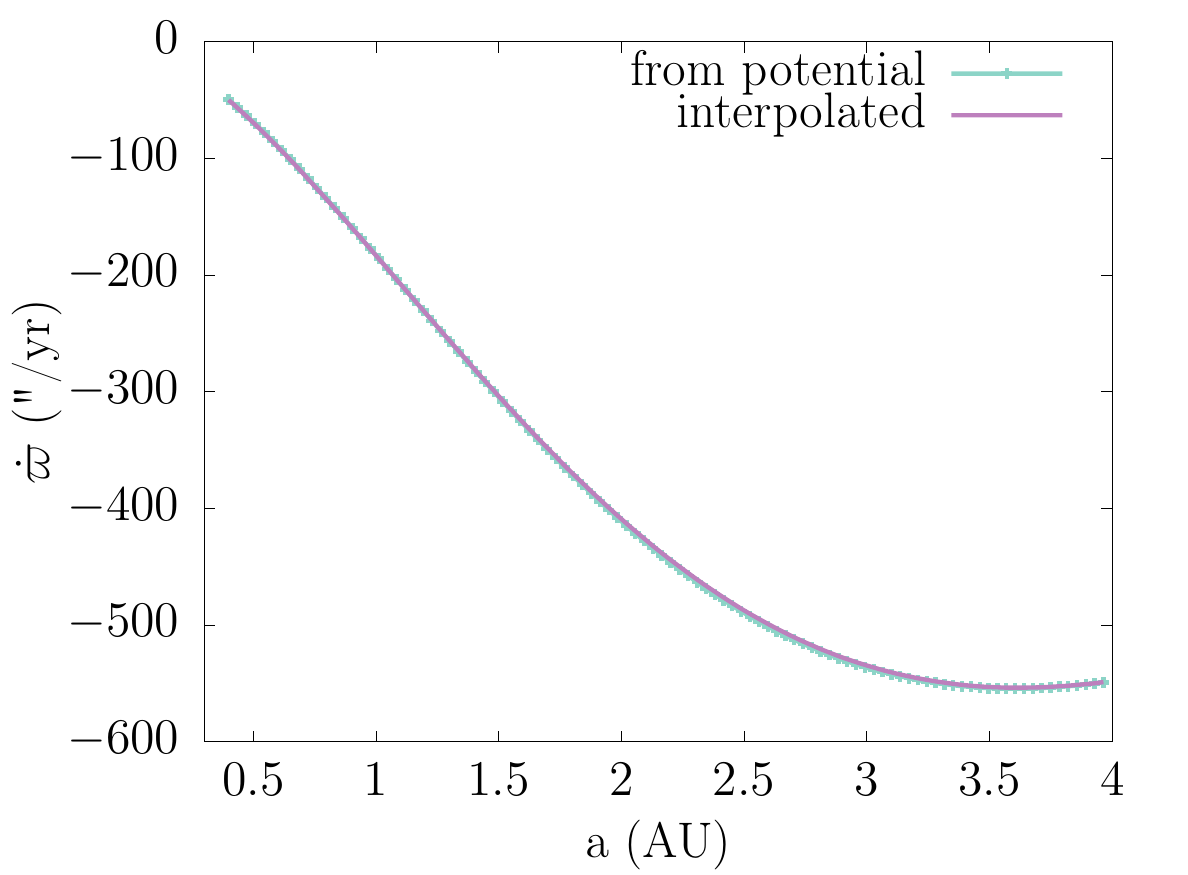}\includegraphics[width=0.4\textwidth]{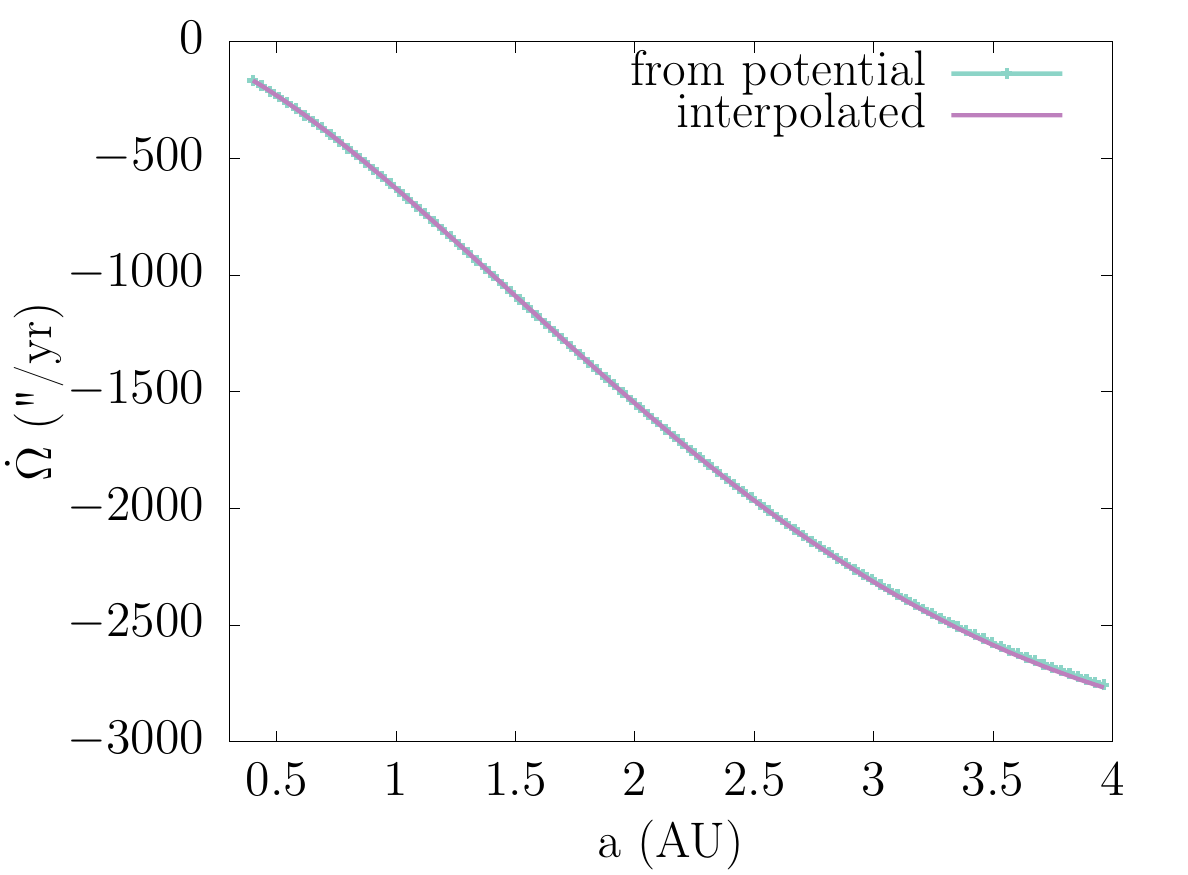}
   \includegraphics[width=0.4\textwidth]{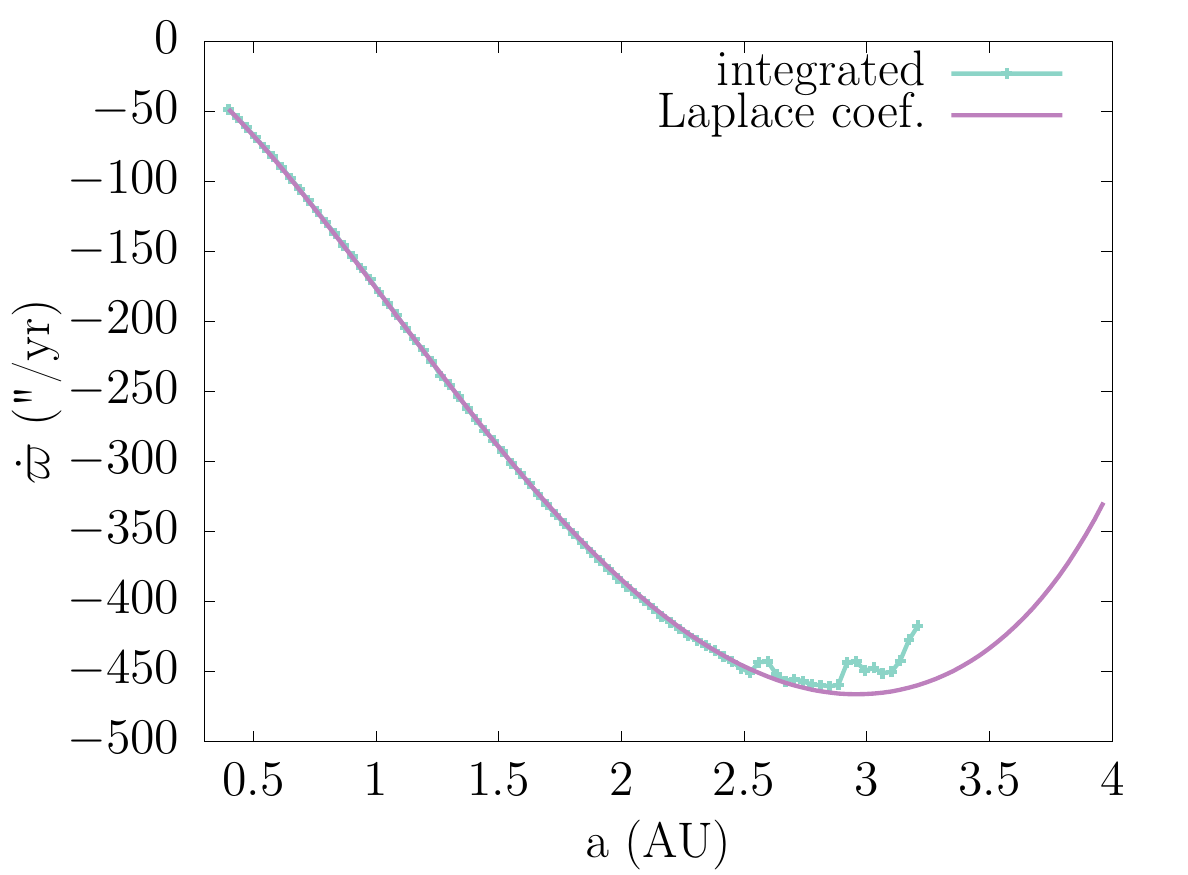}\includegraphics[width=0.4\textwidth]{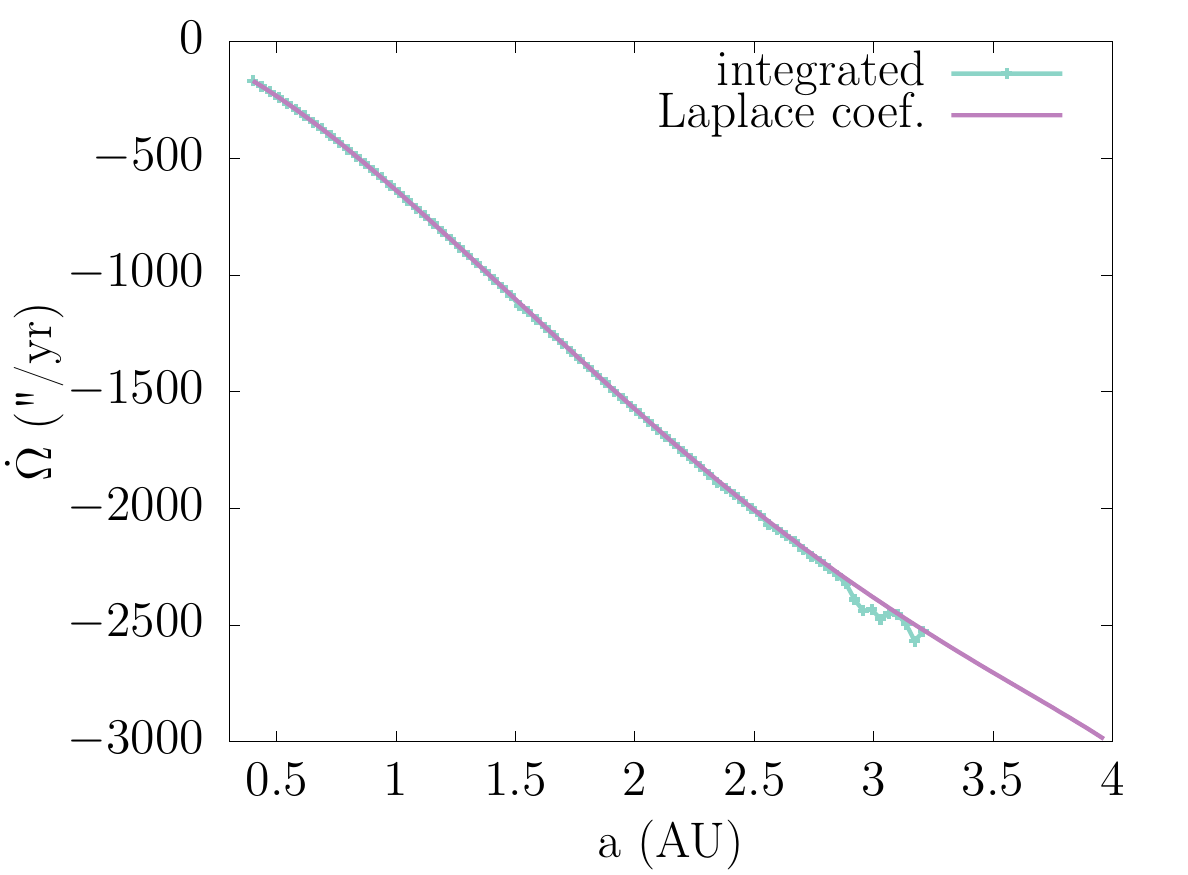}
   \includegraphics[width=0.4\textwidth]{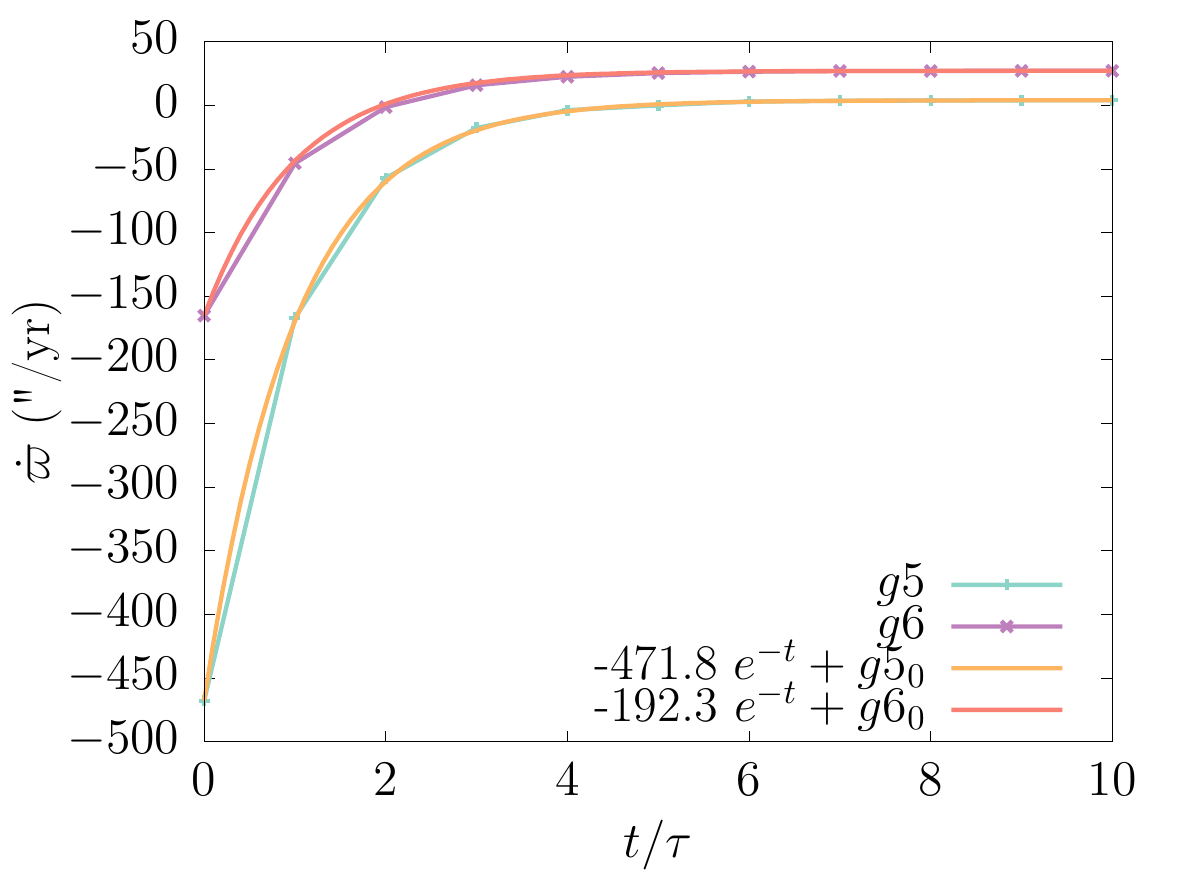}\includegraphics[width=0.4\textwidth]{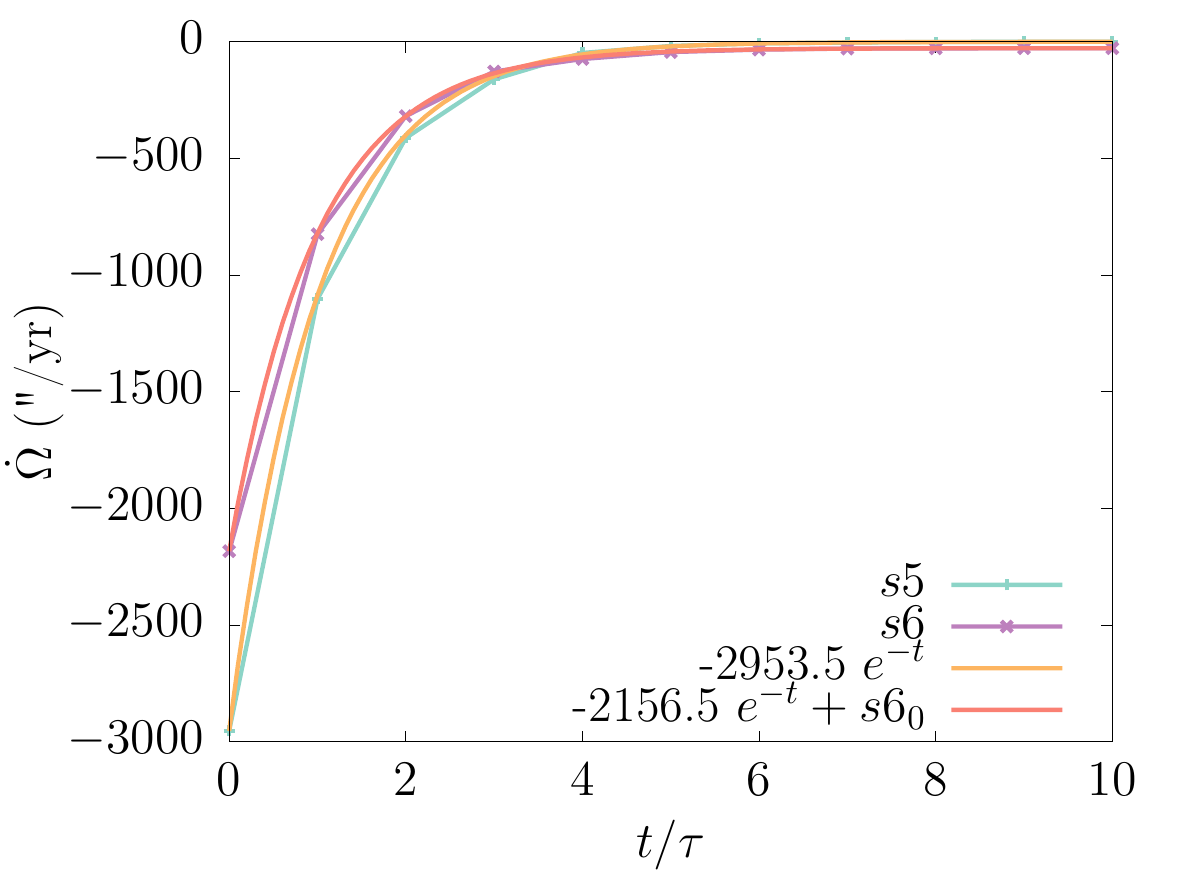} 
   \caption{(1st row) Precession frequencies of test-particles in a disk (no planets), computed by linear theory or numerically, but using the `potential'  representation of the disk. (2nd row) Same as above, but using the `interpolated' representation of the disk. (3rd row) Adding the planet's contribution, in linear theory and in the numerical model. (4th row) Time evolution of the values of the fundamental frequencies of the planets, in an exponentially decaying MN disk.}
 \label{fig:freq_anal_comp1}
\end{figure*}

The accelerations given by the disk are added as usual `velocity kicks' in SyMBA, following the native symplectic scheme. We used our modified version of SyMBA to integrate the orbits of 100 test particles with $0.4\leq a\leq 4$~AU and $e,i\sim 10^{-3}$, assuming a MN disk with $M=\mathcal{M}$, $a=5$~AU and $b=0.5$~AU, for a time corresponding to 1~Myr. The integration was performed twice, using both representations described above; the `potential' one, (a), and the `interpolated' one, (b). In both cases, the secular frequencies of the asteroids were calculated from the orbital elements' time series, using Fast Fourier Transform (FFT)\footnote{We used the code developed by D. $\text{Nesvorn}\grave{y}$ that can be found at \url{https://www.boulder.swri.edu/~davidn/fmft/fmft.html}}. \autoref{fig:freq_anal_comp1} presents the comparison of the frequency functions, computed by either of the two numerical approximations and by linear secular theory, assuming the same disk. The agreement is nearly perfect, which suggests that our code works efficiently. \\

We then added Jupiter and Saturn in our model, placing them near their mutual 2:1 MMR with small eccentricities and inclinations ($a_j\simeq5.4$~AU, $e_j\simeq0.02$, $i_j\simeq0.03^{\circ}$, $a_s\simeq8.6$~AU,  $e_s\simeq0.03$ and , $i_s\simeq0.08^{\circ}$). Repeating the integrations, we calculated again the secular frequencies of the asteroids using FFT and compared the result to that of the complete linear theory (see Equation 5); this is also shown in \autoref{fig:freq_anal_comp1}. As expected, using the linear theory for the planets' contribution (Equation 4) is not valid in the vicinity of MMRs, most notably around $2.5~$AU (3:1 MMR) and beyond $3.2~$AU (2:1 MMR). Note, however, that the difference between the numerical and the analytical solution is not greater than $\approx 10\%$ in the worst case, since the massive disk presents the dominant contribution. Hence, the linear approximation can in fact be used, for the purpose of understanding the time evolution of the frequency functions, as the disk dissipates. We have confirmed that, for a uniform exponential mass decrease, the frequency functions decay also exponentially (see \autoref{subsec:linear}) and converge to the solution of the two-planets problem.     \\

Focusing on the planets, we calculated their fundamental precession frequencies, for various disk masses, using FFT. Placing Jupiter and Saturn close to their current orbits, we took snapshots of the disk, separated by one e-folding time, and integrated the system with a `frozen' disk, in order to compute the planetary frequencies. The time evolution of $g_{5,6}$, and $s_{5,6}$ are shown in the last panel of \autoref{fig:freq_anal_comp1}. As expected, the disk's contribution decays exponentially with time, until the frequencies converge to their current values. As can be easily seen on this plot, the $s_5$ secular frequency is not null, when the disk's mass is non-zero and, hence, it can be very important for the secular dynamics of asteroids during that period. The possible importance of the $s_5$ resonance was also recently noted by \citet{2019Ic..Baguet}, in their study of the effect of a massive {\it planetesimal} disk on the dynamics of primordial Kuiper Belt Objects (KBOs). The presence of the disk has an effect similar to that of an oblate central body, i.e.\ it defines a natural plane of reference and lifts the degeneracy inherent in the two-planets secular theory. In this respect, the secular effect of a ring-like disk, interior to the orbits of the planets, could be approximated by a generalized Laplace-Lagrange theory for oblate central body, with a properly defined `effective' $J_2$ value.  

\begin{figure*}
\centering
\includegraphics[width=0.45\textwidth]{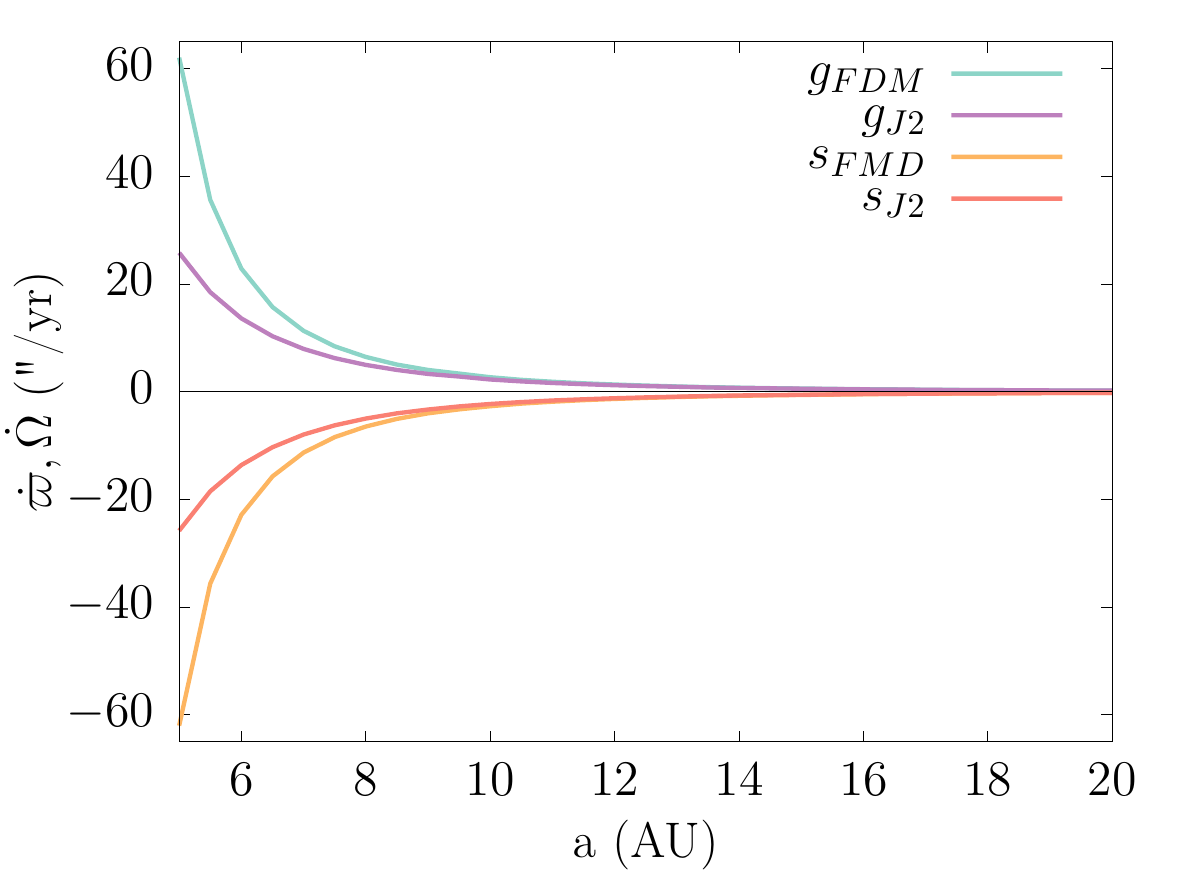}
 \caption{Comparison between the secular frequencies caused by the oblateness of the central body vs. by a truncated FMD.}
 \label{fig:j2}
\end{figure*}

We demonstrate this similarity, using a finite Mestel disk \citep{2012ApJ...747..106S}, with surface density profile 
 \begin{equation}
\Sigma_{\text{FMD}}\left(r\right)=\frac{M}{2\pi\alpha \, r}\arccos{\left(\frac{r}{\alpha}\right)}, \qquad r\le \alpha,
 \end{equation}
 \noindent
where $\alpha$ is the radius of the disk and $M$ its total mass. Using analytical expressions for the forces, given in \citet{2012ApJ...747..106S}, we integrated the orbits of 31 particles with $5\leq a \leq 20$~AU and calculated their secular frequencies, for a disk extending up to 4~AU. For a collection of rings surrounding a point-mass, an effective $J_2$ coefficient can be defined by  

\begin{equation}\label{eq:j2}
J_2=\frac{1}{2}\sum_{i=1}^{k}\frac{m_i a_i^2}{M_{\odot}\alpha^2},
\end{equation} 
where $a_i$ the position of the planet/ring with mass $m_i$ and $k$ the number of rings. We computed $J_2$ using $k=20$ rings for our Mestel disk and integrated again the same 31 particles as before, adding that $J_2$ term to the Sun.\autoref{fig:j2} shows the comparison between the two experiments. As expected, the frequencies differ close to the disk -- as the Mestel potential is quite different from the simple $J_2$ one -- but converge at larger distances.

\section{Determining the paths of secular resonances}\label{sec:maps_n_paths1}

From the analysis presented above, we can conclude that, if the disk decays uniformly in time no matter what its surface density profile, we only need to calculate the precession frequencies of asteroids for one time instant (e.g.\ at $t=0$ when the disk has its full mass) and subsequently multiply their values by the decay function; in the case of exponential decay, this is $\exp(-t/\tau)$. Adding the planetary contribution (linearized) we end up with a simple semi-analytical model of the time evolution of both the planetary and the asteroidal secular frequencies. Hence, we are in a position to derive the paths of secular resonances inside the disk; we restrict ourselves to the part interior to the orbit of Jupiter.   \\

\begin{figure*}
 \centering
 \includegraphics[width=0.5\textwidth]{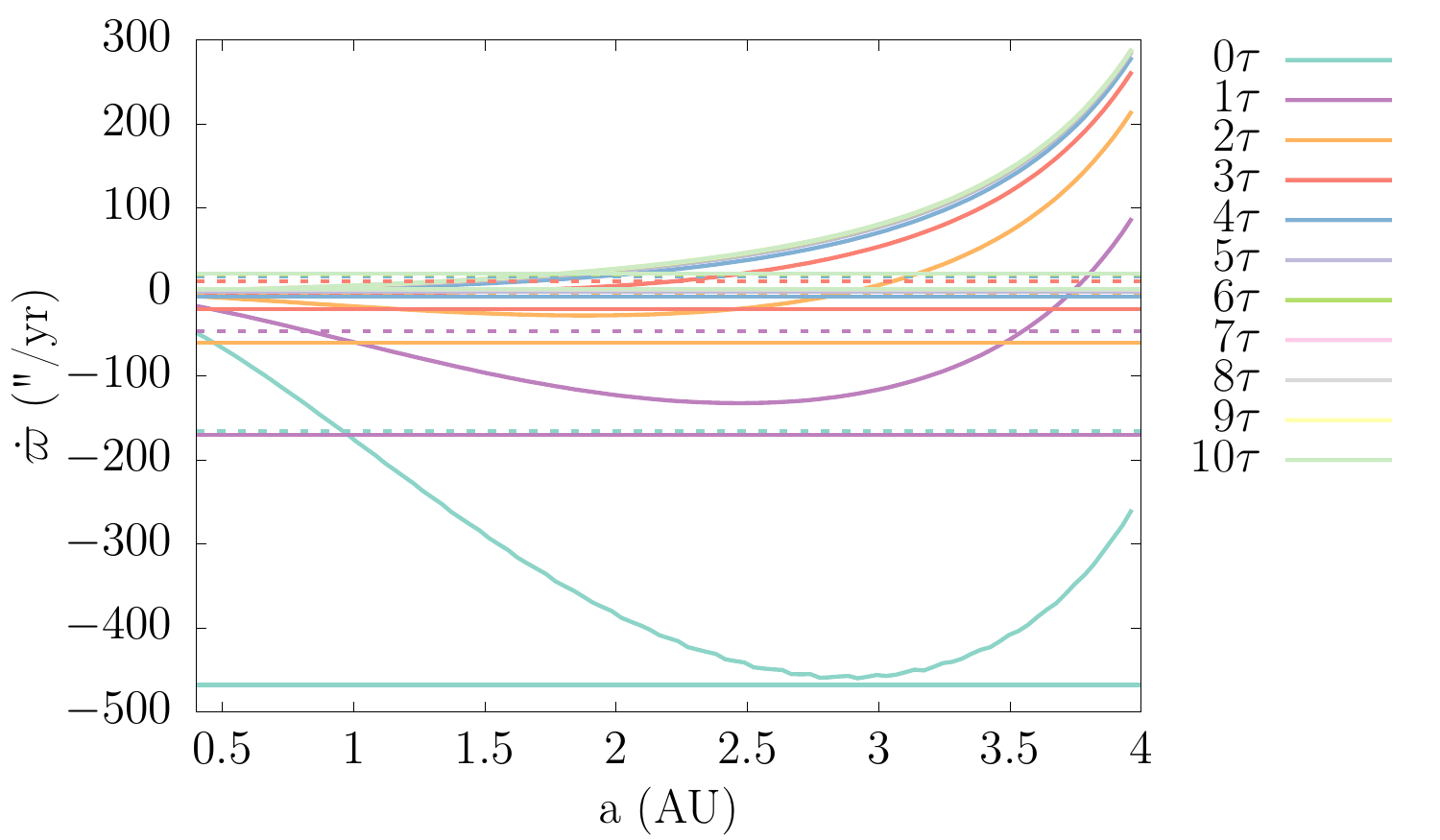}\includegraphics[width=0.39\textwidth]{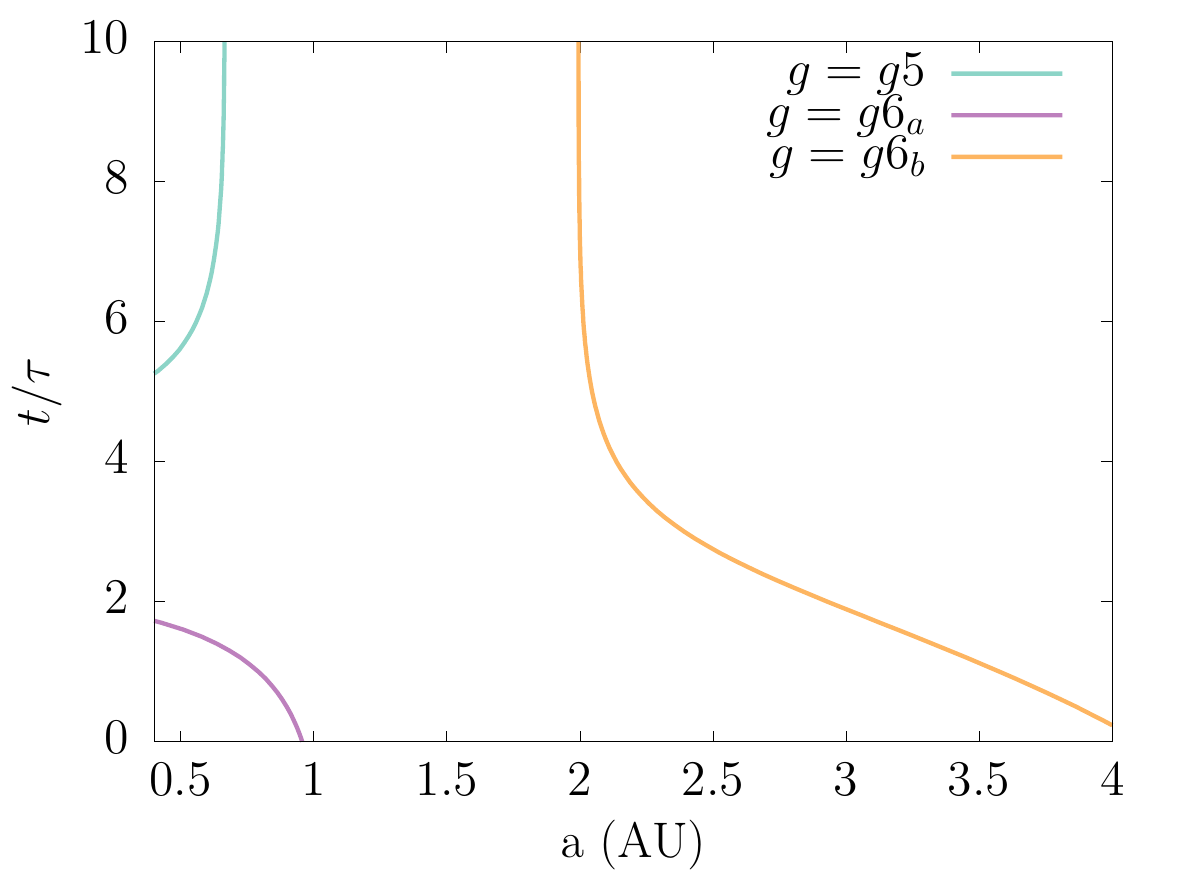}
 \includegraphics[width=0.5\textwidth]{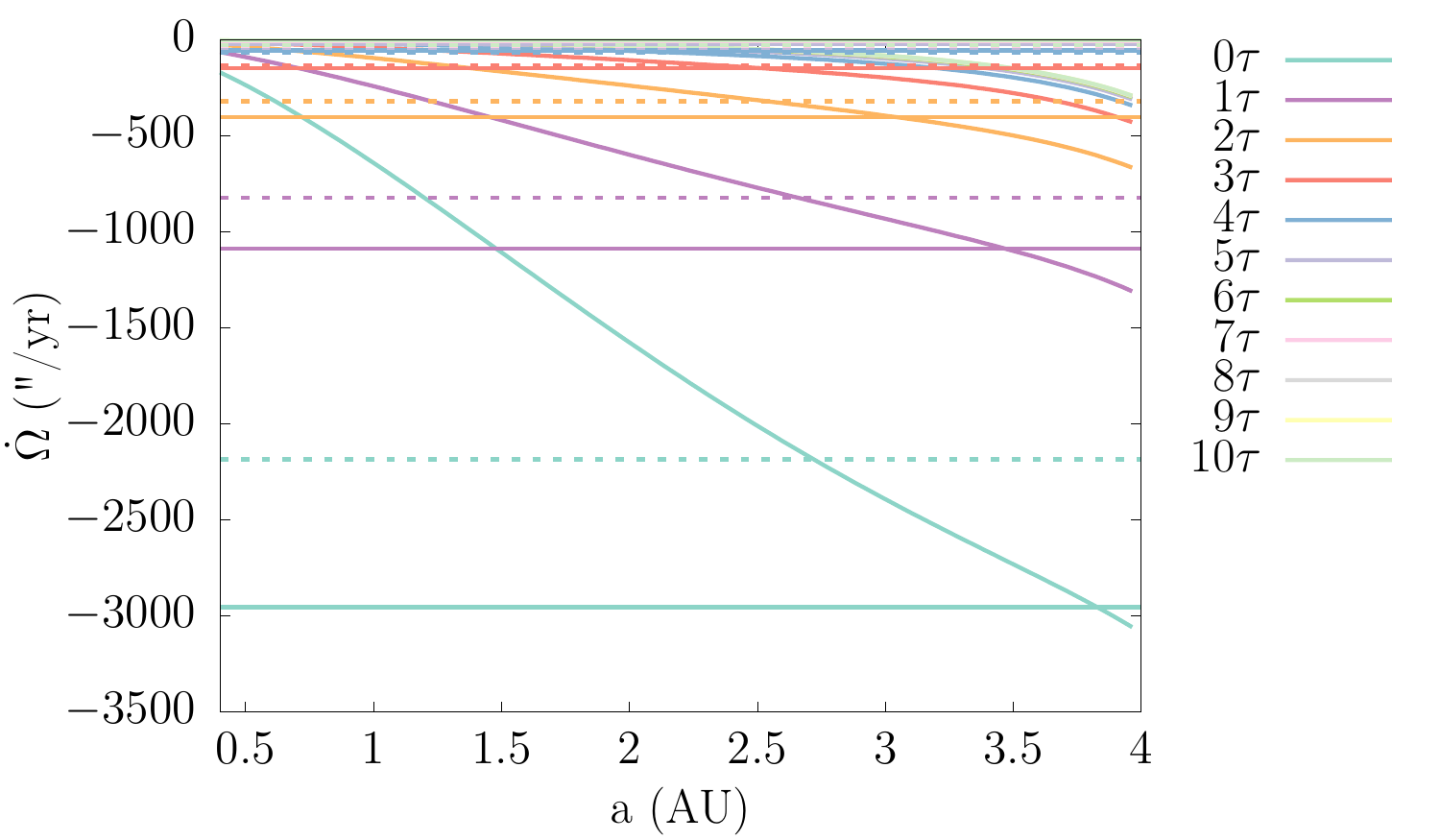}\includegraphics[width=0.39\textwidth]{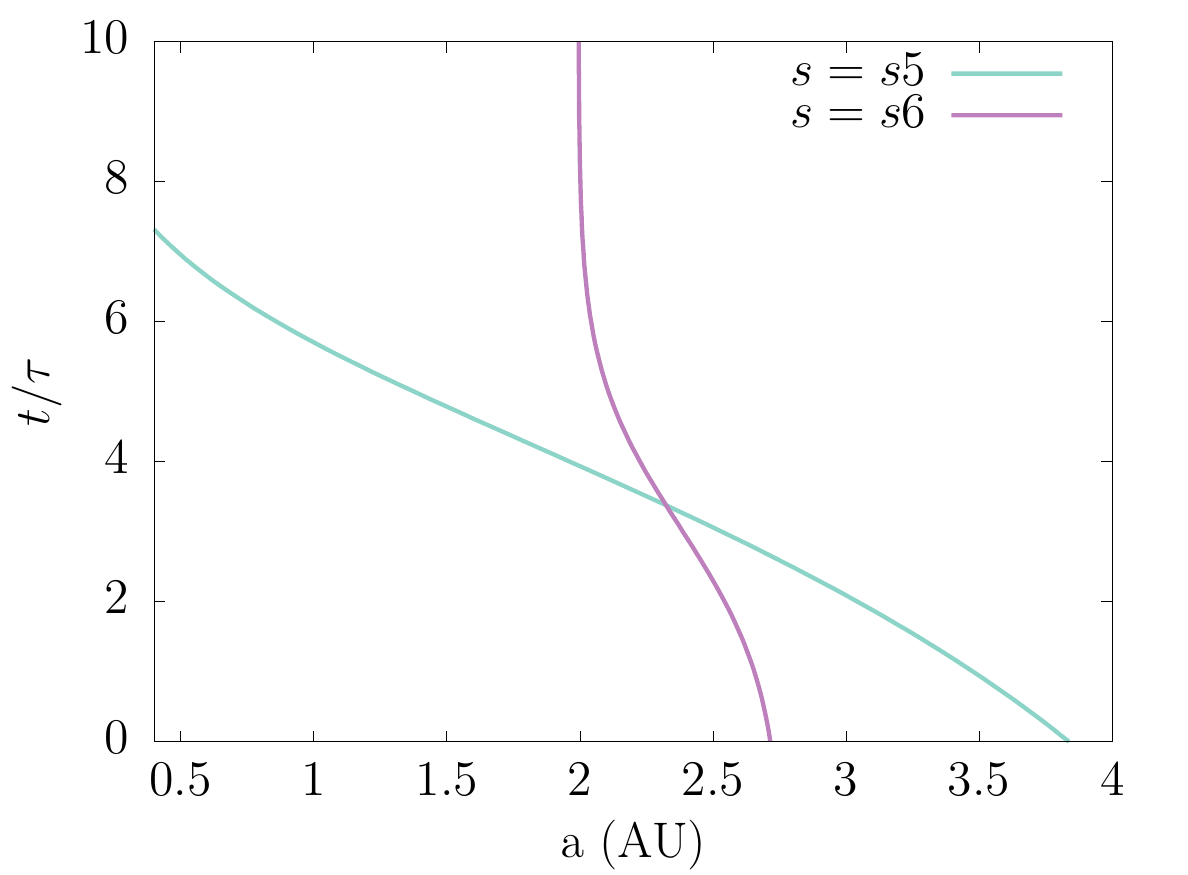}
 \caption{(left) Secular precession frequencies and their time evolution, in an exponentially decaying MN disk. Horizontal lines denote the planetary frequencies (dashed for $g_5$ and $s_5$, full for $g_6$ and $s_6$). Each color corresponds to a given time instant. Jupiter and Saturn are assumed to follow their current orbits. (right) The paths of $g=g_i$ (top) and $s=s_i$ (bottom) secular resonances. The $g=g_6$ secular resonance occurs at two heliocentric distances simultaneously (a and b indices).}
 \label{fig:maps-insitu_obar_omega}
\end{figure*}

We apply this recipe, using the same 100 asteroids as in \autoref{sec:sec_freq} embedded in a MN disk. \autoref{fig:maps-insitu_obar_omega} is a map of secular resonance occurrences.  This happens when the asteroids' frequency function intersects a horizontal line (a planetary frequency) and both curves have the same color, which signifies the time instant, according to the color scale shown in the plot legend. The disk undergoes uniform exponential decay, and hence the time is measured in e-folding times ($\tau$). Multiple crossings are seen on the map. To better understand the paths of the resonances, we find the set of points $(a_*,t_*)$ for which $g(a_*,t_*)=g_{5,6}(t_*)$ or $s(a_*,t_*)=s_{5,6}(t_*)$ and plot them on a corresponding diagram, also shown in \autoref{fig:maps-insitu_obar_omega}. Note that, in this example the two nodal frequencies collapse to one at $a=2.4~$AU for $t\approx 3\tau$, before they separate again.   \\

\begin{figure*}
\centering
 \includegraphics[width=0.5\textwidth]{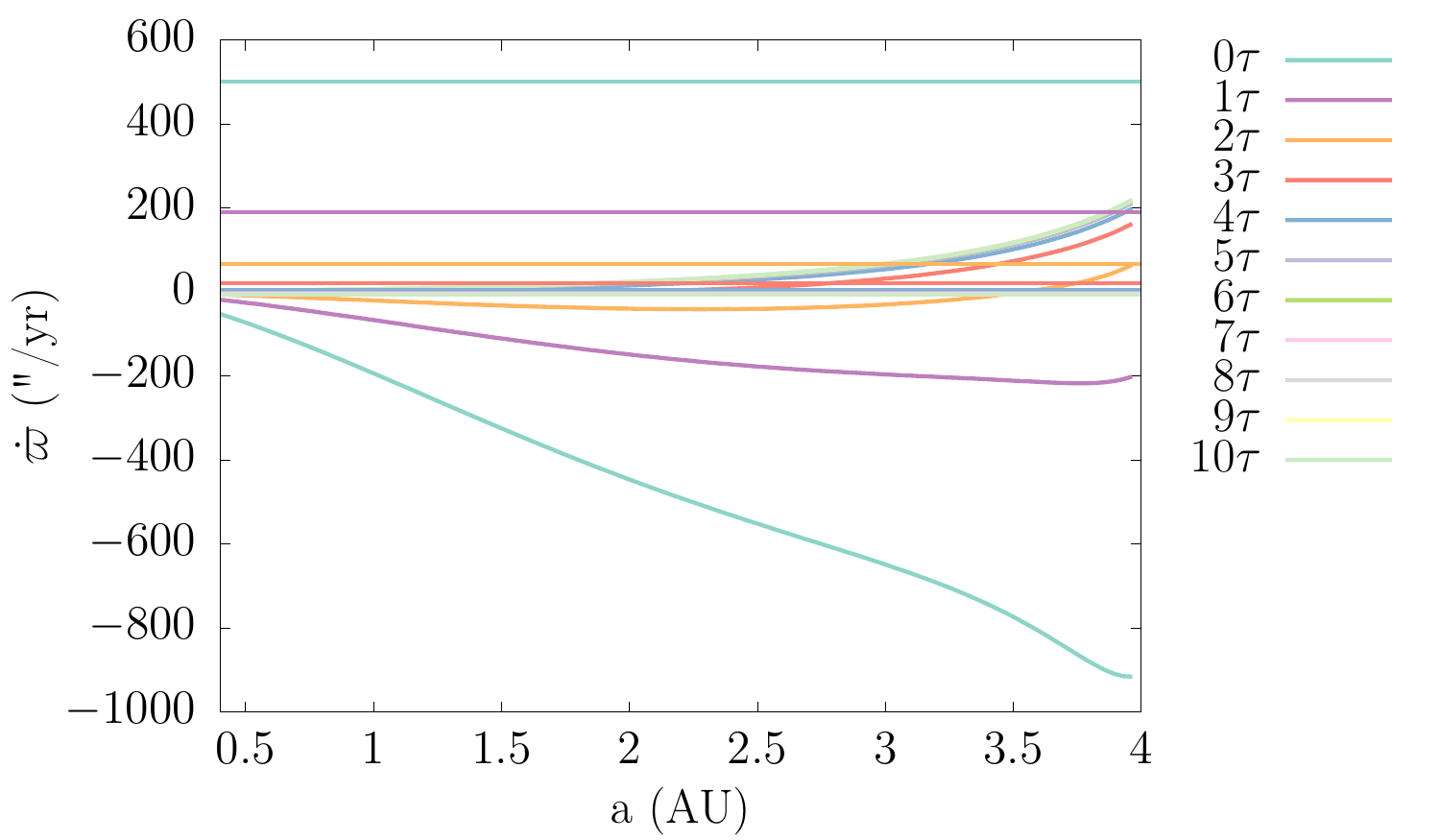}\includegraphics[width=0.39\textwidth]{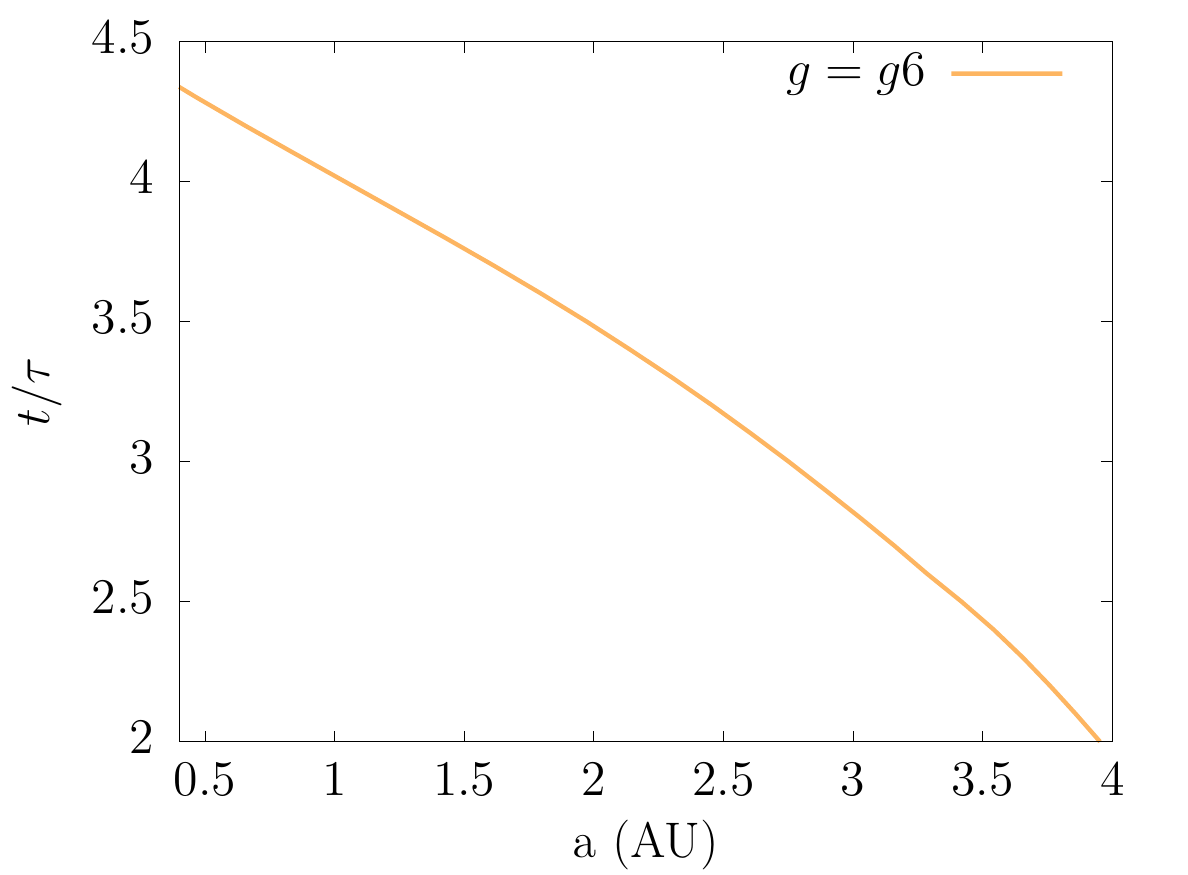}
 \includegraphics[width=0.5\textwidth]{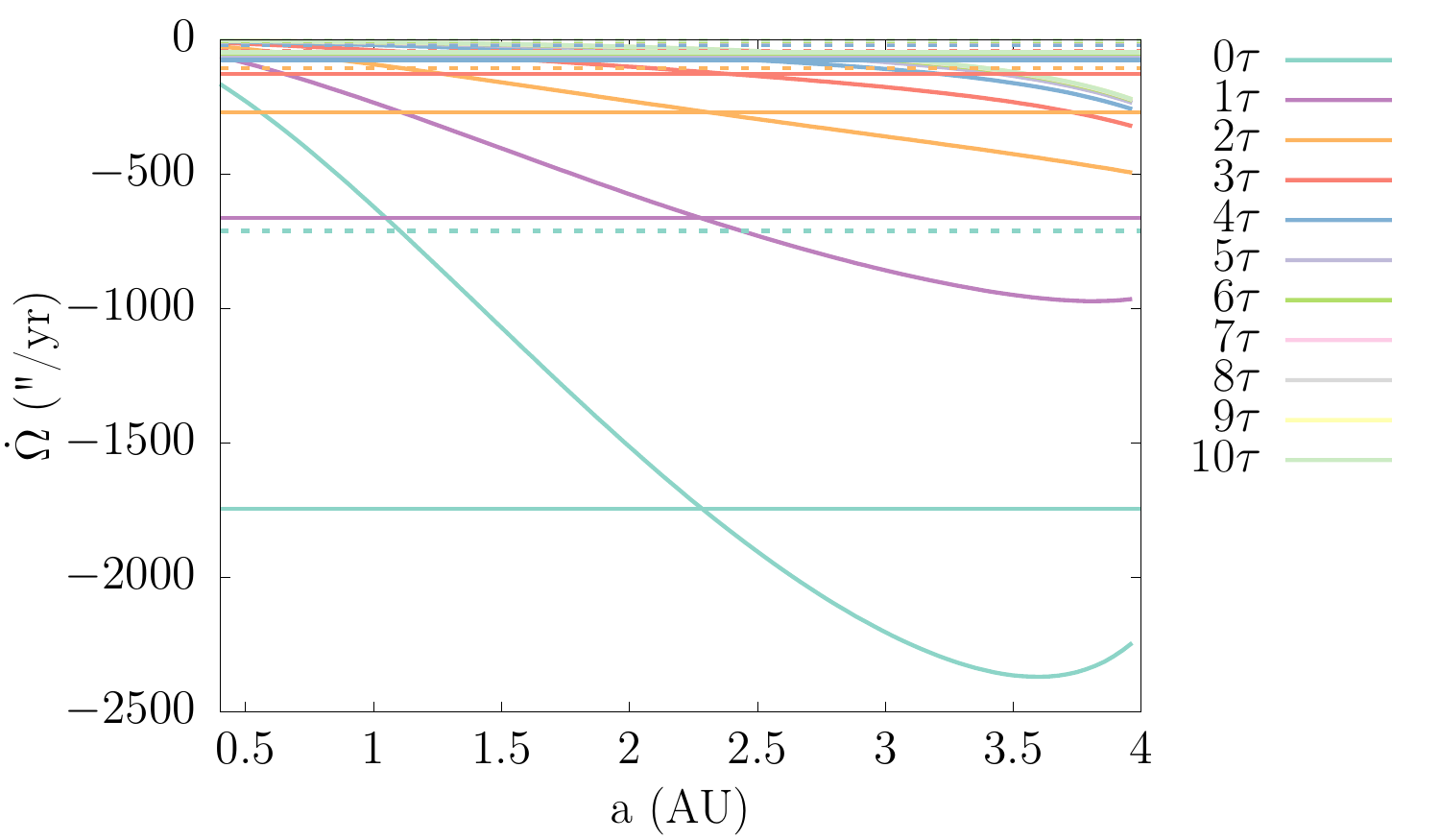}\includegraphics[width=0.39\textwidth]{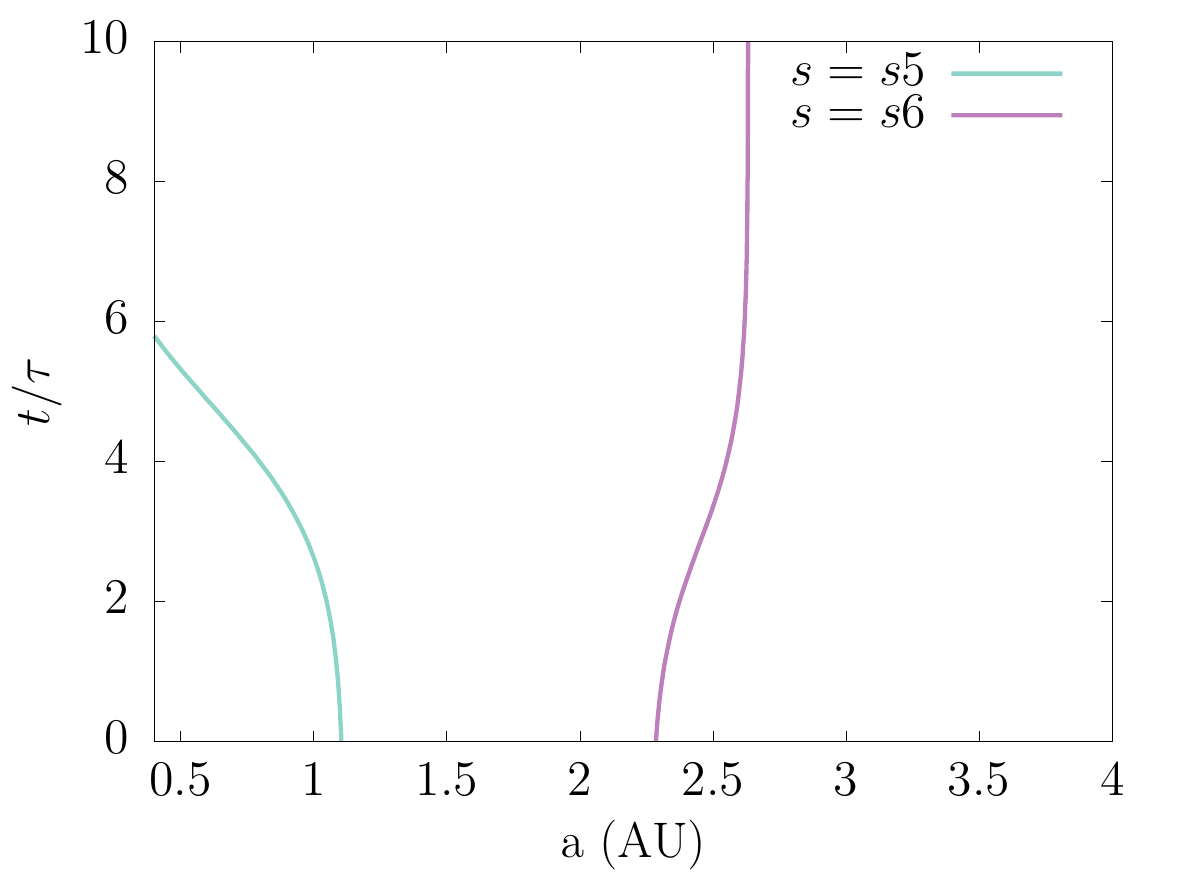}
 \caption{Same as \autoref{fig:maps-insitu_obar_omega}, for a modified MN disk (with gaps) and a 2:1-resonant configuration of Jupiter and Saturn. }
 \label{fig:maps-21obar_omega_gaps}
\end{figure*}

If we assume different orbits for Jupiter and Saturn and a different surface density profile, the picture changes significantly. We integrated again the same 100 particles, but we assumed Jupiter and Saturn to be in a 2/1 MMR and also modify the MN density profile, by taking into account for the gaps around the planets (see \autoref{sec:sec_freq}). The corresponding maps are shown in \autoref{fig:maps-21obar_omega_gaps}. As we see, only the $g_6$ resonance crosses the domain of interest, while the two nodal resonances span distant, small portions of the asteroid belt. These examples highlight the importance of using appropriate disk profiles and planetary configurations.

\section{Results on exponentially decaying protoplanetary disks}\label{sec:maps_n_paths2}
So far, we have only studied MN disks. However, protoplanetary gas disks are typically assumed to have density profiles of the form (\citet{1974MNRAS.168..603L}; \citet{1998ApJ...495..385H})

\begin{equation}\label{eq:Sigma_good}
\Sigma ( r ) \, =\, ( 2- \gamma ) \left( \frac{M}{2 \pi R_c^2}\right) \left( \frac{r}{R_c} \right)^{-\gamma} \exp \left[ -\left( \frac{r}{R_c}\right)^{\left( 2-\gamma\right) } \right], 
\end{equation}
\noindent
where $R_c$ is a characteristic scaling radius, $M$ the disk mass and $\gamma$ a characteristic exponent associated with the disk viscosity. This type of profile agrees well with observations \citep{2010ApJ...723.1241A}. For a solar-system-like model, we adopted $M=\mathcal{M}$, $\gamma=1$ and $R_c=15$~AU and $H=0.05$ as scale height. Following our method, we constructed several disks with gaps around the planets, assuming the latter to be on orbits (a) similar to their current ones, (b) close to their 2:1 MMR and (c) close to their 3:2 MMR ($a_j\simeq5.4$~AU, $e_j\simeq0.01$, $i_j\simeq0.03^{\circ}$, $a_s\simeq9.6~AU$,  $e_s\simeq0.035$ and , $i_s\simeq0.08^{\circ}$). In addition, we considered another case (d) for two Jupiter-like planets orbiting in the same 2:1 resonance, as a possible extrasolar system planetary configuration. In all cases the disk is considered to be depleted uniformly in an exponential fashion.\\

\begin{figure*}
\centering
\includegraphics[width=0.5\textwidth]{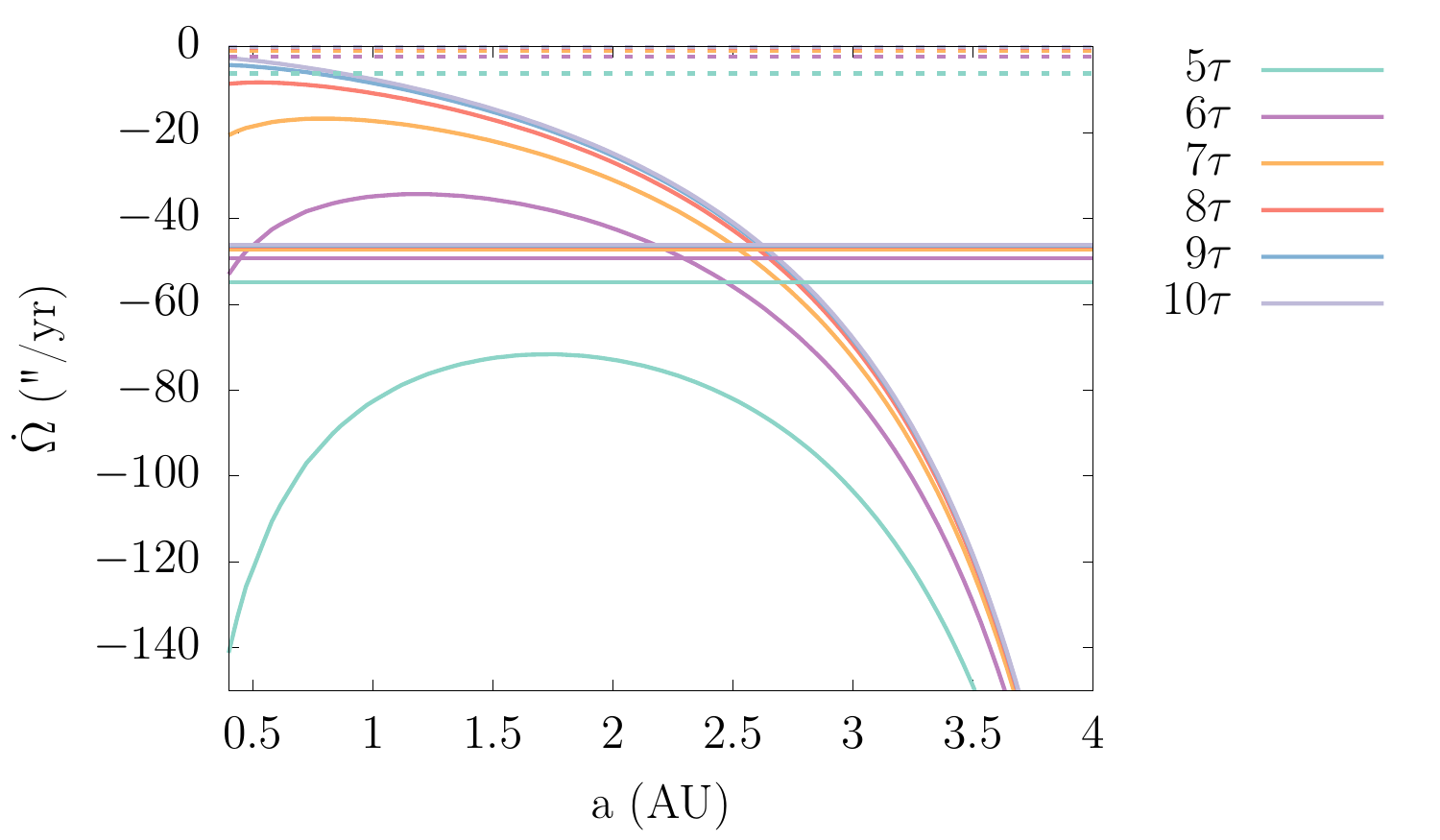}
\includegraphics[width=0.39\textwidth]{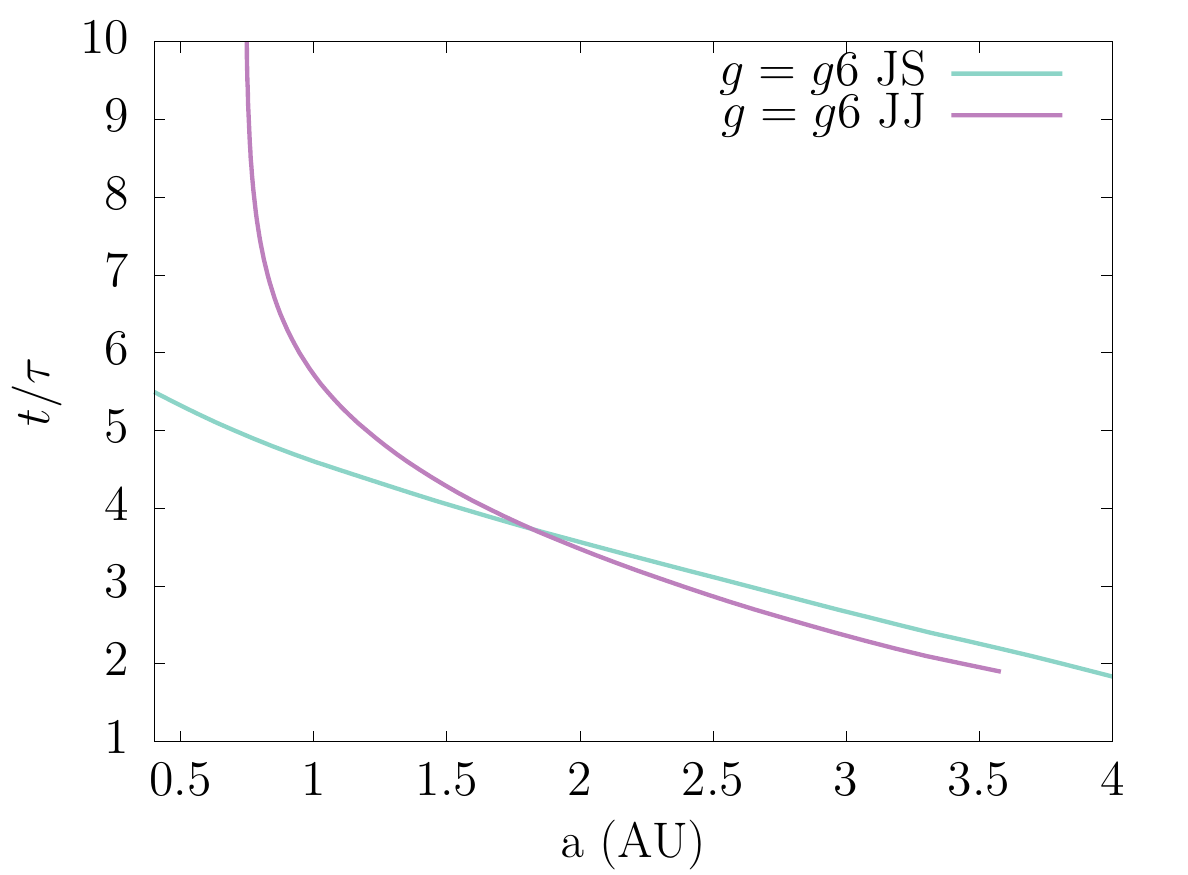}
\includegraphics[width=0.5\textwidth]{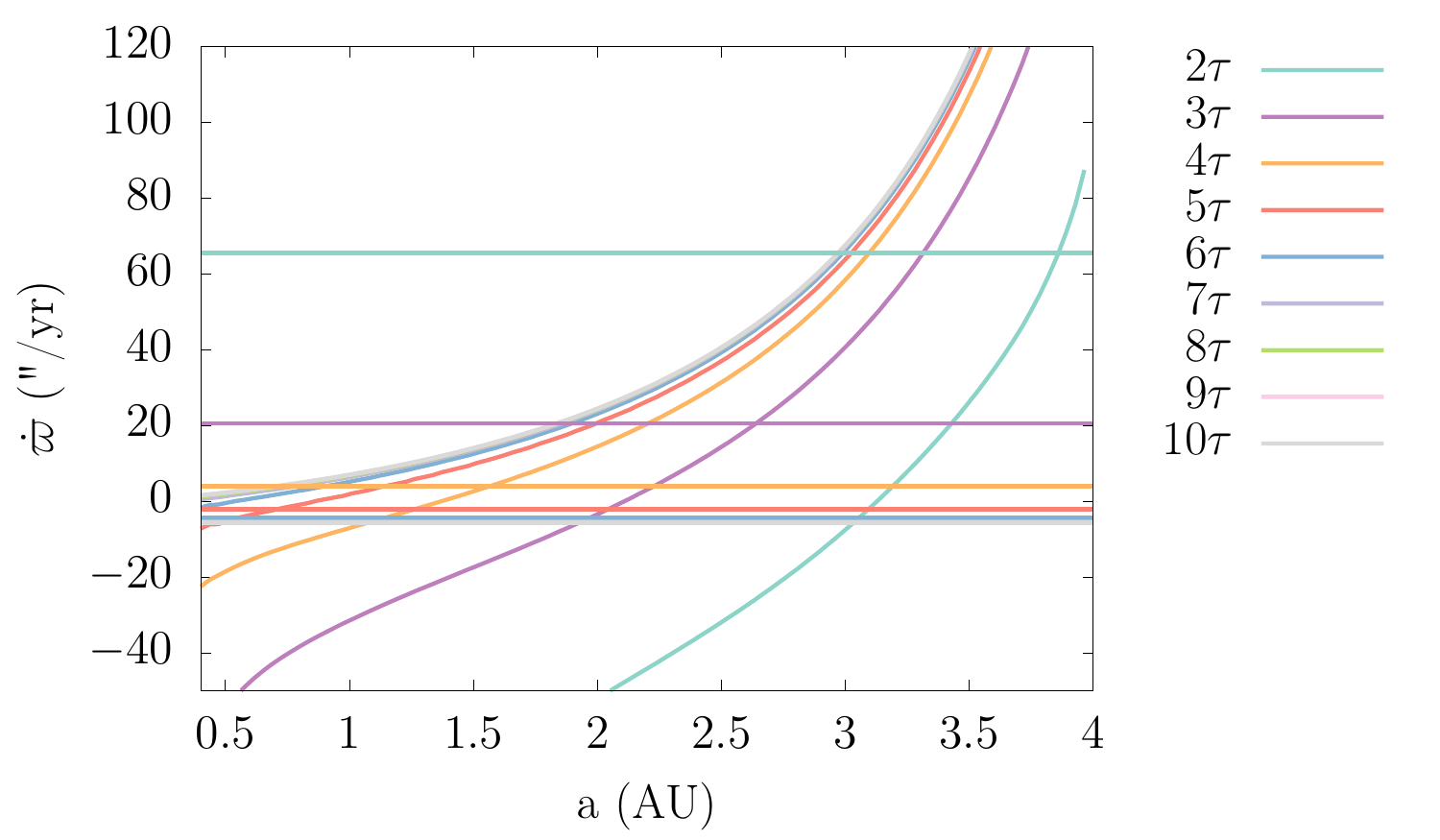}\includegraphics[width=0.39\textwidth]{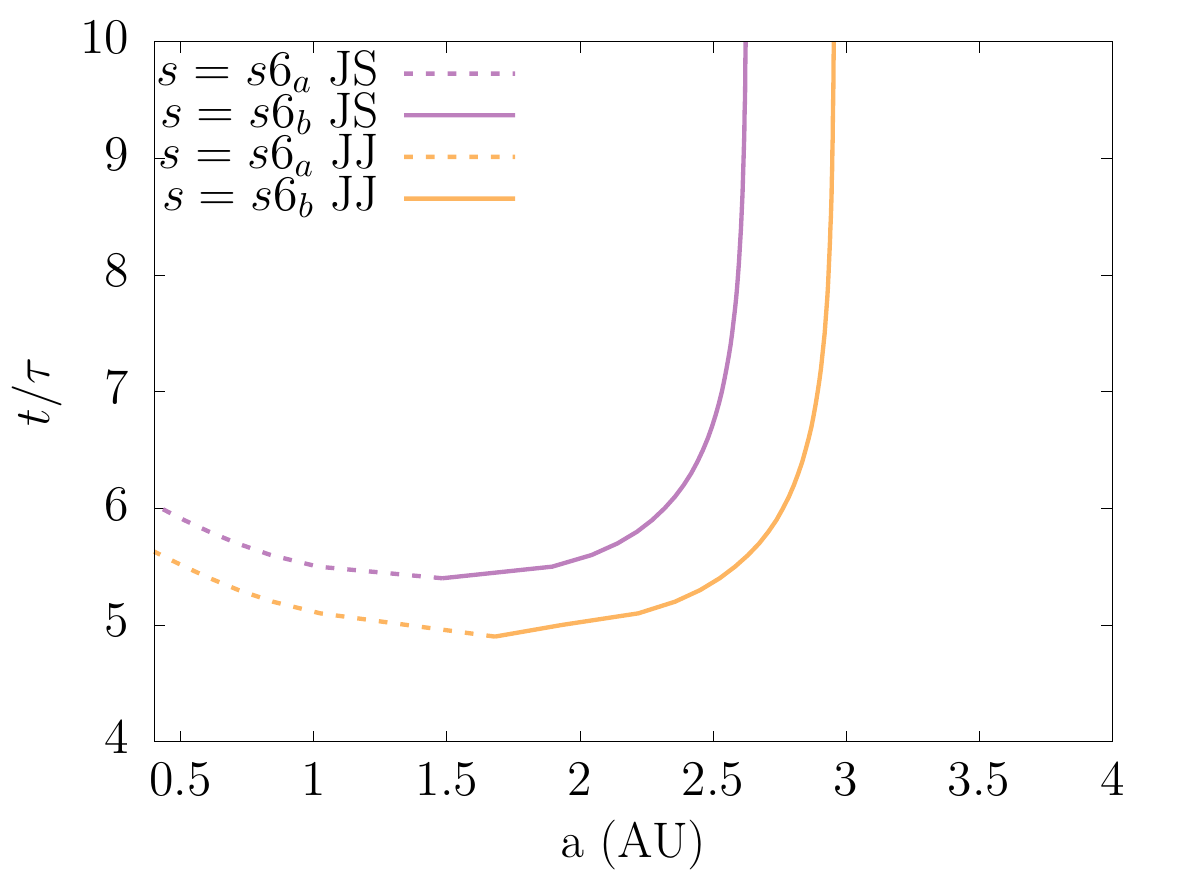}
 \caption{ Same as \autoref{fig:maps-insitu_obar_omega}, for a realistic protoplanetary disk with a power-law surface density profile. The maps of the secular resonance occurrences (left) consider a 2:1-resonant configuration of Jupiter and Saturn. The paths of the resonances (right) show both the cases of the solar-system-like model (case b) and in addition a 2:1-resonant configuration consisting of two Jupiter-like planets (case d).}
 \label{fig:crossing-gs_expo_JS-JJ}
\end{figure*}

\begin{figure*}
\centering
\includegraphics[width=0.33\textwidth]{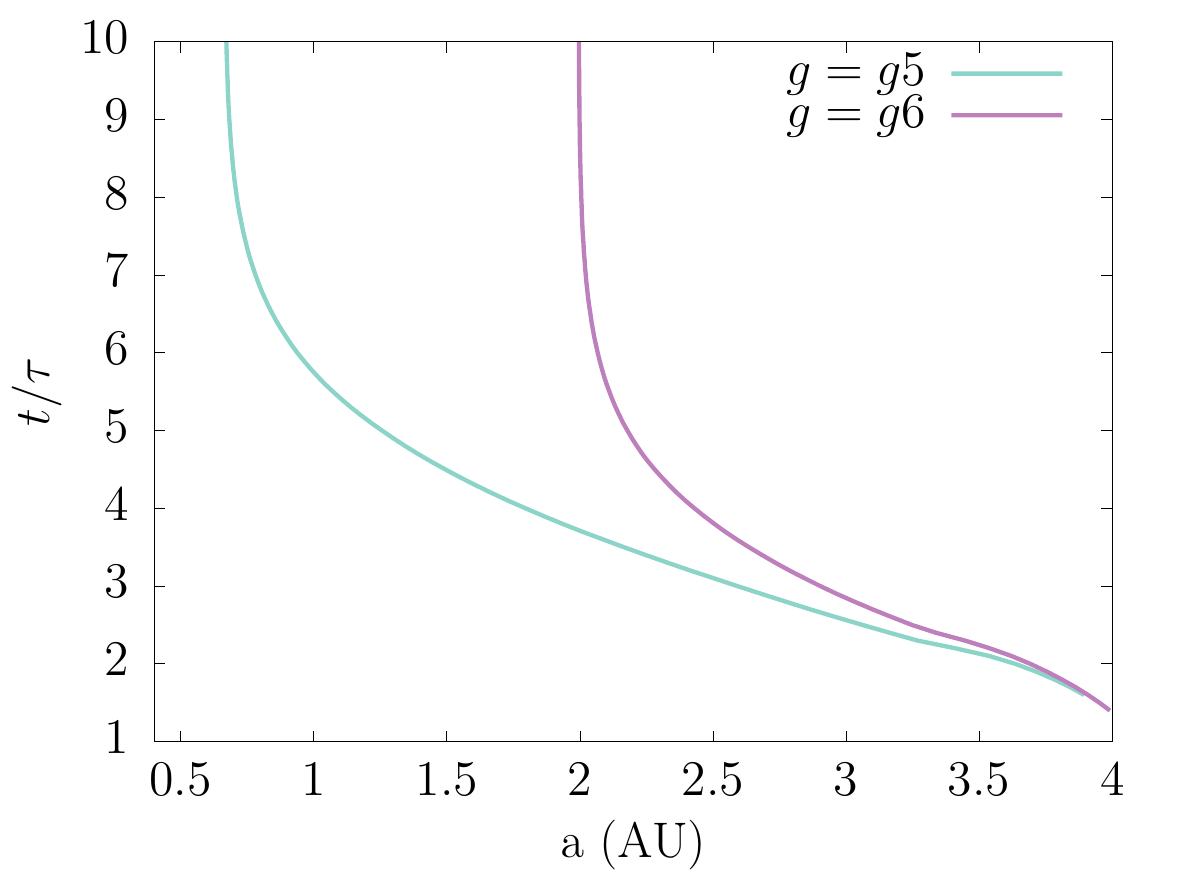}\includegraphics[width=0.33\textwidth]{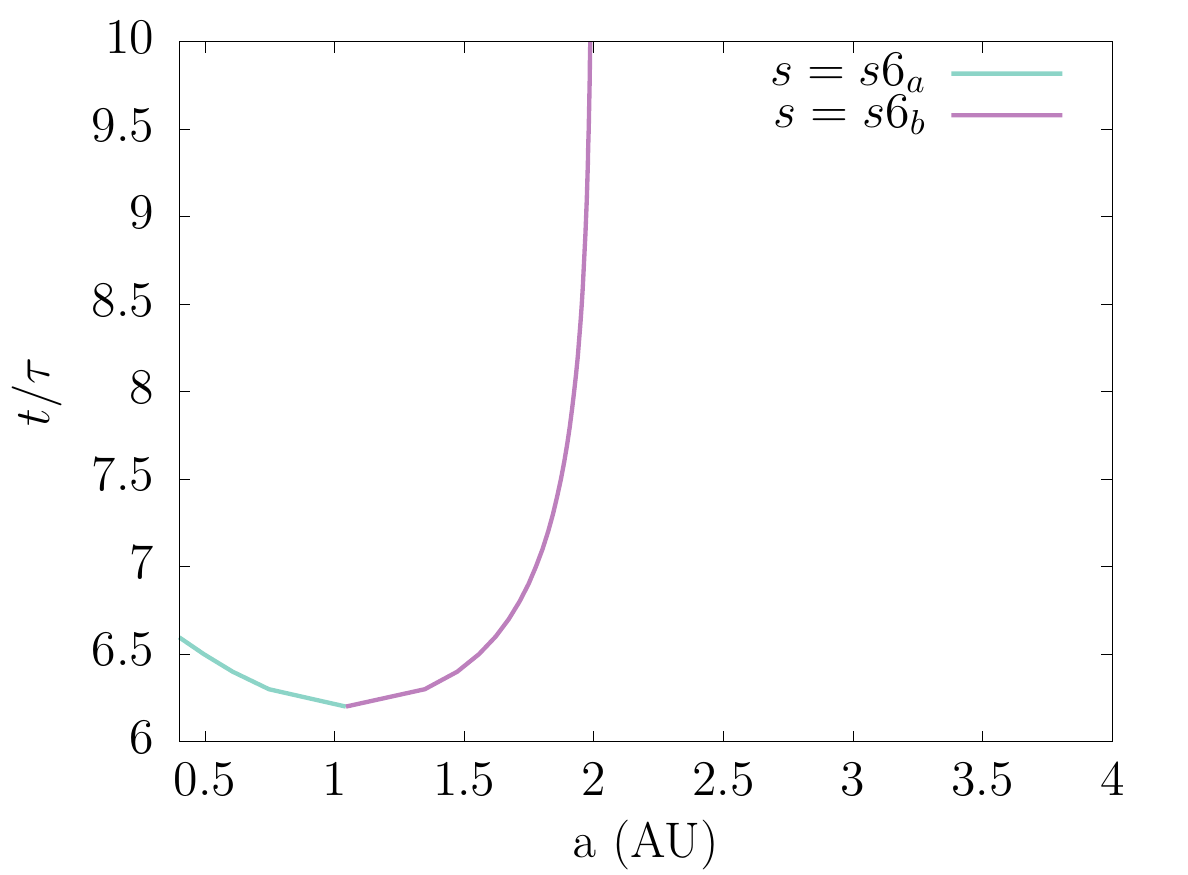}
\includegraphics[width=0.33\textwidth]{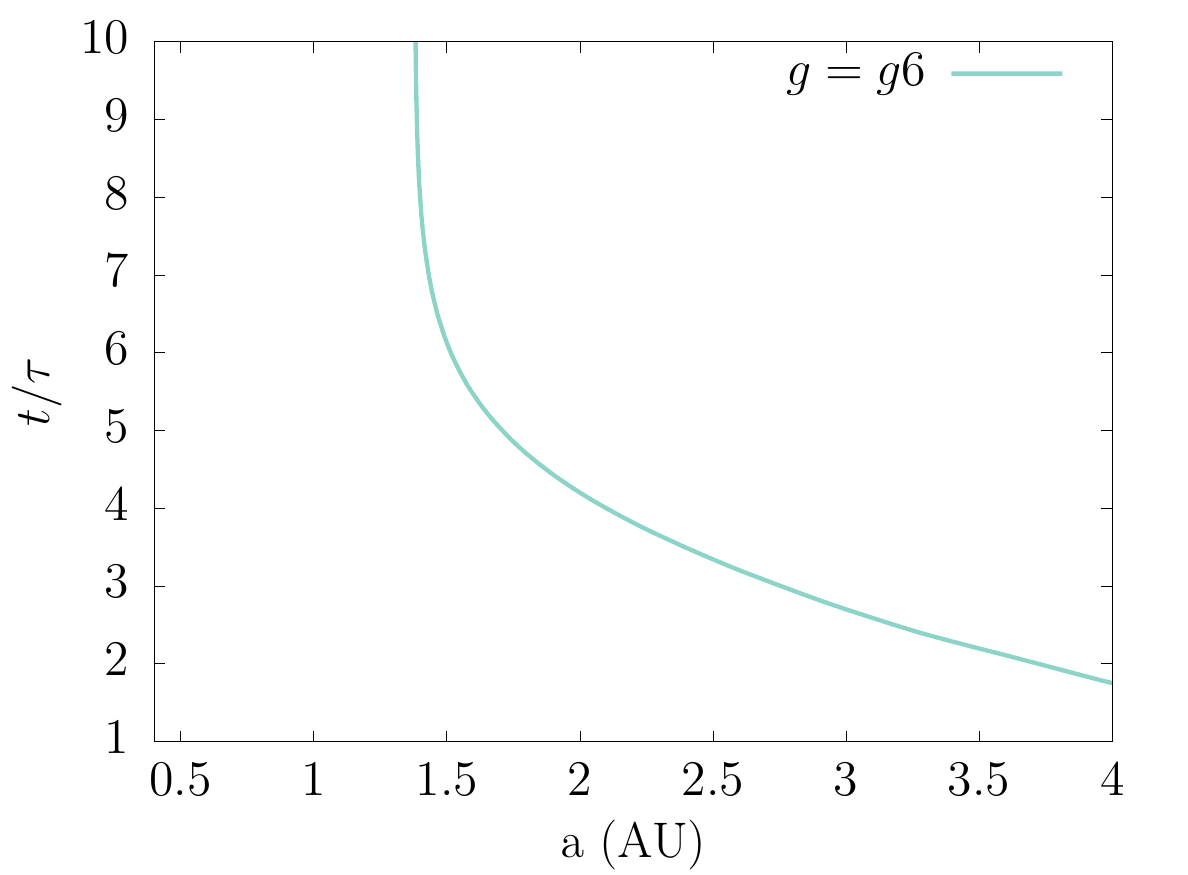}\includegraphics[width=0.33\textwidth]{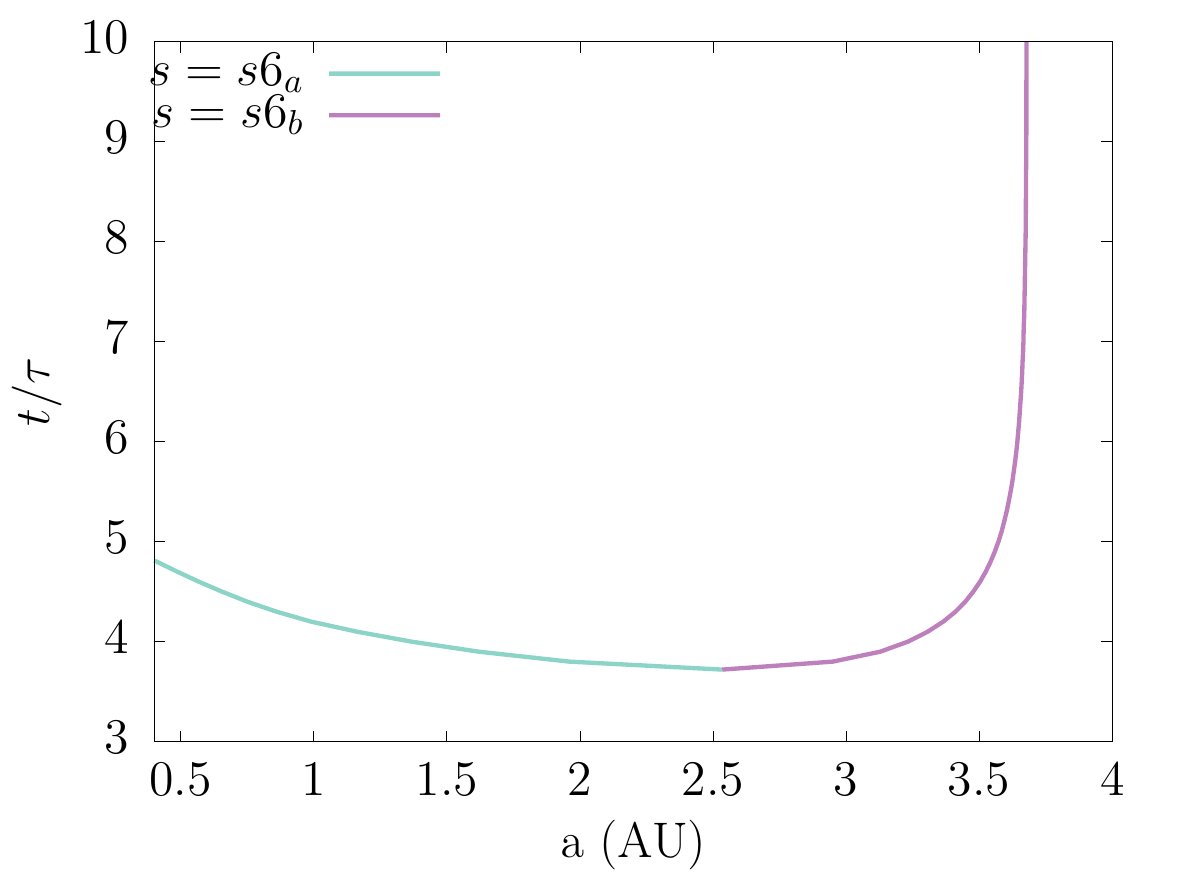}
 \caption{The paths of secular resonances in the two different planetary configurations: (top) current orbits and (bottom) 3:2 resonance}
 \label{fig:crossing-gs_expo}
\end{figure*}

Integrating the same 100 particles as before in the four planetary configurations described above and in disks with $\gamma = 0.5$ we derive the secular maps shown in \autoref{fig:crossing-gs_expo_JS-JJ} and \autoref{fig:crossing-gs_expo}. In cases (b) and (d) we notice that, qualitatively, the secular resonances follow similar trends, although the $s=s_6$ secular resonance occurs earlier in the extrasolar case. In case (a), both the $g=g_5$ the $g=g_6$ resonances cross the computational domain; $g=g_6$ sweeps through the main belt region while $g=g_5$ sweeps through almost the entire inner Solar System. In (b) and (c), only the $g=g_6$ resonance is relevant. Most notably, as we go from case (a) to case (c), the $s=s_6$ resonance settles at larger and larger heliocentric distances. 

\begin{figure*}
\centering
\includegraphics[width=0.33\textwidth]{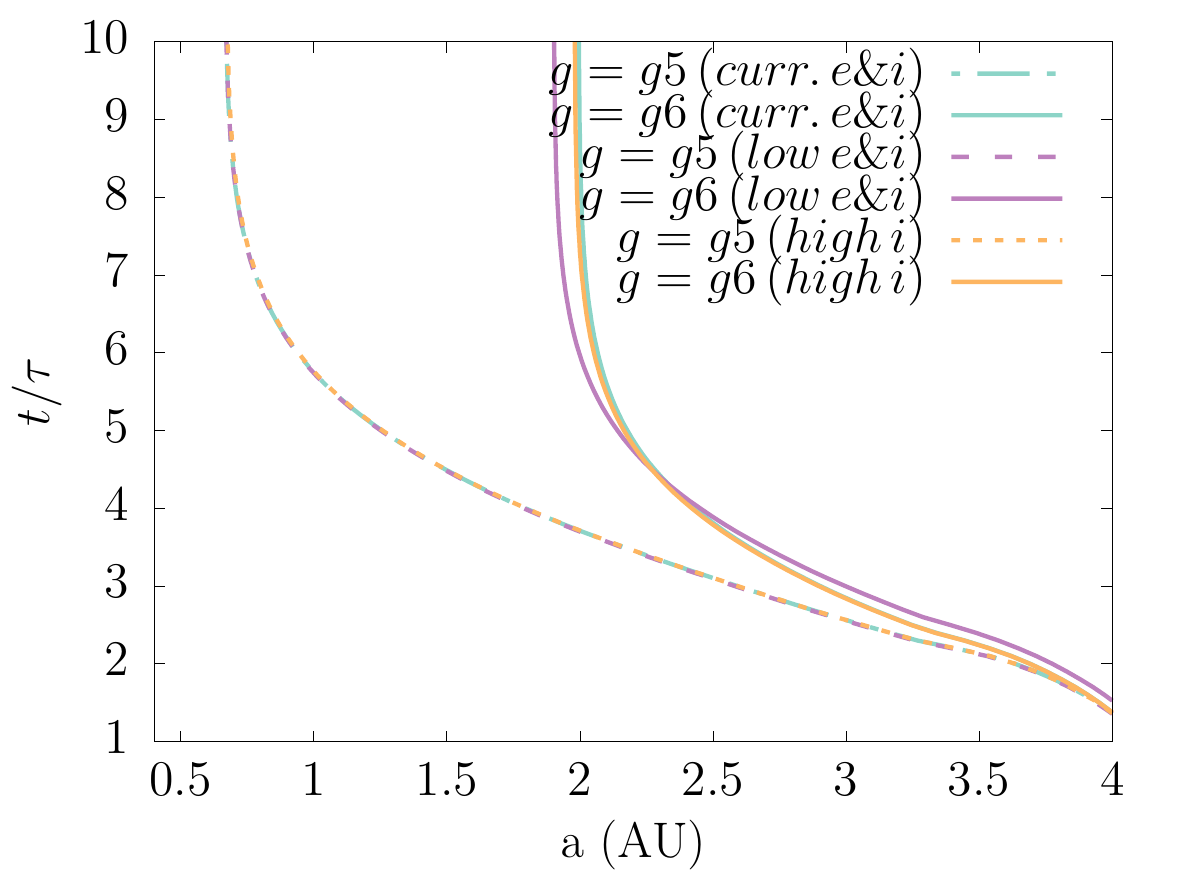}\includegraphics[width=0.33\textwidth]{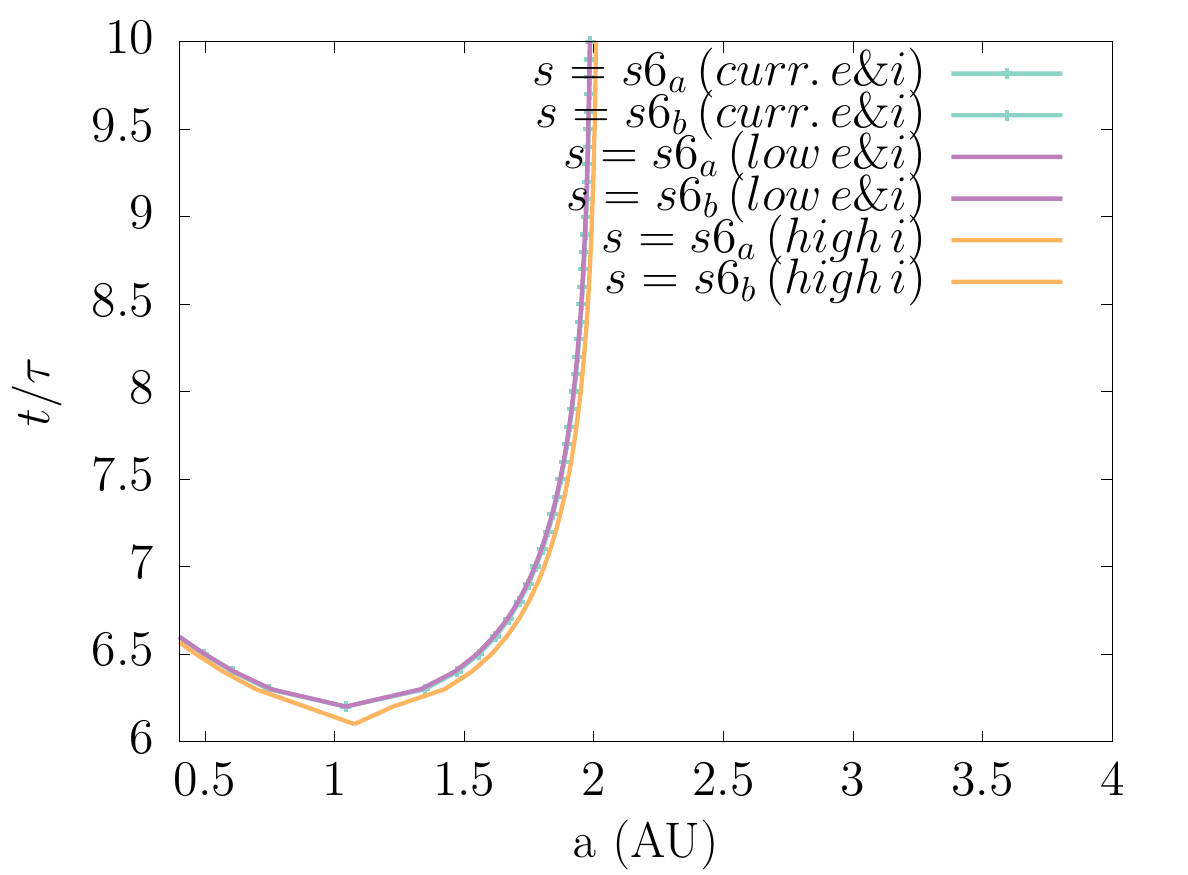}
\includegraphics[width=0.33\textwidth]{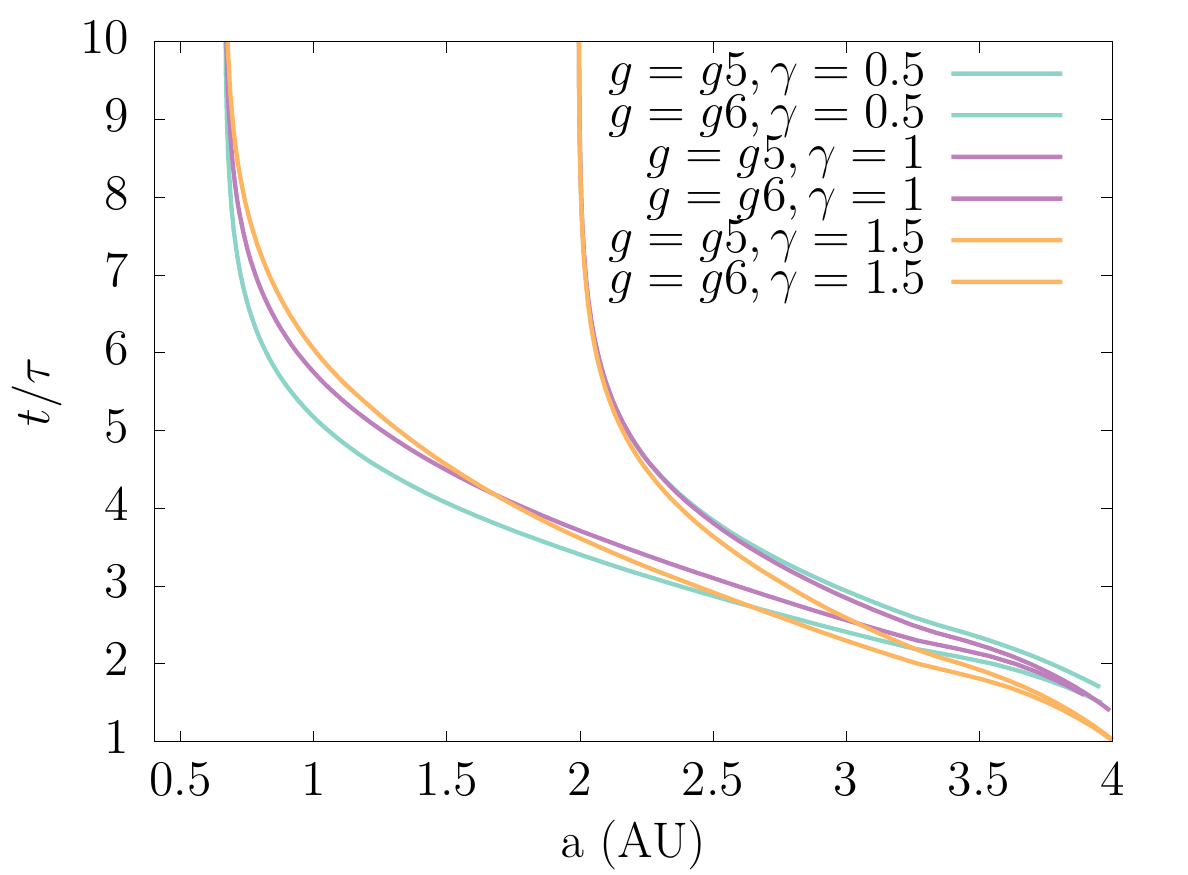}\includegraphics[width=0.33\textwidth]{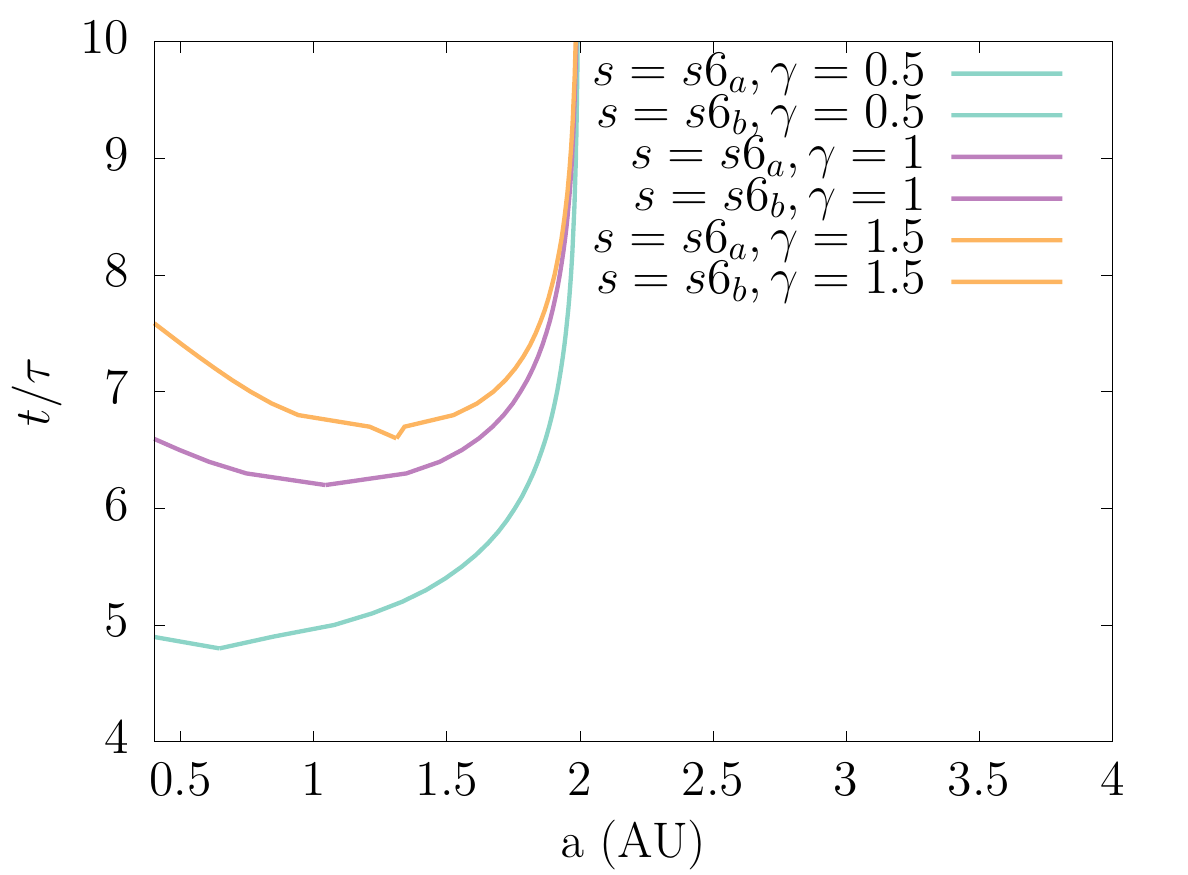}
 \caption{Same as \autoref{fig:crossing-gs_expo}, for varying eccentricities and inclinations of Jupiter and Saturn (top), or for the same planetary configuration as in \autoref{fig:crossing-gs_expo_JS-JJ}, but with different slope ($\gamma$) of the surface density profile (bottom).}
 \label{fig:crossing-gs-insitu_ei}
\end{figure*}

It is evident that if Jupiter and Saturn had settled on their current orbits, when their nascent disk was still around, the $s=s_6$ resonance would not sweep at all through the main belt region (it stops at $a\sim2$~AU). We found that the situation does not change much, even if we consider giant planets on mildly eccentric or inclined orbits  (see \autoref{fig:crossing-gs-insitu_ei}). The $s=s_6$ resonance practically follows the same path, since it occurs late, when the disk has only a minor contribution to the precession frequencies of asteroids, compared to the planets. We repeated these tests using disks with $\gamma=1.5$ (steeper profile) and another $\gamma=0.5$ (shallower profile). As can be seen in \autoref{fig:crossing-gs-insitu_ei}, the results are practically unchanged. Hence, we can conclude that, in this model, no inclination excitation by resonance sweeping could have occurred in the asteroid belt, if Jupiter and Saturn had already settled on their final orbits. 

\begin{figure*}
\centering
 \includegraphics[width=0.4\textwidth]{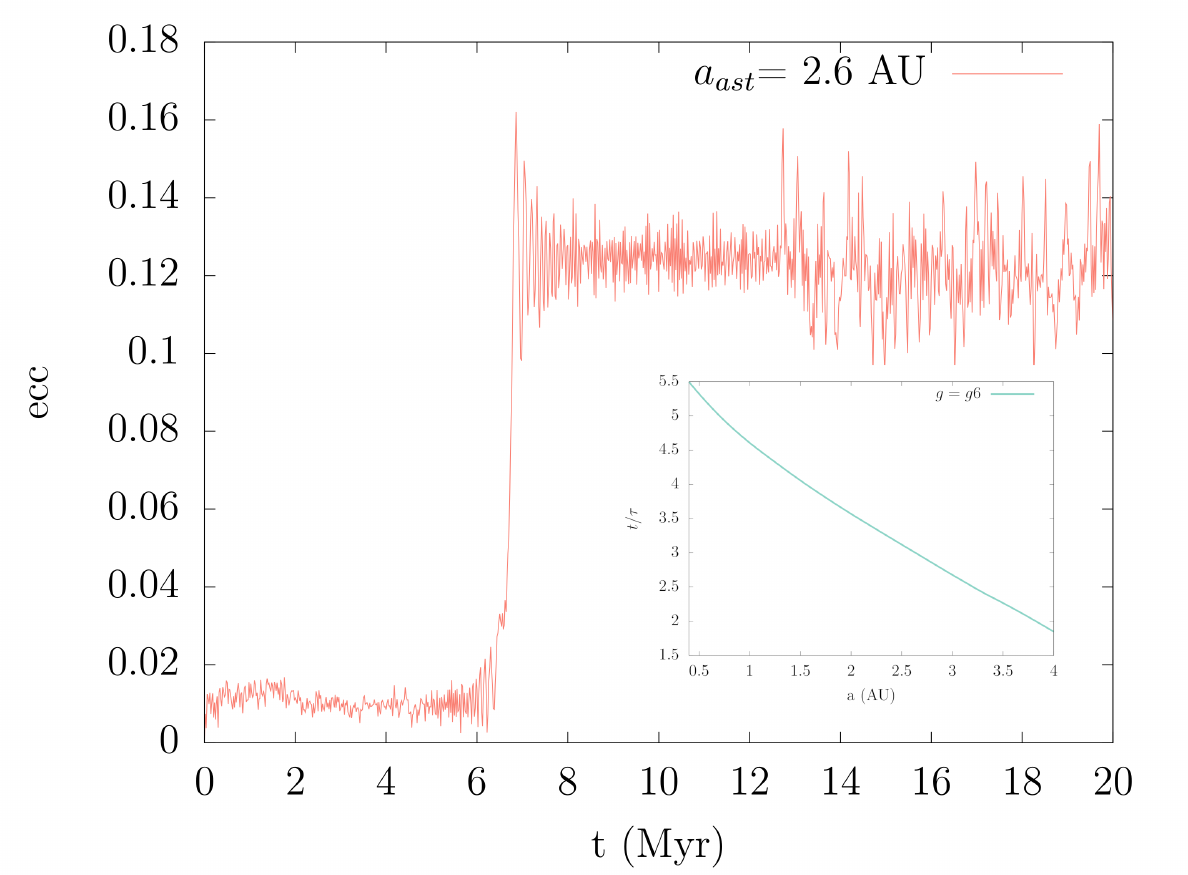}\includegraphics[width=0.4\textwidth]{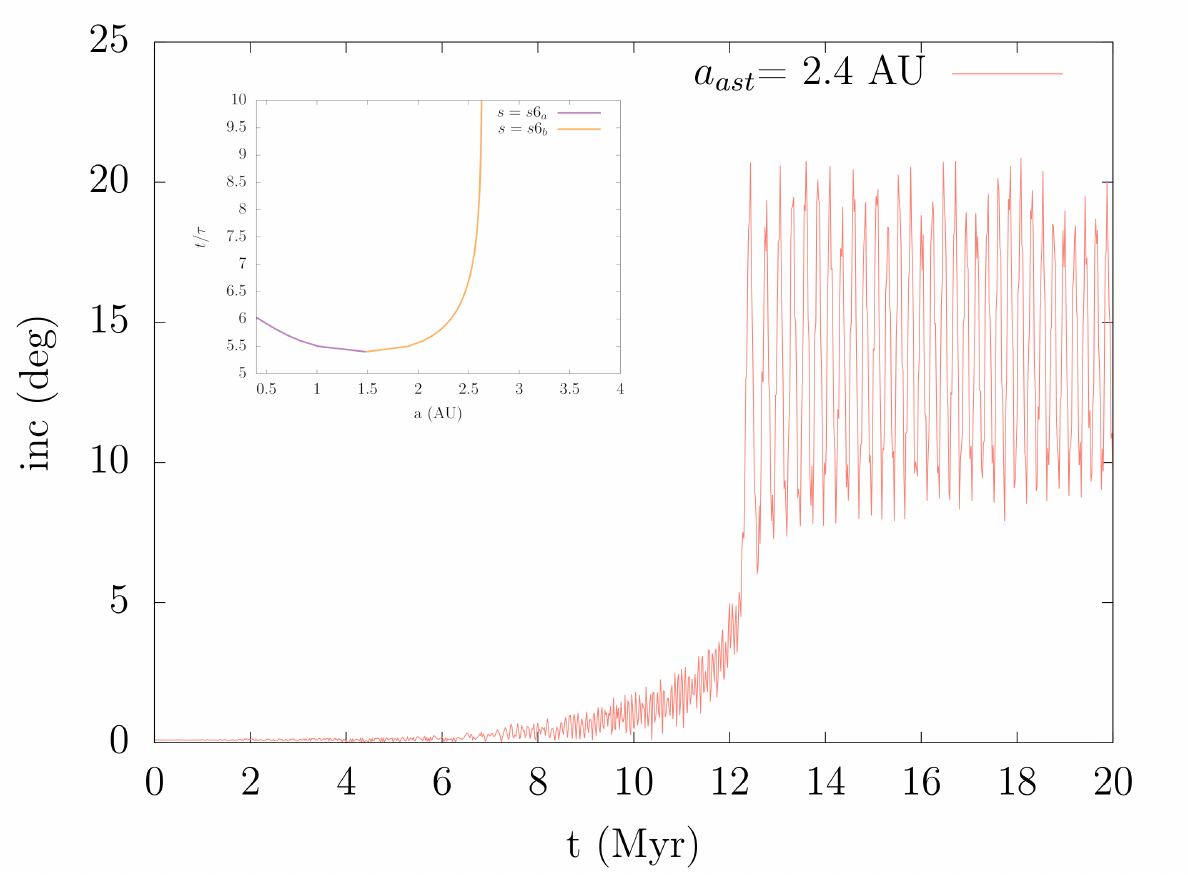}
 \caption{The evolution of eccentricity and inclination for two particles during the simulation of the exponential decay of the disk.}
 \label{fig:maps-particles}
\end{figure*}

Carrying out the integration of 100 particles (Jupiter and Saturn in 2:1 MMR) embedded in the exponentially decaying disk for a total time of $20$~Myr and using $\tau=2$~Myr, we can check whether the secular resonance crossings found in our semi-analytical model above really occur at the correct time and place. In \autoref{fig:maps-particles} we show the eccentricity evolution for a particle orbiting at $a=2.6$~AU. We can observe that the eccentricity jumps at $t\approx 6\ \text{Myr}=3 \tau$, in agreement with what was expected from the corresponding map (see inlet). The same holds for the inclination of a particle orbiting at $a=2.4$~AU, which exhibits a jump at $t\approx 12\ \text{Myr}=6 \tau$. We have confirmed that this holds throughout the region of interest.  

Our secular resonance maps do not provide information about the magnitude of variation of a particles' $e$ and $i$. Both \citet{2002Icar..159..322N} and \citet{2011ApJ...732...53M} derived estimates of the variation in the action $J\simeq\frac{1}{2}\sqrt{a}e^2$ conjugate to $\varpi$, owing to crossing of a $g=g_i$ secular resonance. In \citeauthor{2011ApJ...732...53M}'s notation

\begin{equation}\label{eq:Je}
\Delta J \, = \, \frac{\pi\epsilon^2}{2|\lambda|}+\epsilon\sqrt{\frac{2\pi J_{\text{init}}}{|\lambda|}}\cos \varpi,
\end{equation}

\noindent
where $\epsilon$ is the forced eccentricity at the resonance location, approximated by

\begin{equation}
\epsilon=\frac{1}{4a^{5/4}}\sum_{j=1}^2 \left(a/a_j\right)^2 b_{3/2}^{(2)}(a/a_j) \, m_j \, E_j^{(i)},
\end{equation}

\noindent
where $b_{3/2}^{(2)}$ is the Laplace coefficient and  $E_j^{(i)}$ the amplitude of the $g_i$ mode in the eccentricity spectrum of the $j$-th planet. 
The constant $\lambda$ is associated with the rate of change of $g_i$ (for a linear decay, $\dot{g_i}=2\lambda$). This estimate is based on linear secular theory. Since, to second degree in $e$ and $i$, the secular Hamiltonian is formally identical (save for a sign) for the eccentric $(J,\varpi)$ and the inclined $(P,\Omega)$ part, a similar formula can be found for $P\simeq \frac{1}{2}\sqrt{a} i^2$, namely

\begin{equation}\label{eq:Pi}
\Delta P \, = \, \frac{\pi\epsilon'^2}{2|\lambda|}+\epsilon'\sqrt{\frac{2\pi P_{\text{init}}}{|\lambda|}}\cos \Omega,
\end{equation}

\noindent 
but with the forced inclination, $\epsilon '$, given by

\begin{equation}
\epsilon'=-\frac{1}{4a^{5/4}}\sum_{j=1}^2 \left(a/a_j\right)^2 b_{3/2}^{(2)}(a/a_j)\, m_j \, I_j^{(i)},
\end{equation}

\noindent
where $I_j^{(i)}$ is the amplitude of the $s_i$ mode of in the $j$-th planet's inclination spectrum.

The dependence of $\Delta J$ (resp. $\Delta P$) on $\varpi$ (resp. $\Omega$) means that $e$ (resp. $i$) can either increase or decrease upon resonance crossing, depending on its secular phase. However, for almost circular orbits very close to the invariant plane, $e$ and $i$ can only increase. Also, the variations are functions of $a$, following the forced terms. \\

\begin{figure*}
\centering
\includegraphics[width=0.4\textwidth]{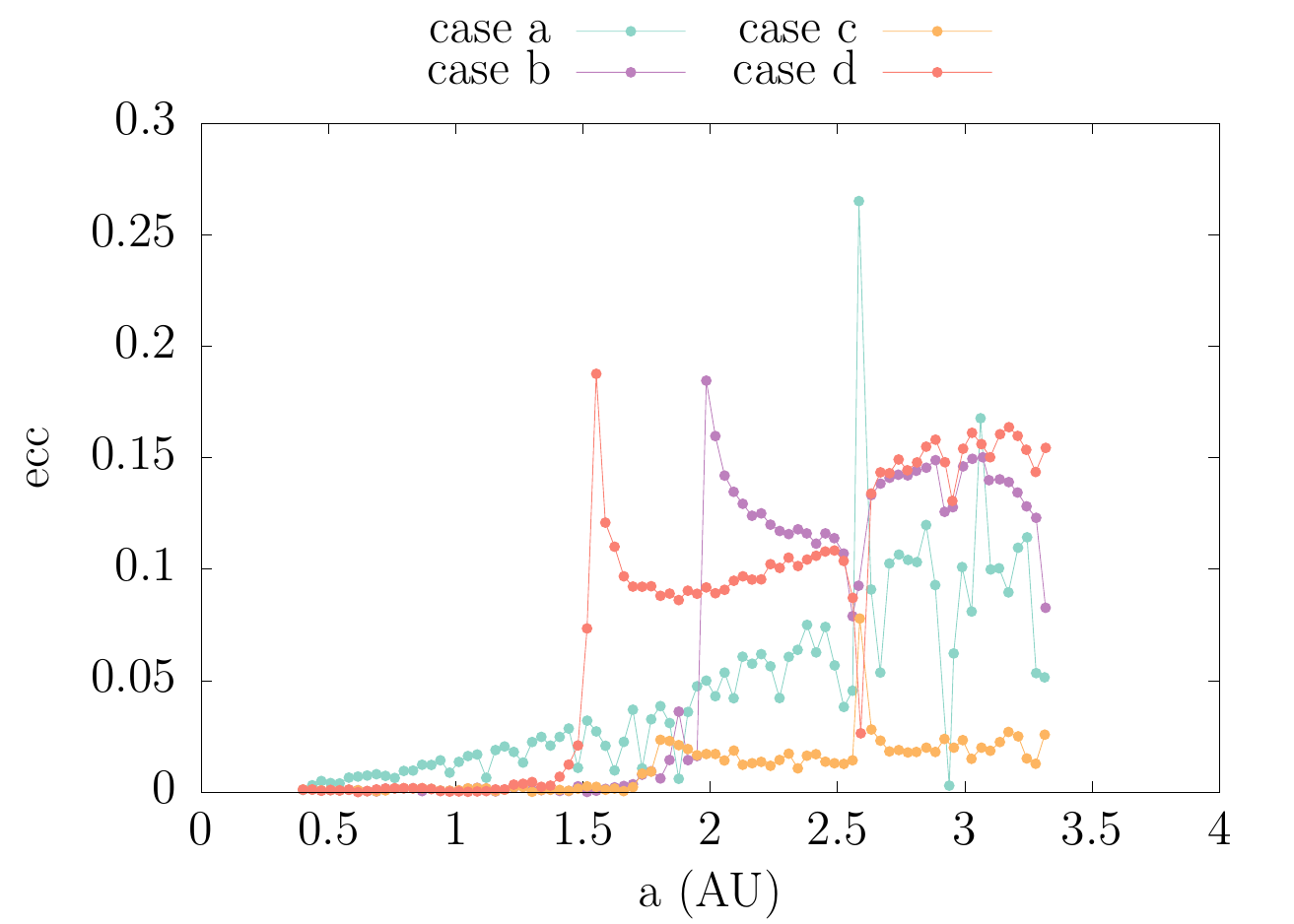}\includegraphics[width=0.4\textwidth]{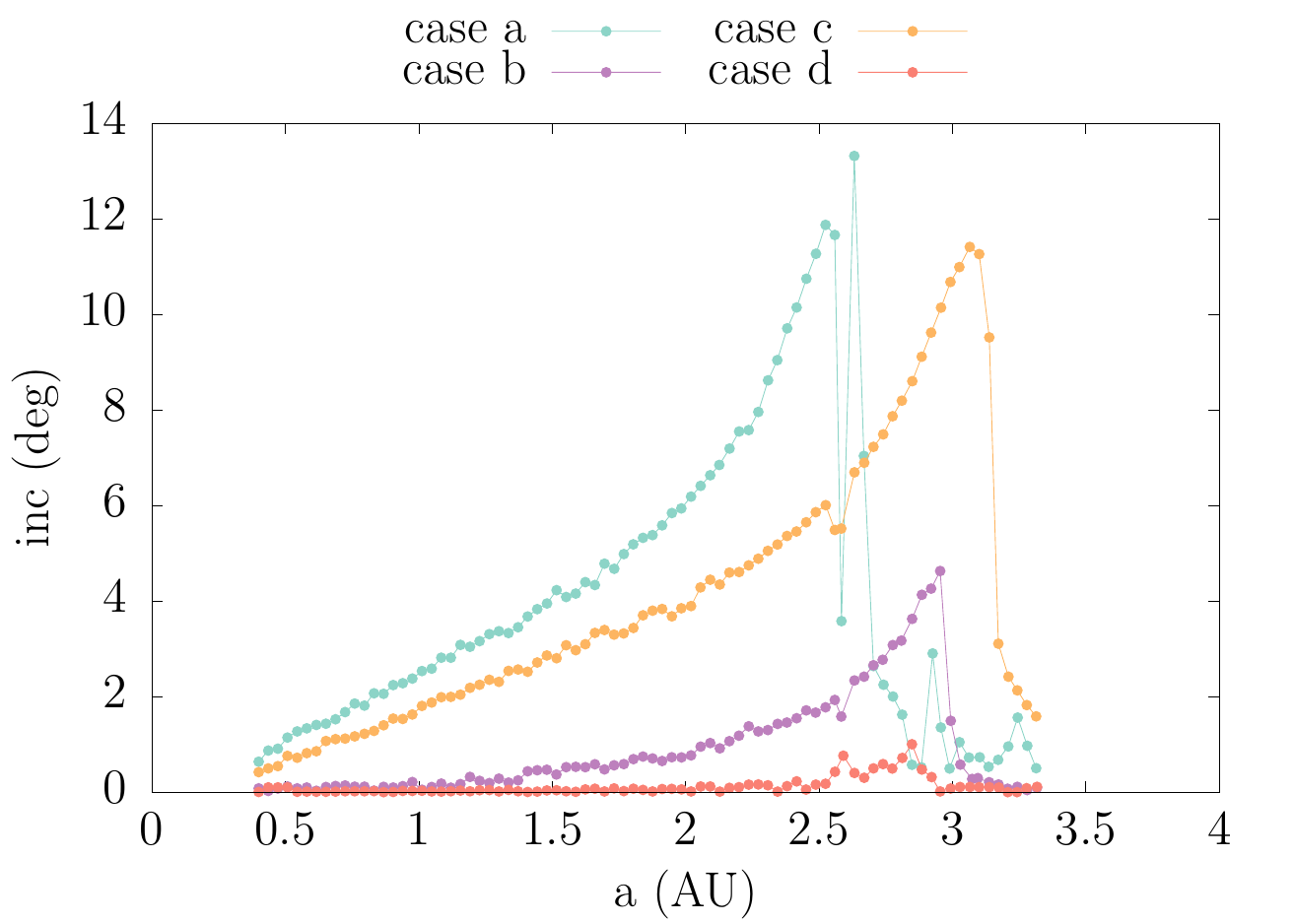}
\includegraphics[width=0.4\textwidth]{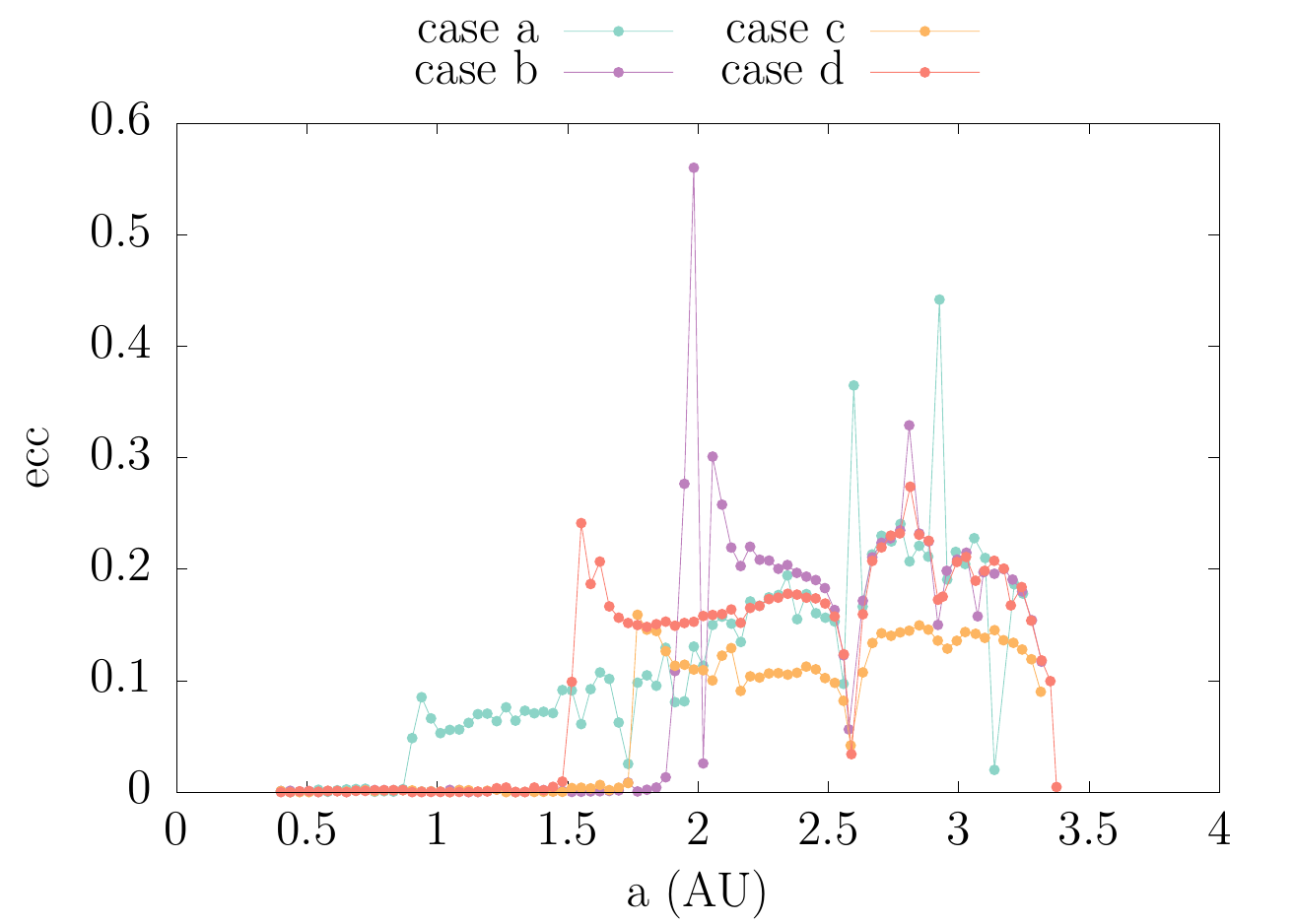}\includegraphics[width=0.4\textwidth]{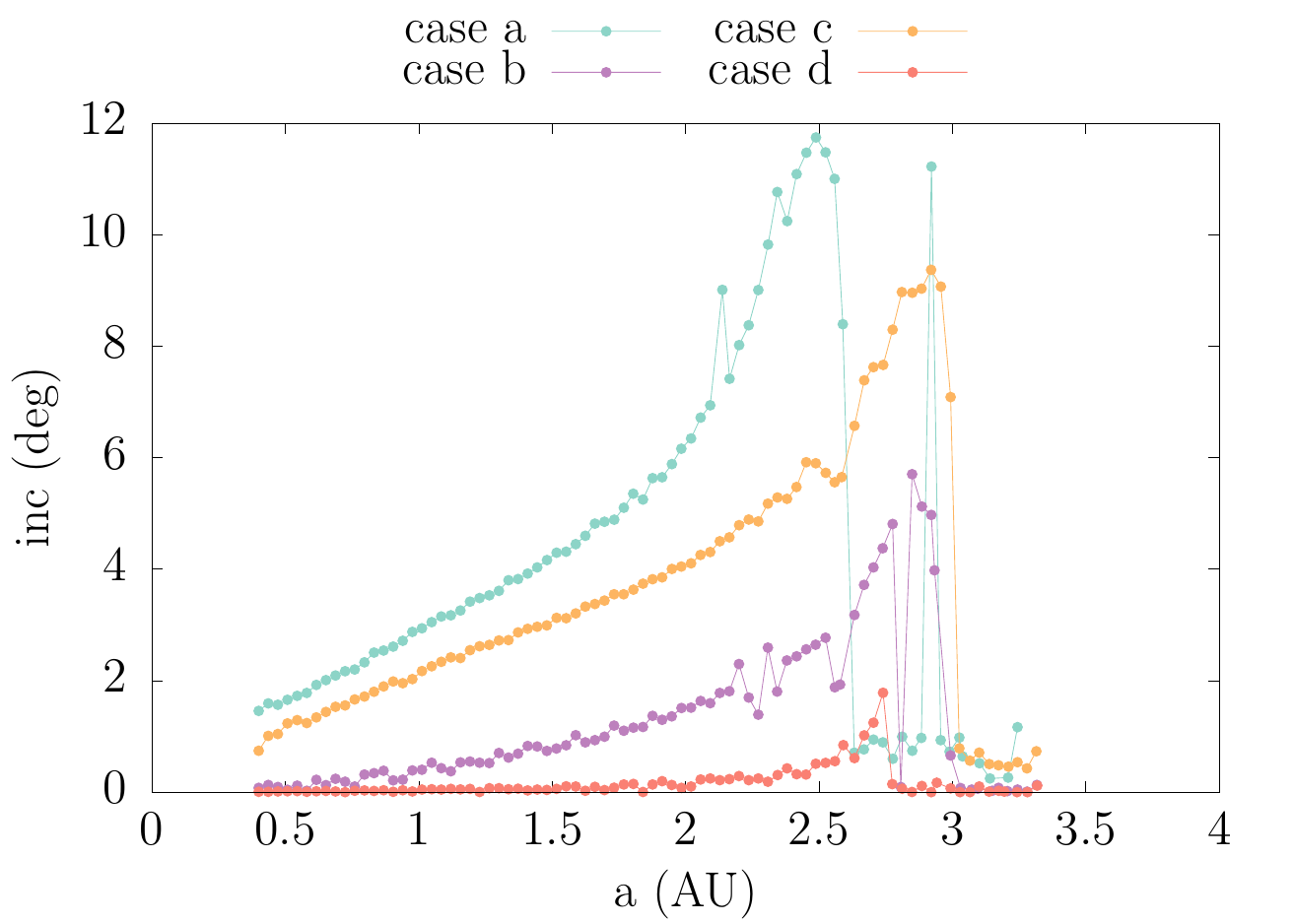}
\includegraphics[width=0.4\textwidth]{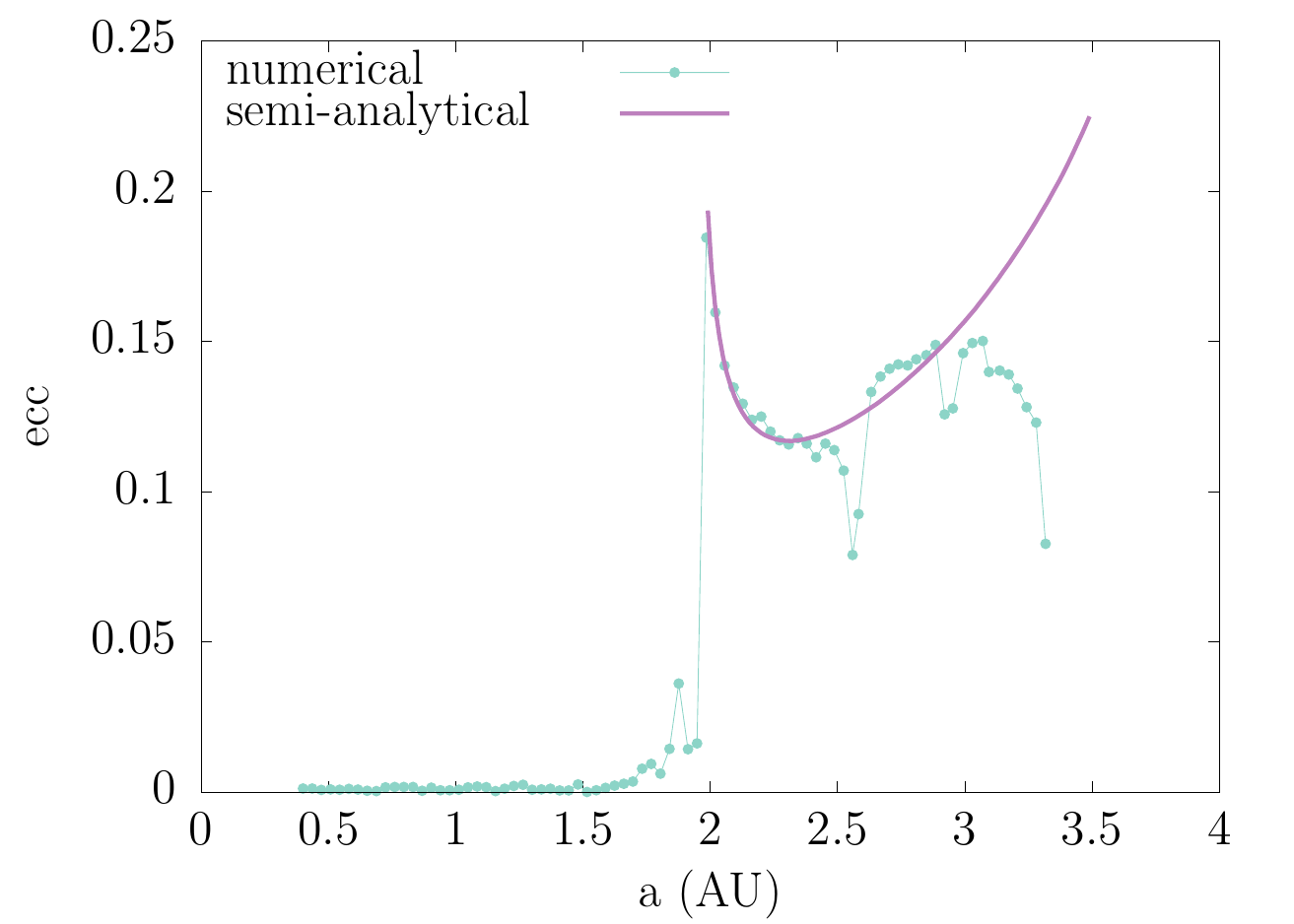}\includegraphics[width=0.4\textwidth]{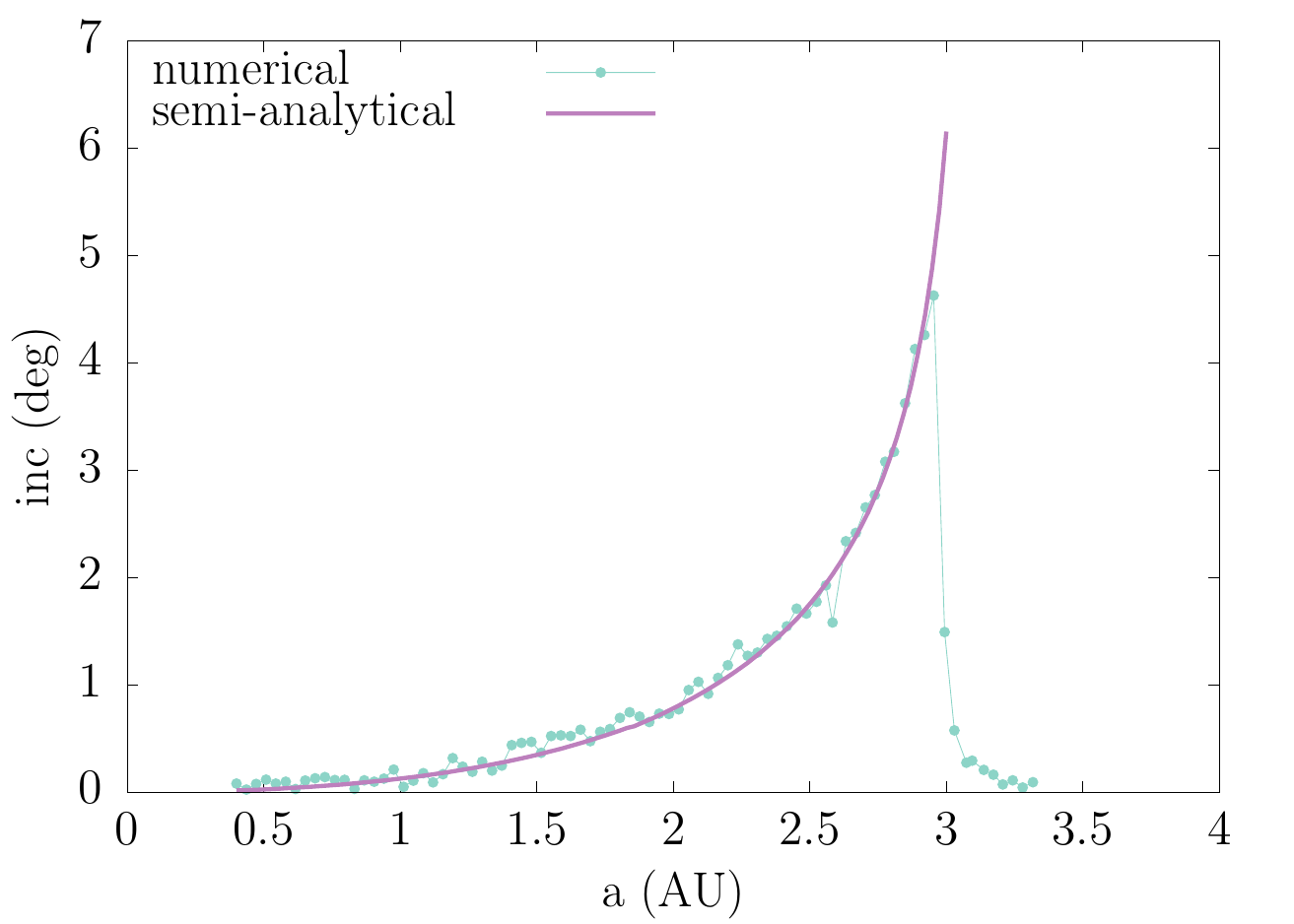}
\caption{The jumps in eccentricity $\Delta e$ and inclination $\Delta i$ suffered by the simulated particles for all four giant planets configurations and $\tau=1$~My (top) and $3$~My (middle). The last row focuses on Case (b), where fits, based on Equations (12) and (14), are superimposed.}
\label{fig:jumps-ei}
\end{figure*}

To check the effect of sweeping secular resonances in terms of excitation in $e$ and $i$ and test the validity of the simple analytical estimates given above, we studied 12 different systems. We considered Jupiter and Saturn to be near their 2:1 MMR, having (a) `high' eccentricities and `high' inclinations ($e_j\simeq0.025$, $e_s\simeq0.2$,  $i_j\simeq0.76^{\circ}$ and , $i_s\simeq2.1^{\circ}$), (b) `high' eccentricities and `low' inclination ($e_j\simeq0.015$, $e_s\simeq0.025$,  $i_j\simeq0.08^{\circ}$ and , $i_s\simeq0.2^{\circ}$), (c) `low' eccentricities and `high' inclinations ($e_j\simeq0.009$, $e_s\simeq0.014$,  $i_j\simeq0.76^{\circ}$ and , $i_s\simeq2.1^{\circ}$) and (d) `low' eccentricities and `low' inclinations ($e_j\simeq0.006$, $e_s\simeq0.01$,  $i_j\simeq0.01^{\circ}$ and , $i_s\simeq0.027^{\circ}$). Configuration (d) is, of course, closer to what simulations of resonance capture usually give. The giant planets, together with 100 particles that have initially circular and co-planar orbits, are embedded in a protoplanetary disk with $M=\mathcal{M}$, $\gamma=1$ and $R_c=15$~AU (with gaps). We used three realistic decay timescales: $\tau_1=1$~Myr, $\tau_2=2$~Myr and $\tau_3=3$~Myr and evolved the systems for 10 e-folding times. The time and magnitude of excitation in $e$ and $i$ was computed for each individual particle and are shown in \autoref{fig:jumps-ei}. 

As seen in \autoref{fig:jumps-ei}, more eccentric/inclined planets give larger jumps, as their forced elements are larger. The decay time scale does not seem to affect the results much, especially in $i$. Note that, in these models, secular resonances occur late, when the variation of the planetary frequencies is very small (i.e. $\lambda$ is very small) and a change by a factor of $\sim 2$ should not have a large effect. The change in $i$ is roughly monotonic with $a$ and the location of the peak in $\Delta i$ depends on the final value of the (common) $s_6$ frequency of the two planets. This value is $s_{6_a}=-44.9$~(``/yr), $s_{6_b}=-73.6$~(``/yr), $s_{6_c}=-68.4$~(``/yr), $s_{6_d}=-69.3$~(``/yr) respectively, which places the `finish line' of resonance crossing at $a_{f_a}=2.6$, $a_{f_b}=3.1$, $a_{f_c}=3.01$ and $a_{f_d}=3.03$~AU, respectively. These results are in excellent agreement with our semi-analytically derived resonance sweeping maps. In $e$, a nearly smooth variation with $a$ is interrupted at $\simeq 2.5$ and $\simeq 3$~AU, where mean-motion resonances affect particles more than the sweeping secular resonances. Let us also note that, during the first few e-folding times, the disk strongly perturbs the resonant planetary orbits, such that they exhibit chaotic behavior. This likely leads to an increased dispersal in $\Delta e$. \\

Also in \autoref{fig:jumps-ei} we present a comparison of our numerical results to the simple analytical estimates given above. Note that, in our case, the variation of the planetary frequencies is exponential and not linear, so Equations \ref{eq:Je} and \ref{eq:Pi} are not strictly valid. To overcome this issue we compute the derivative at the time of resonance crossing and assume this to be equal to $2 \lambda$. While the functional forms of these semi-analytically computed functions, $\Delta e (a)$ and $\Delta i (a)$, seem to fit well the numerically computed ones -- except of course near the major MMRs and the vicinity of Jupiter -- we had to actually multiply by a factor $\sim 2.5$-4 for the curve to fall on the data points. This may reflect a poor estimate of the forced amplitudes $\epsilon$ and $\epsilon '$ or of the local $\lambda$. However, we are pleased with the qualitative agreement, which actually allows us to use our semi-analytical sweeping maps and quasi-linear estimates to `predict' the time, the location and (within a small factor) the magnitude of $\Delta e$ and $\Delta i$, in different disk models and planetary configurations. This, however, holds for a uniformly decaying disk.

\section{Results on photoevaporating disks}\label{sec:photoevaporation}

Recent astrophysical models \citep{2001MNRAS.328..485C, 2006MNRAS.369..216A, 2006MNRAS.369..229A} focus on the competition between the slow viscous evolution and accretion of the disk on the star and its tendency to deplete swiftly by photoevaporation, as UV photons and X-rays ionize and heat the inner part of the disk, forcing it to escape as wind. Initially, accretion wins. However, sooner or later, a ``hole'' opens at a specific location in the disk, where the accretion and escape rates are equal, since the inner disk cannot be supplied anymore with gas form the outer parts. In the Solar System, this critical radius was at $\approx 2~$AU. \citep{2010MNRAS.401.1415O}. Once the gap opens, the dispersal timescale of the disk depends solely on photoevaporation. The inner part accumulates on the Sun within $\sim 10^5$~yr and the outer disk, exposed to high-energy photons, empties from the inside out. \citet{2011MNRAS.412...13O} suggest that the total time taken from the formation of the gap until the full depletion of the disk amounts to $10-20\%$ of the total disk lifetime.

\begin{figure*}
 \centering
 \includegraphics[width=0.45\textwidth]{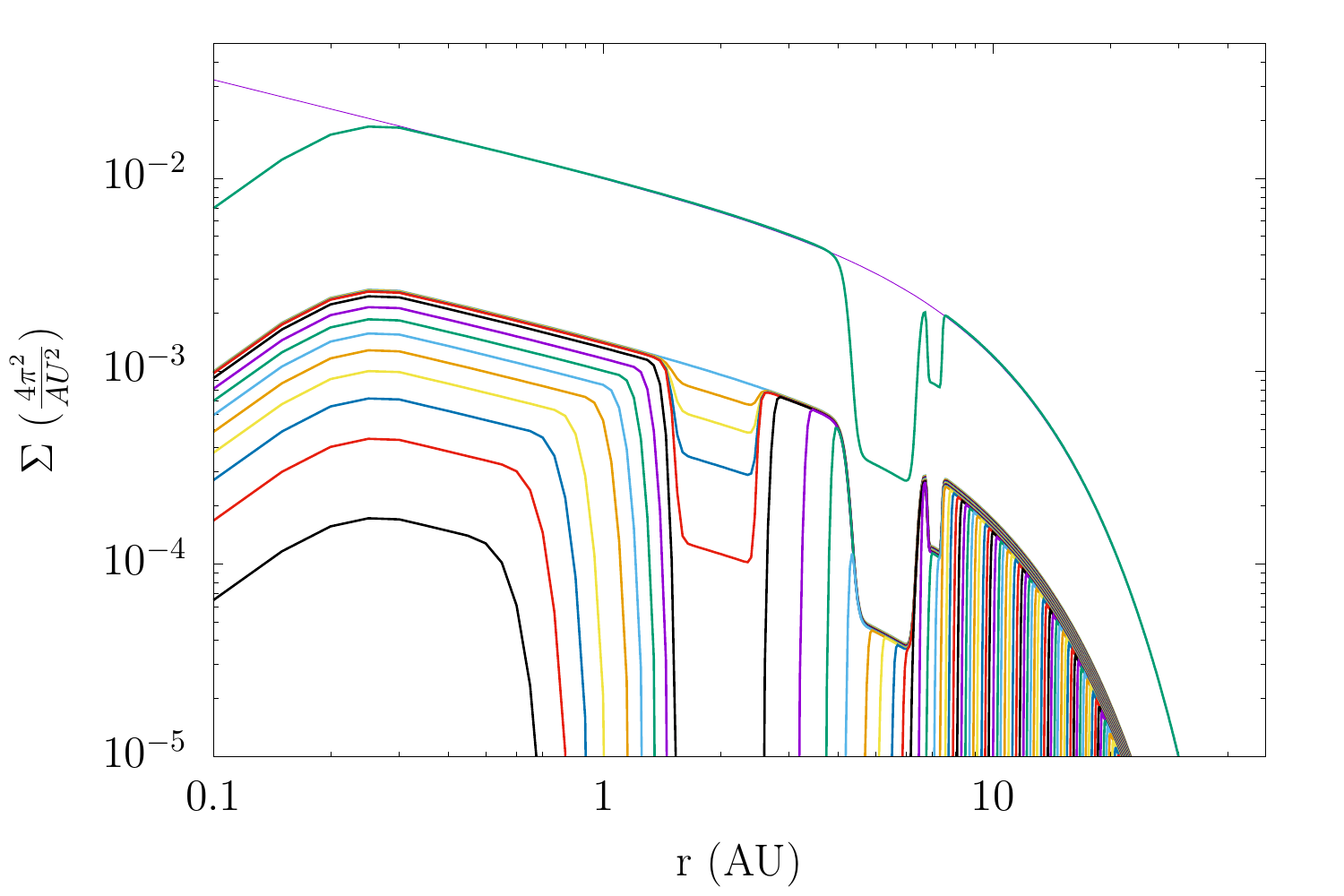}
 \caption{The surface density of a disk, evolving under the effect of photoevaporation. The upper line corresponds to the unperturbed $\Sigma$ profile. The second line represents the profile, modified by the presence of the giant planets (here in the 3:2 MMR). The remaining lines show $\Sigma(r,t)$ after the moment that the gap opens at 2~AU and are separated by $10^4$~yr.}
\label{fig:sigmas_pe}
\end{figure*}

We model the photoevaporation of a disk, by manipulating its surface density profile, according to the time scales discussed above (see \autoref{fig:sigmas_pe}). We construct numerically three sets (one for each giant planet configuration assumed) of 1,500 snapshots of the disk profiles, separated by $10^3$~yr. According to the models presented above, we assume that the gap at 2~AU starts forming at $t_{\text{pe}}=2\cdot \tau$ and so, assuming $\tau=2$~Myr, the clearing of the disk begins at $t=4$~Myr. We compute the acceleration of a body in our modified SyMBA, using linear interpolation for $\rho(r,z)$, between the recorded successive snapshots. To be more consistent with recent models of giant planet formation in the Solar System \citep{2015Icar..258..418M}, we adopt a shallow density profile with $\gamma=0.5$, such that specifically $\Sigma(1~AU)=2400~\text{gr}/\text{cm}^2$. \\

\begin{figure*}
\includegraphics[width=0.33\textwidth]{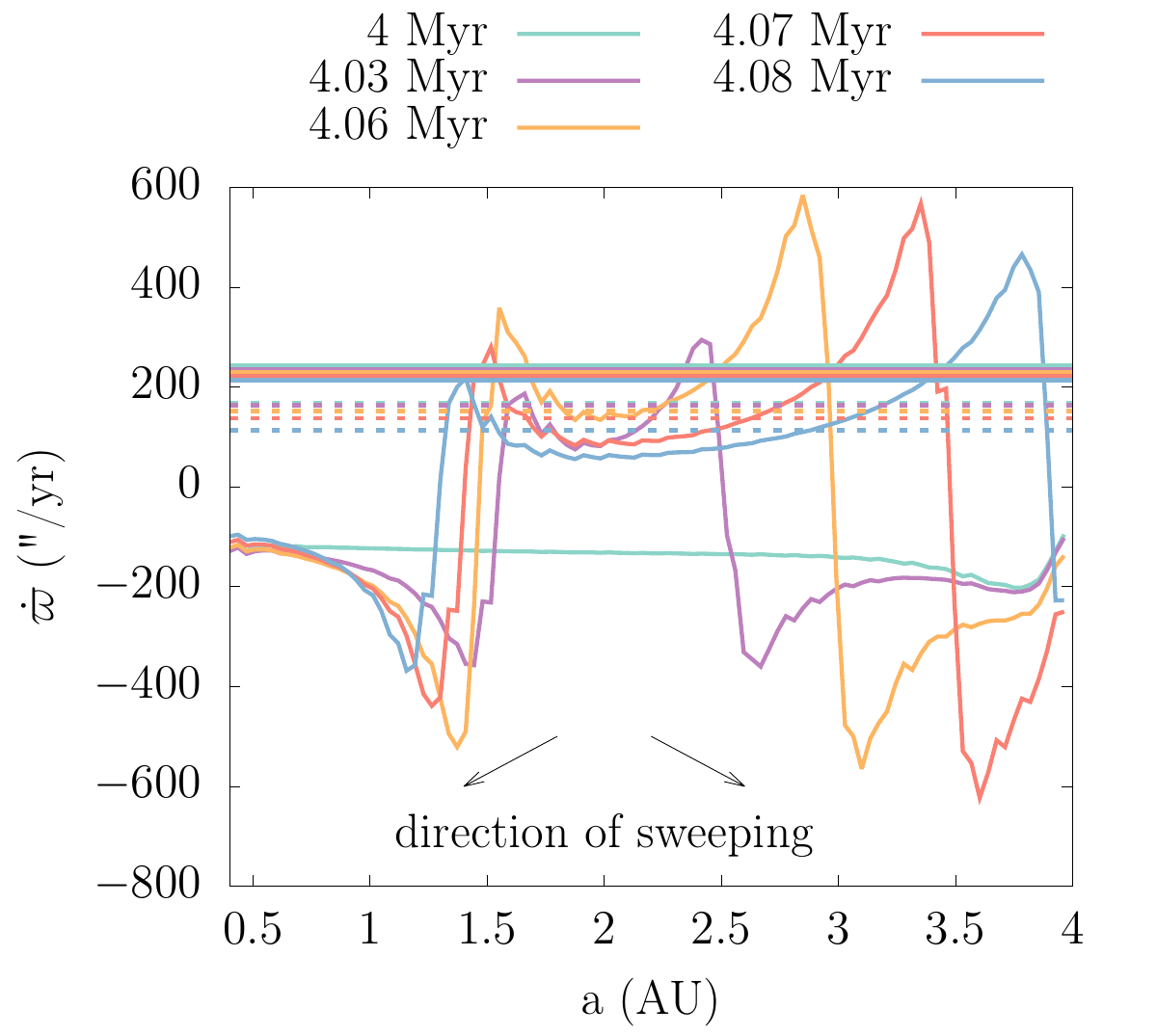}\includegraphics[width=0.33\textwidth]{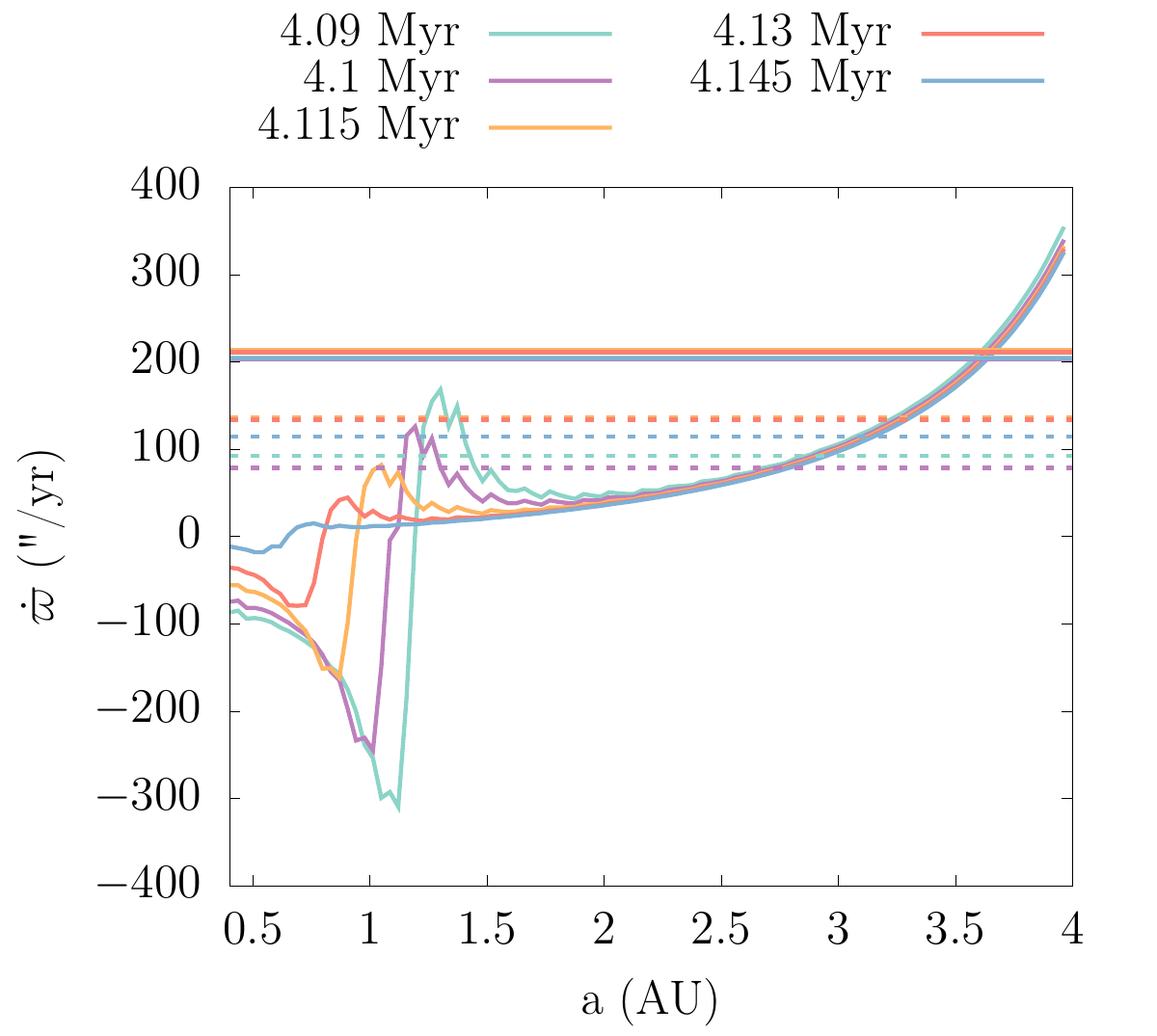}\includegraphics[width=0.33\textwidth]{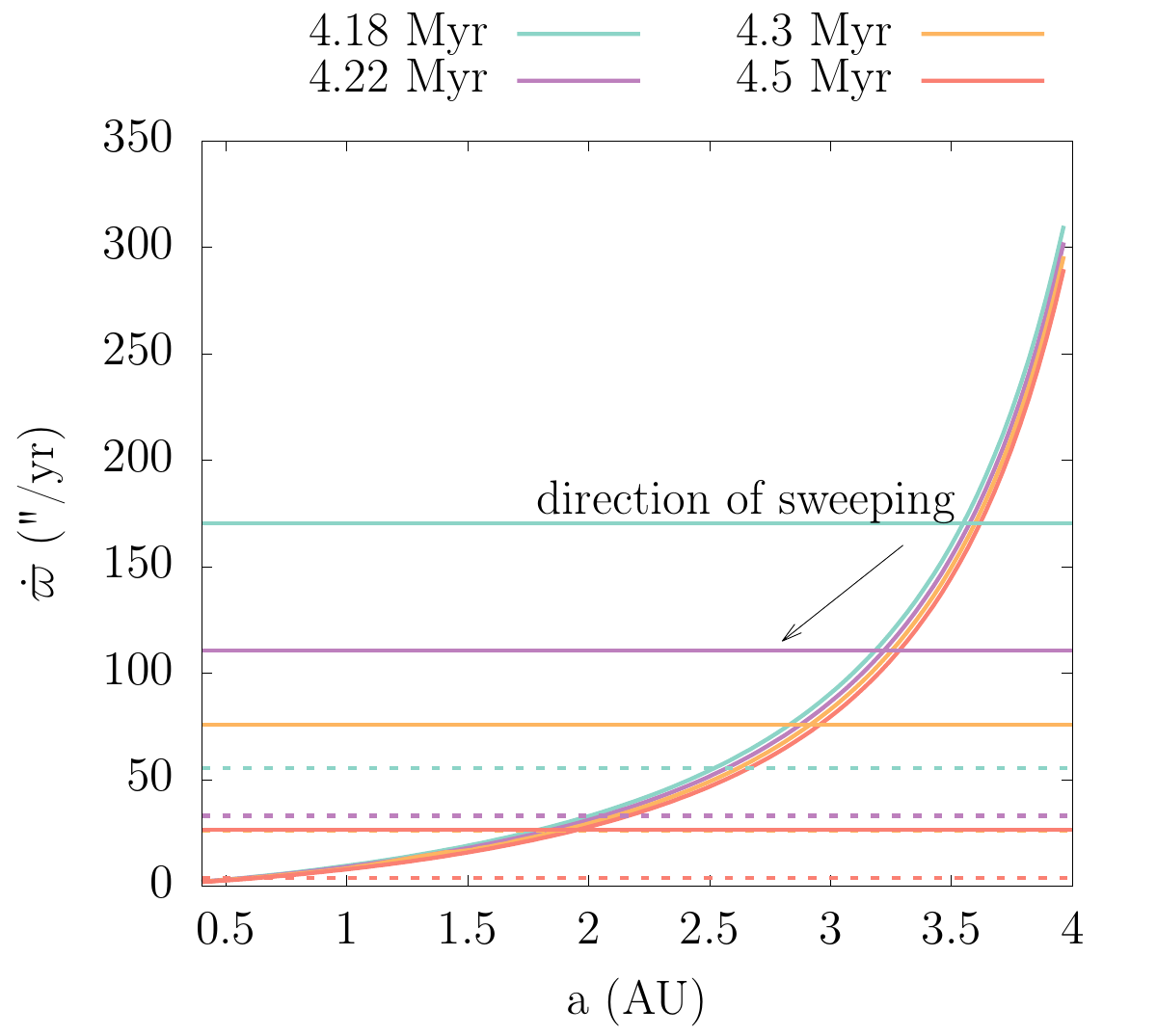}
\caption{Maps of secular resonance occurrences for the pericentric frequency (Jupiter and Saturn on their current orbits). (top) first stages of photoevaporation of the disk (4-4.08~Myr). The sweeping proceeds outwards for $a>2$~AU and inwards for $a<2$~AU. (bottom) Last stages of evolution (4.18-4.5~Myr), when the sweeping proceeds inwards. (middle) reversal in the direction of sweeping, occurring for $t\sim $4.09-4.145~Myr. The dashed line corresponds to the $g_5$ frequency and the solid line to the $g_6$ frequency.}
 \label{fig:maps-g-insitu_pe}
\end{figure*}

\begin{figure*}
\centering
\includegraphics[width=0.4\textwidth]{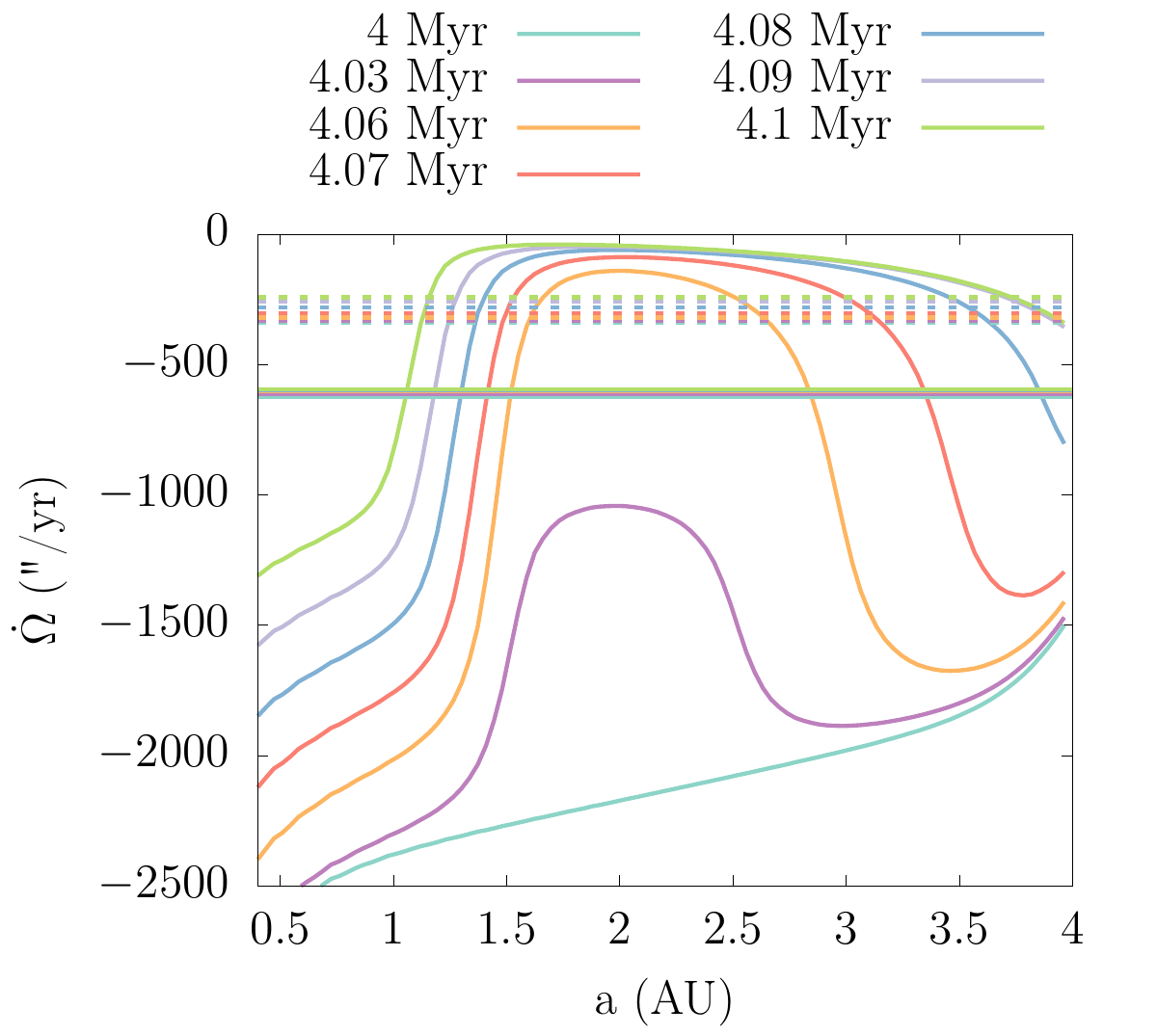} \includegraphics[width=0.4\textwidth]{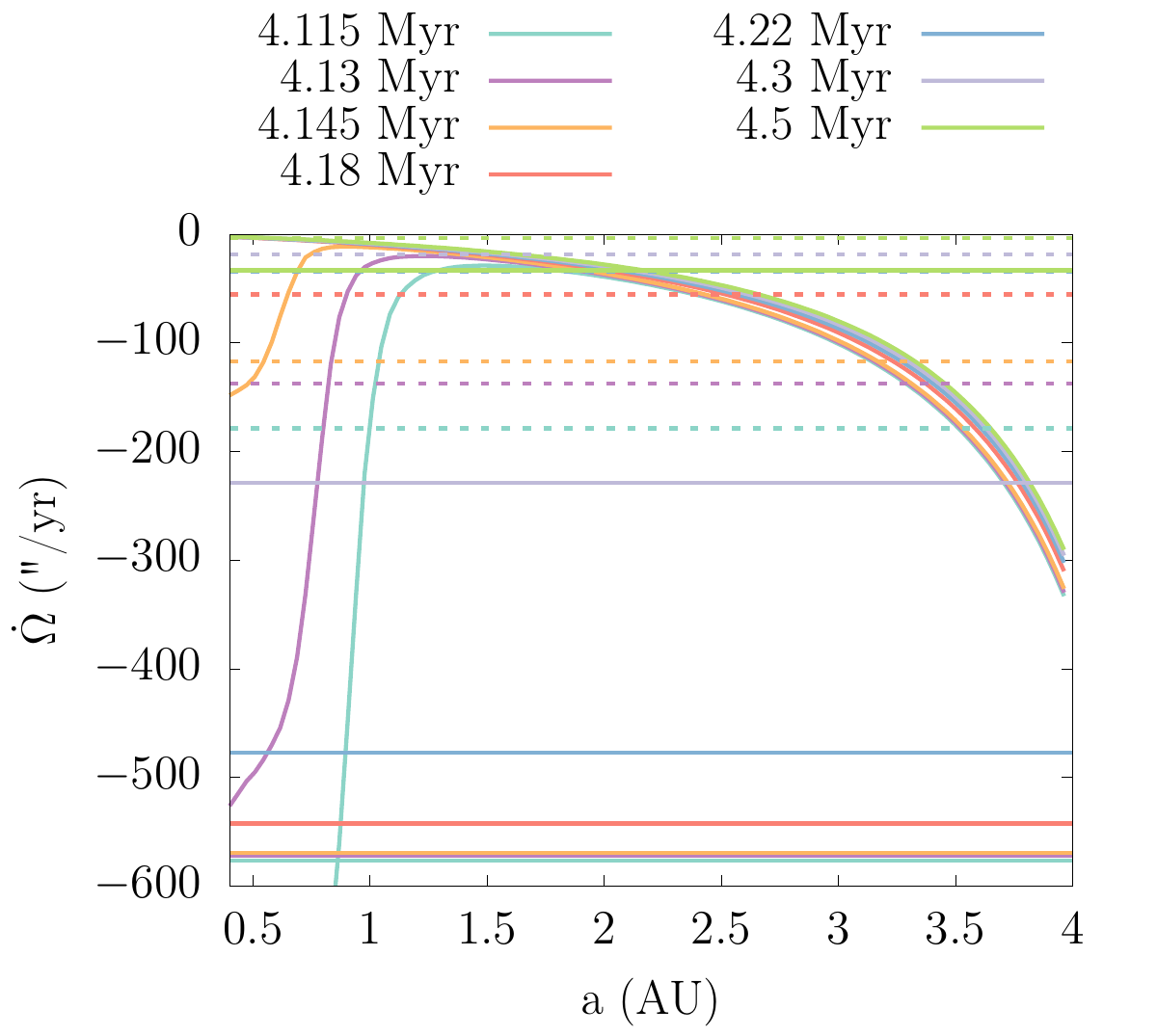}
\caption{Maps of secular resonance occurrences between the $s$-frequencies of the particles and those of the giant planets, that have orbits like their current ones. The top panel shows the crossing of the resonances during the first stages of the photoevaporation of the gas disk (4-4.1~Myr) and the bottom panel during the last stages (4.115-4.5~Myr). The dashed line corresponds to the $s_5$ frequency and the solid line to the $s_6$ frequency.}
 \label{fig:maps-s-insitu_pe}
\end{figure*}

\begin{figure*}
 \includegraphics[width=0.33\textwidth]{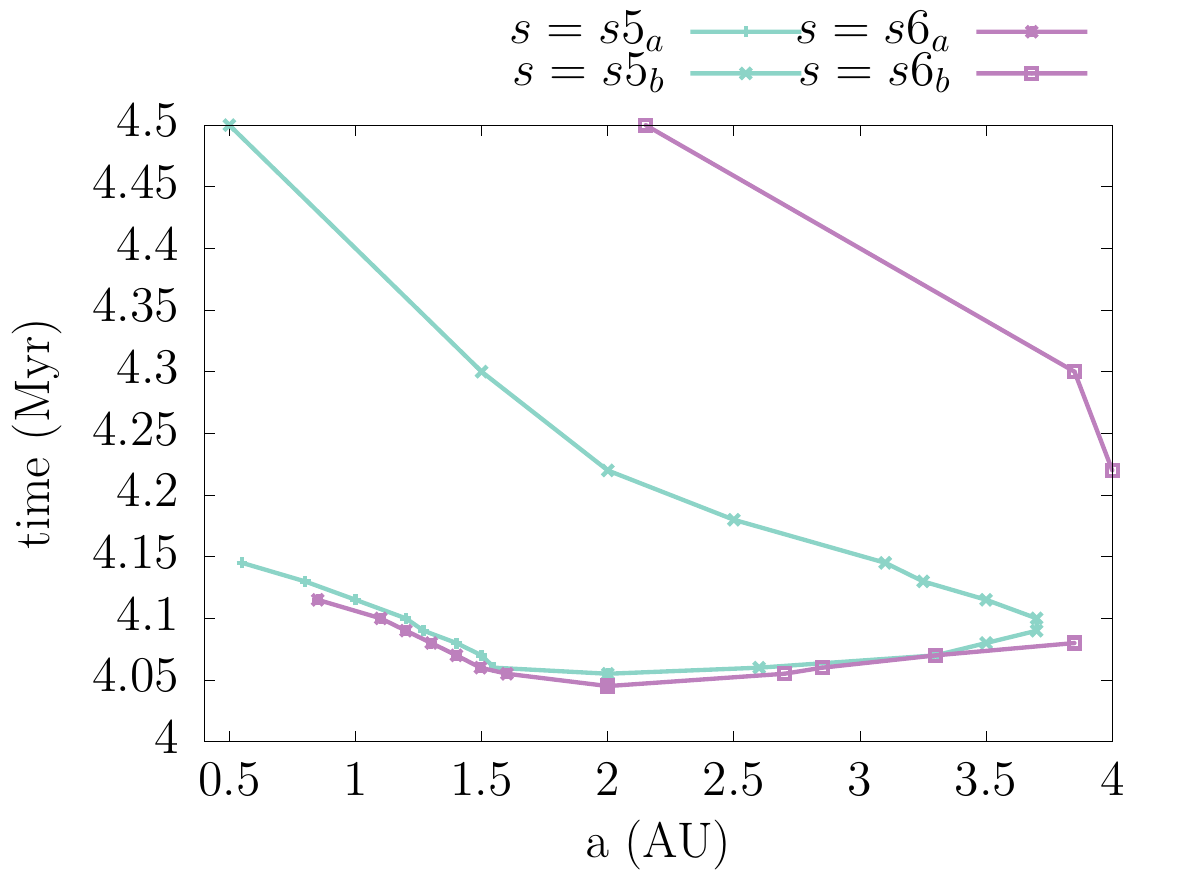}\includegraphics[width=0.33\textwidth]{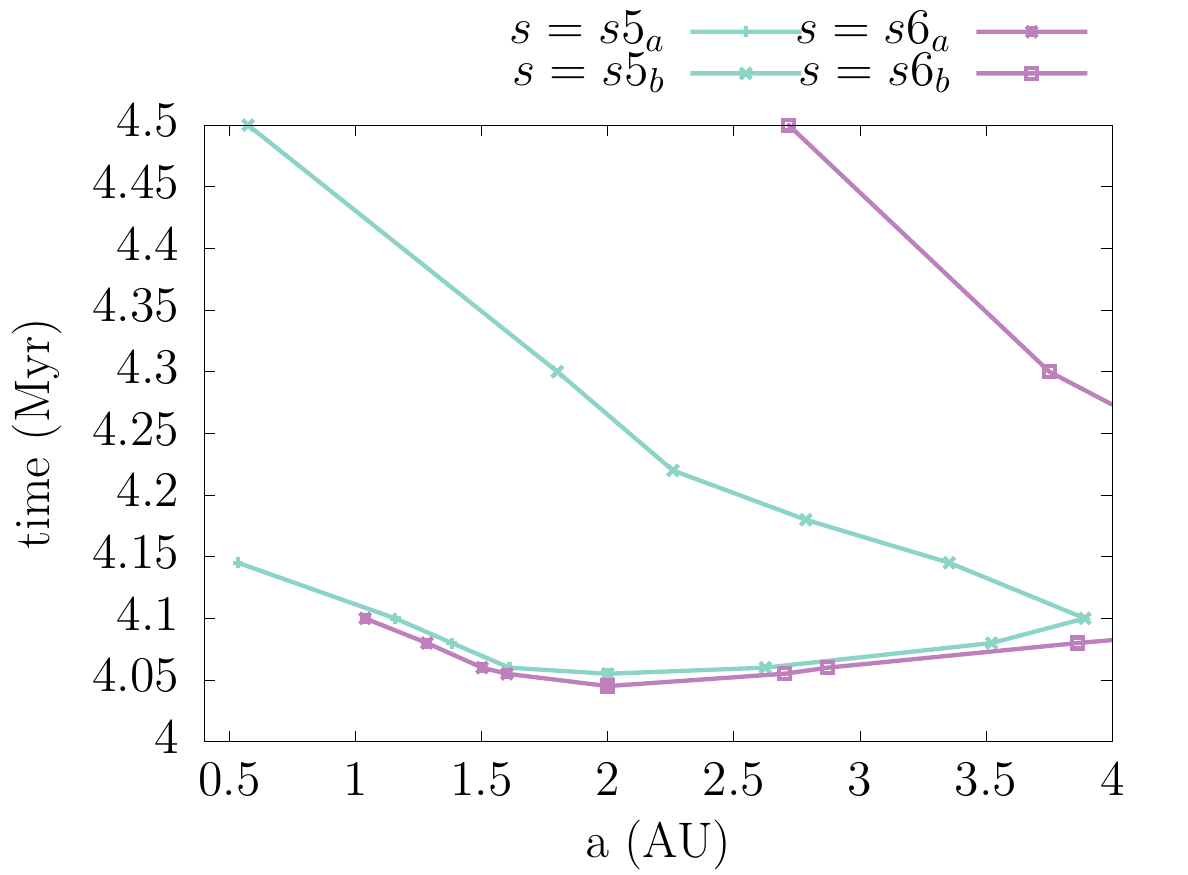}\includegraphics[width=0.33\textwidth]{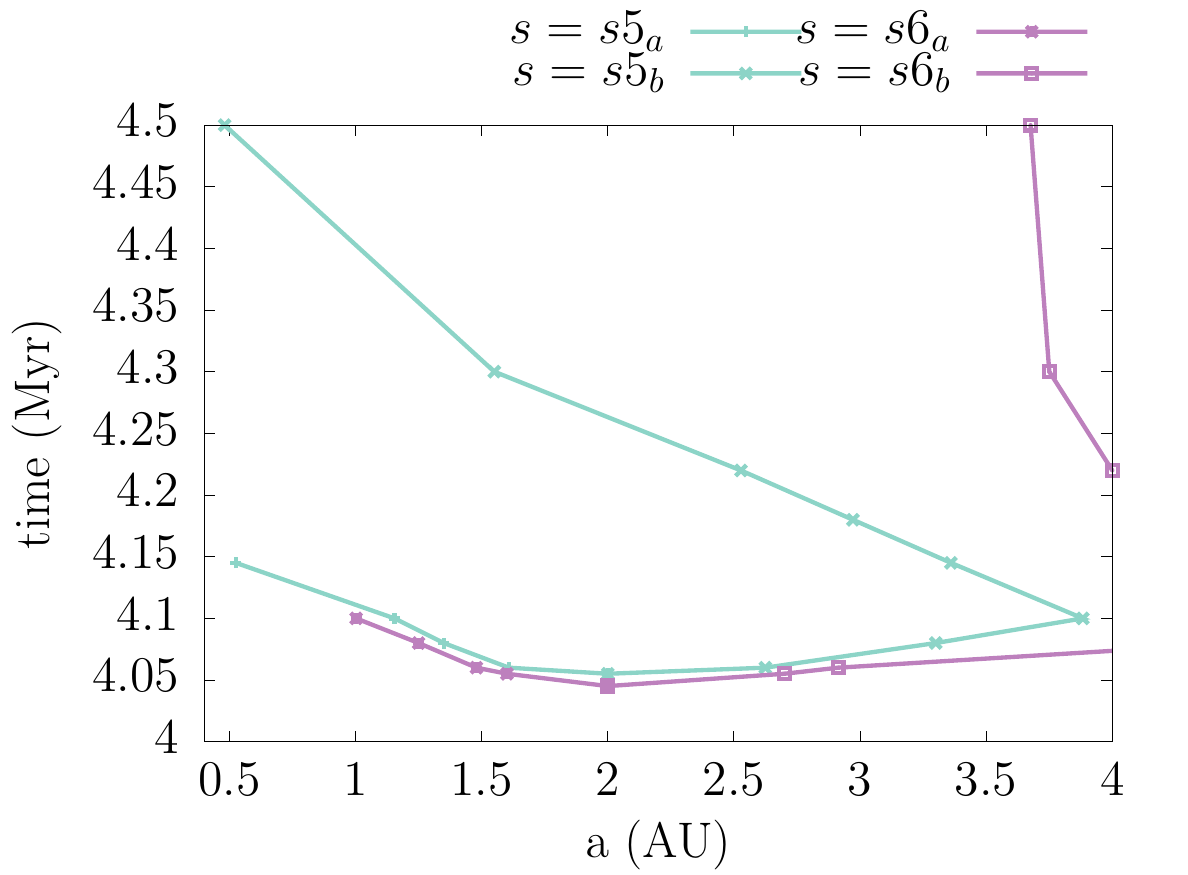}
\caption{Paths of the $s=s_5$ and $s=s_6$ secular resonances, for the same setting as in \autoref{fig:maps-g-insitu_pe}, for Jupiter and Saturn in their current positions, and captured in a 2:1 and 3:2 MMR. The ``a'' and ``b'' subscripts denote the two intersection points of each resonance.}
 \label{fig:crossing-s-all_pe}
\end{figure*}

Unfortunately, in the photoevaporating case, we can no longer have a simple semi-analytical model to follow the paths of secular resonances, as the contribution of the disk does not decay uniformly. Hence, we have to compute by FFT the precession frequencies of both planets and asteroids for every snapshot of the disk's evolution, which means integrating the system of planets and asteroids for 4,500 different `static' disks. The corresponding resonance occurrence maps and sweeping paths are shown in Figures \ref{fig:maps-g-insitu_pe}, \ref{fig:maps-s-insitu_pe} and \ref{fig:crossing-s-all_pe}. 

It is evident that all resonances occur and sweep through the inner Solar System, as the disk is depleted. The frequency functions are not so smooth anymore, given the way we are forced to follow the non-unifrom depletion of the disk. However, we can still deduce from \autoref{fig:maps-g-insitu_pe} that both resonances spread from $\sim2$~AU inwards and outwards, as the disk evolves. After the outer branches of the resonances sweep through the main belt and beyond, at $t\approx 4.08$~Myr, they reverse their sweeping direction and start moving again inwards, crossing the belt for the second time.

The same picture -- though, with smoother curves -- emerges also for the $s=s_5$ and $s=s_6$ secular resonances. While one branch moves inwards as time progresses, the other one moves outwards, but at some point in time, it switches direction and sweeps through the main belt again. In terms of orbital excitation, this is a favorable situation, given that the first sweep of the resonance can excite a dynamically ``cold '' distribution of asteroids, while the second sweep can ``stretch'' the distribution to both lower and higher values, depending on the secular phase. In \autoref{fig:crossing-s-all_pe} we can see precisely paths of the $s=s_i$ secular resonances, for the three different configurations of the giant planets.

\begin{figure*}
\centering
 \includegraphics[width=0.4\textwidth]{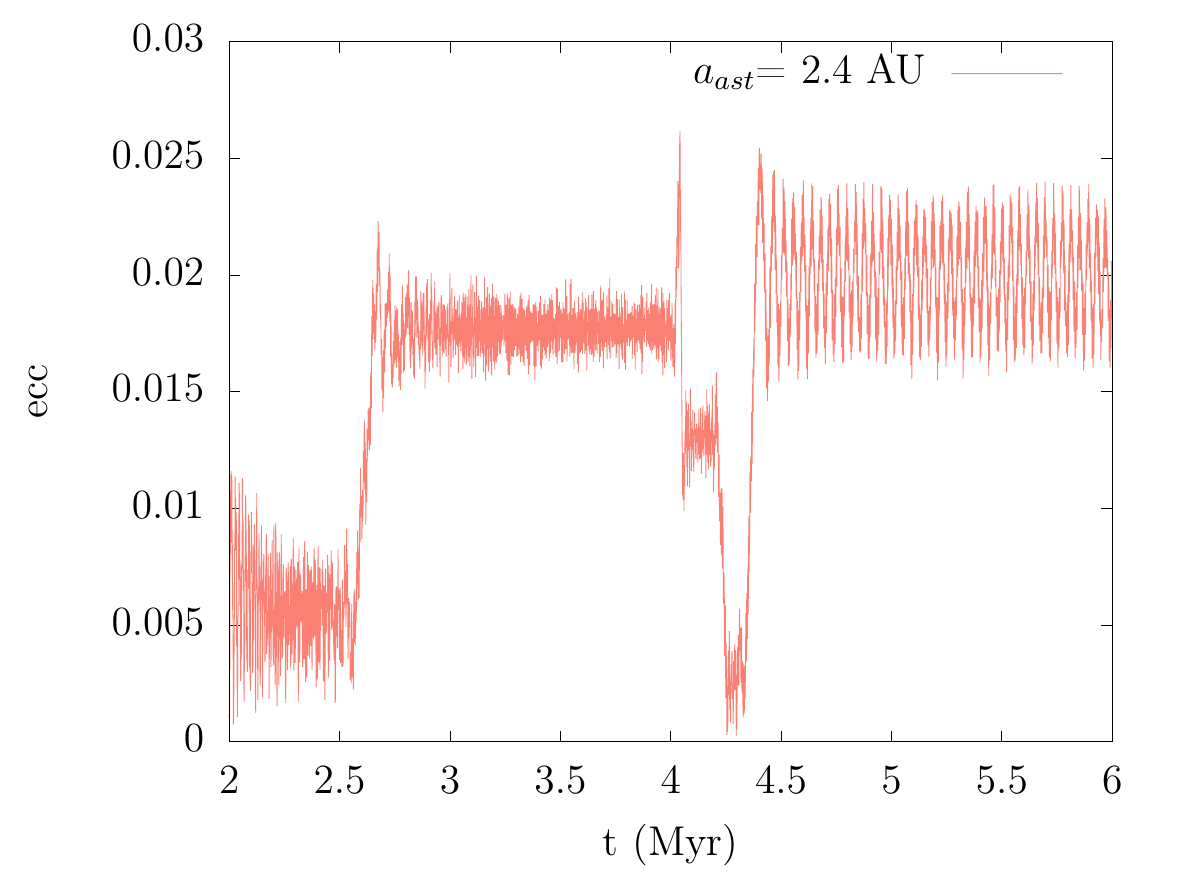}\includegraphics[width=0.4\textwidth]{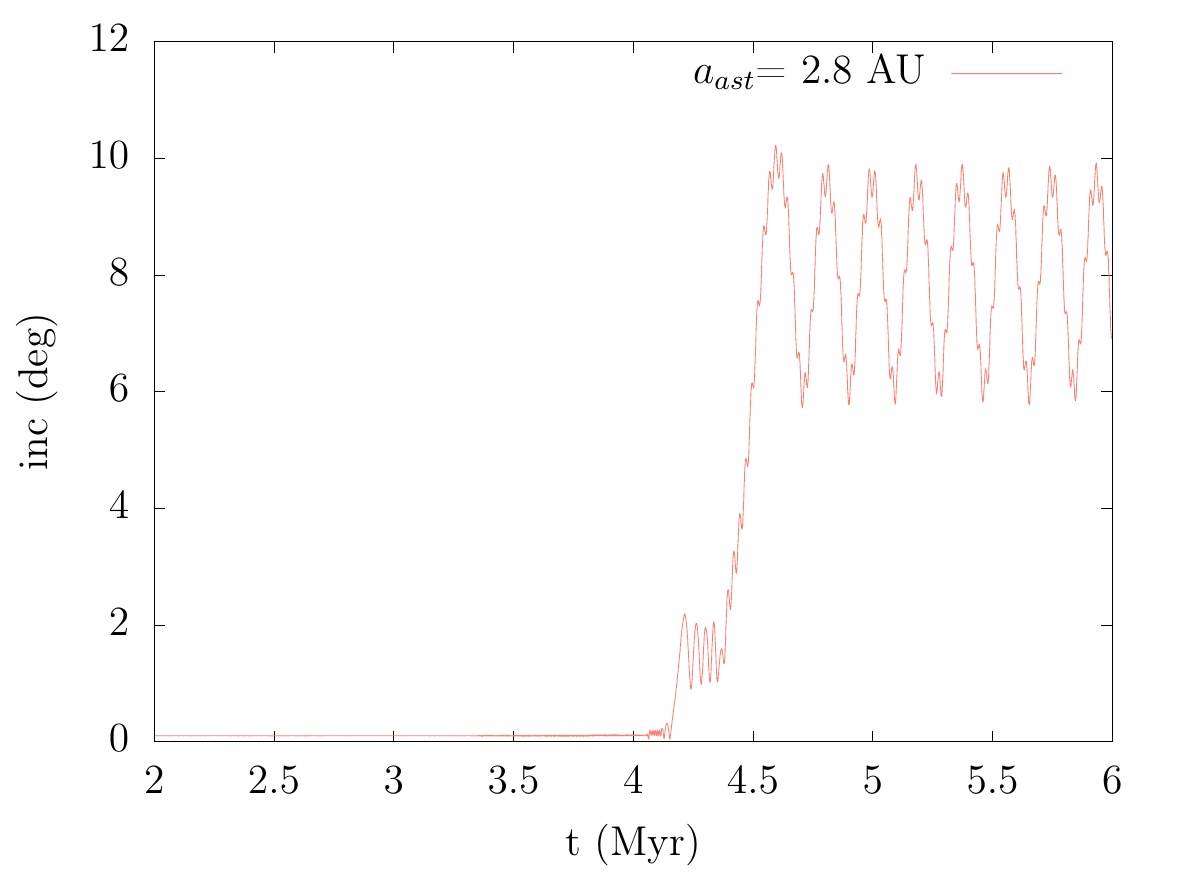}
 \caption{The evolution of eccentricity and inclination of two particles in the simulation of a photoevaporating disk.}
 \label{fig:particles2}
\end{figure*}

\begin{figure*}
\centering
\includegraphics[width=0.5\textwidth]{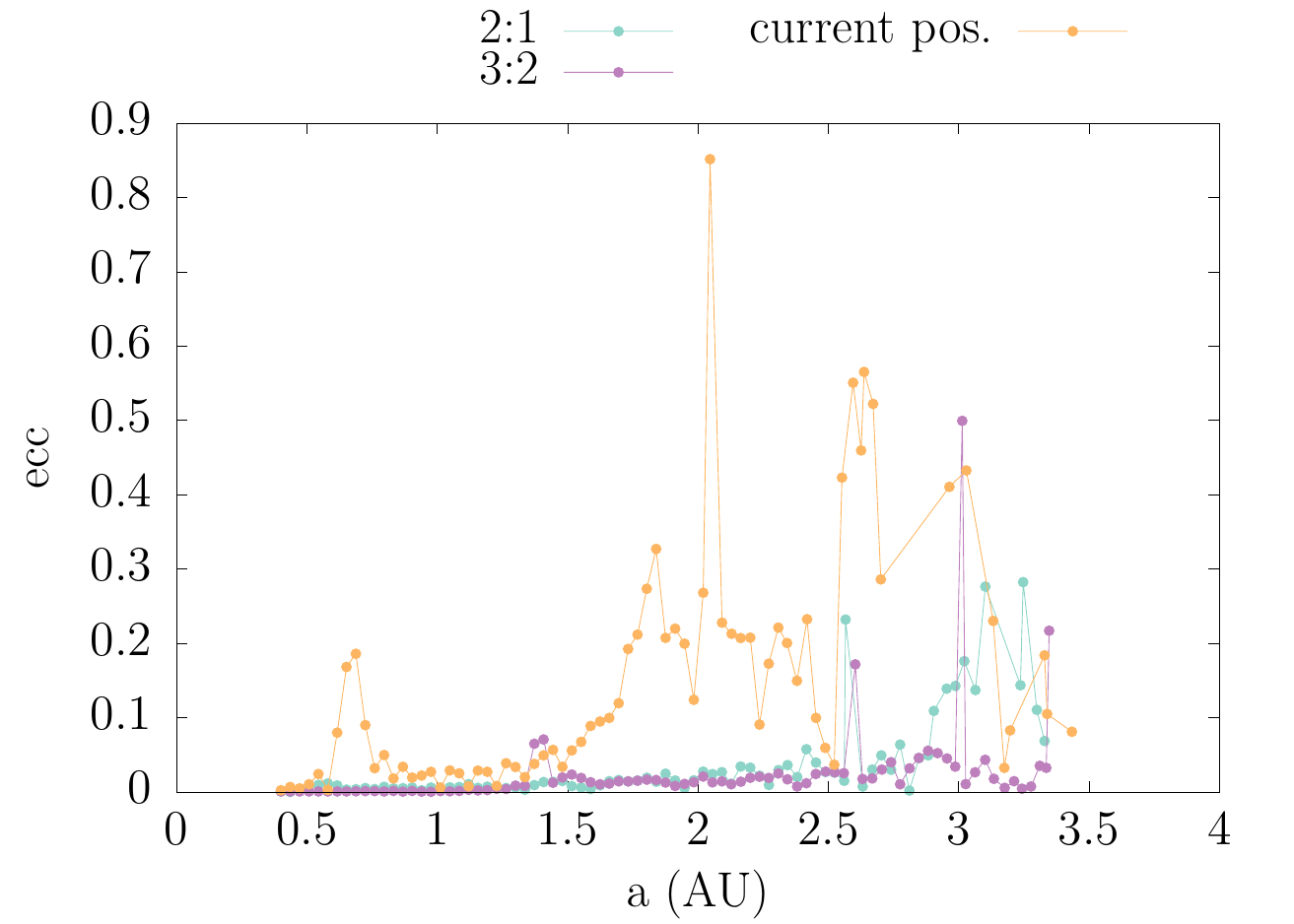}\includegraphics[width=0.5\textwidth]{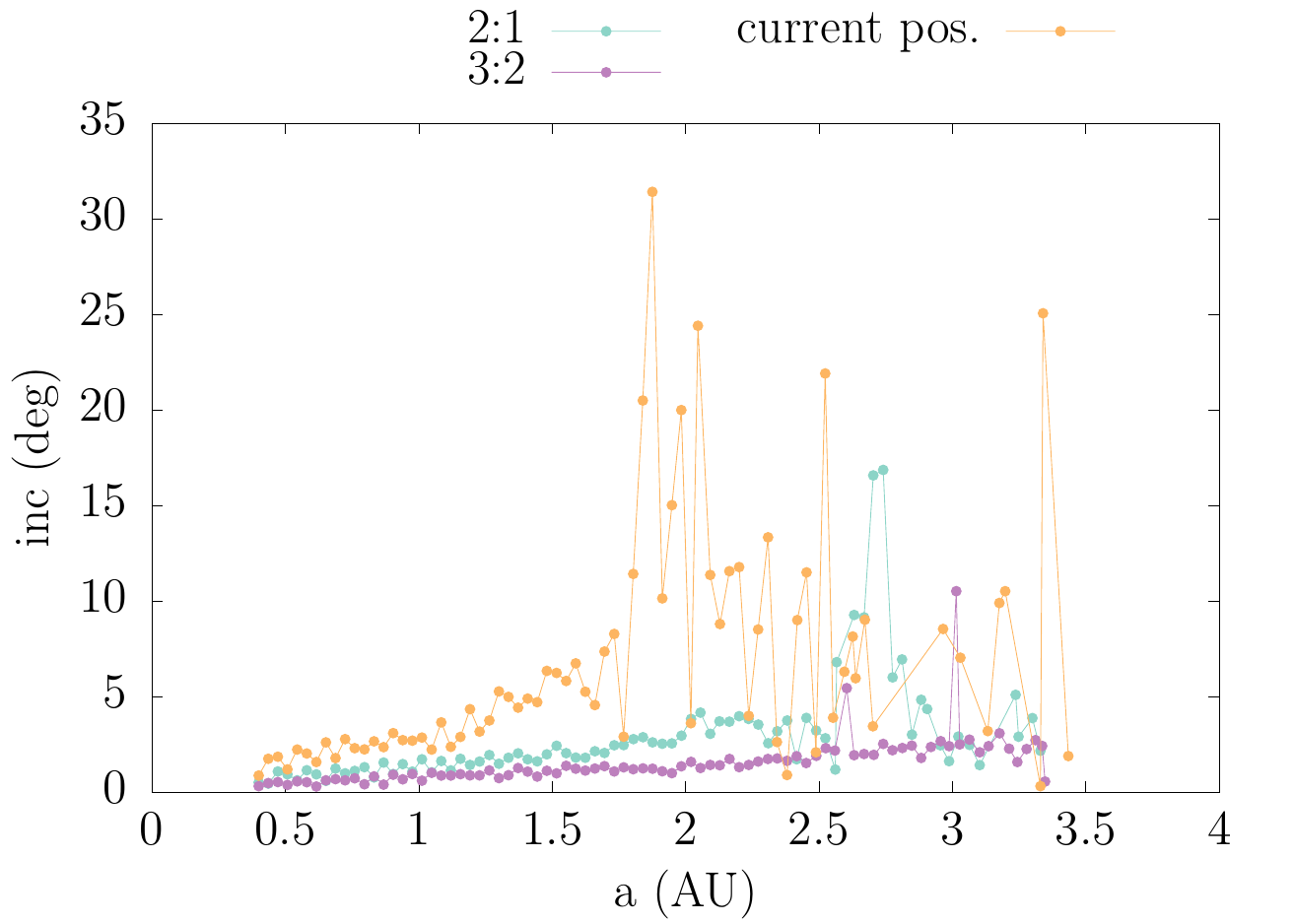}
\caption{The jumps in eccentricity $\Delta e$ and inclination $\Delta i$ suffered by the simulated particles for three giant planets configurations (2:1, 3:2 and in their current positions) for a photoevaporating protoplanetary gas disk.}
\label{fig:jumps-ei_pe}
\end{figure*}

We integrated our usual 100 test particles in all three giant planet configurations, adding the effect of the non-uniformly decaying disk, computing again the excitation in eccentricity and inclination for each particle. Two examples of evolution are shown in \autoref{fig:particles2}, where multiple resonance crossings can be deduced from the consequtive jumps. The final states for all particles in all three simulations is given in \autoref{fig:jumps-ei_pe}. In the previous section we saw how the final position of the $s=s_6$ secular resonance, depends on the final common $s$-frequency of the two giant planets, which implied crossing of the complete belt in a 3:2 MMR configuration of Jupiter and Saturn. However, in the photoevaporating case studied here, we see that the reversal in the direction of the sweeping of the $s=s_5$ and $s=s_6$ resonances works in favor of giant planet configurations that have lower frequency values. Consequently, the inclination excitation is larger in the current orbital configuration of Jupiter and Saturn.

\section{Conclusions and Discussion}\label{sec:concl}

We have revisited the problem of secular resonance sweeping in a decaying protoplanetary disk and its effect on the dynamics of a pair of giant planets and test particles. The motivation behind our study came from recent developments in solar system evolution models (dynamics) and the related astrophysical context (protoplanetary disks). More specifically, this paper contributes to the understanding of the sweeping process and its effects on the orbital excitation of an initially flat disk of circular orbits -- the presumed state of the primordial asteroid belt -- assuming minimal orbital migration of the giant planets during this stage of Solar System formation. The methods used here can be readily applied and the results easily generalized for the dynamics of primordial KBOs or (near-)resonant extrasolar systems with two giant planets or more.\\

We studied a large number of models (planetary configurations, disk profiles, depletion scenarios), which allows us to reach some general conclusions. To achieve this, we first developed a numerical code that computes the gravitational acceleration, experienced by a planet or test-particle, from a uniformly or not decaying, massive disk of arbitrary surface density profile. Note that the axisymmetric conditions can essentially be relaxed, when we use the time interpolation feature of the code, through a series of snapshots of the disk. We tested our code and confirmed that linear secular theory can be used, together with a model of time evolution of the fundamental frequencies, to derive a simple semi-analytical model that tracks the location and occurrence time of secular resonances. An important element of this theory is the existence of the $s=s_5$ secular resonance; a direct consequence of the fact that the disk essentially defines the invariant plane. The deviations of our semi-analytical model from an accurate numerical model are minor. However, in non-uniform depletion models, the semi-analytical approach can no longer be used.\\

In the uniform depletion (exponential) models, we found that no inclination excitation would occur on the region of today's asteroid belt, if Jupiter and Saturn always occupied their current orbits. On the other hand, a resonant planetary configuration with a low $s_6$ frequency (e.g.\ 3:2 MMR) would lead the $s_6$ resonance to sweep through the entire belt, leading to excitation that could exceed $\sim 10^{\circ}$ in the outer parts. Conversely, only the $g_6$ resonance would cross the belt in resonant planetary configurations, as opposed to their current orbital configuration. However, an excitation of $\Delta e\sim 0.1-0.15$ is easily achievable in all cases.\\

In the non-uniform depletion scenario (disk photoevaporation) secular resonances follow a more complex and more interesting evolution. Because of the cavity that develops in the disk around $2~$AU -- a number that depends on the properties of the host star and the disk -- and the subsequent outwards depletion of the disk, secular resonances can sweep throughout the belt several times. In particular, both inclination resonances can sweep the belt twice, which could produce a larger spread in inclnation; again, this depends on the planetary configuration through the final value of $s_6$. \\

Note that, in all scenarios, a flat disk of asteroids cannot be excited such that it develops a nearly uniform distribution of inclinations with a spread $> 20^{\circ}$, which is the current state of the asteroid belt. However, our model still does not take into account important dynamical aspects of the early stages of solar system formation, most notably the process of terrestrial planets formation itself. At that time, a large number of planetary embryos would have co-existed with primordial asteroids in the belt. Scattering of asteroids by embryos, which would also excite each other and likely migrate radially in the depleting disk, would certainly be a ``game-changer''. Numerical simulations show that scatterring by Mars-sized embryos results into a uniform spread of at least $\sim 5^{\circ}$ in inclination, which would in turn imply a larger spread upon resonance sweeping, as indicated even by the simple analytical estimate given in Section \ref{sec:photoevaporation}. Now that we have developed the necessary tools and understood the basic dependences of resonance sweeping to model parameters, we intend to report soon the interplay of these processes, in an effort to produce a final verdict on the relevance of secular resonance sweeping in the excitation of the asteroid belt.


\bibliographystyle{spbasic}


\end{document}